\pdfoutput=1

\documentclass[preprint,3p]{elsarticle}

\usepackage{lineno,hyperref}
\usepackage{bm}
\usepackage{amsmath}
\usepackage{mathtools}
\usepackage{color}
\usepackage{algorithm}
\usepackage{algorithmic}
\usepackage{multirow}
\modulolinenumbers[5]

\usepackage{amssymb}

\usepackage{subfigure}
\usepackage{array}
\newcolumntype{C}[1]{>{\centering\arraybackslash}p{#1}}

\usepackage{bigints}

\usepackage{bbding}

\usepackage{xcolor}

\newcommand{\Div}[1]{\nabla \cdot {#1}}
\newcommand{\Grad}[1]{\nabla {#1}}

\newcommand{\boundary}[1]{\Gamma^{\mathrm{#1}}}
\newcommand{\hboundary}[1]{\Gamma_{h}^{\mathrm{#1}}}

\usepackage{stmaryrd} % für sprungoperator

\newcommand{\avg}[1]{\{\!\{#1\}\!\}}

\newcommand{\jump}[1]{\llbracket {#1} \rrbracket }
\newcommand{\jumporiented}[1]{\left[ {#1} \right] }

\newcommand{\intdomain}[2]{ \left( {#1},{#2} \right)_{\Omega_{h}} }
\newcommand{\intinteriorfaces}[2]{ \left( {#1},{#2} \right)_{\Gamma_{h}^{\mathrm{int}} }}

\newcommand{\intele}[2]{ \left( {#1},{#2} \right)_{\Omega_{e}} }
\newcommand{\inteleface}[2]{ \left( {#1},{#2} \right)_{\partial\Omega_{e}} }
\newcommand{\intelefaceInterior}[2]{ \left( {#1},{#2} \right)_{\partial\Omega_{e}\setminus\Gamma_h }}

\newenvironment{remark}[1][Remark]{\begin{trivlist}
\item[\hskip \labelsep {\bfseries #1}]}{\end{trivlist}}

\usepackage{enumitem}
\setlist[enumerate]{label*=\roman*),ref=\roman*)}

\begin{document}

\begin{frontmatter}

\title{Robust and efficient discontinuous Galerkin methods\\ for under-resolved turbulent incompressible flows}

\author{Niklas Fehn}
\ead{fehn@lnm.mw.tum.de}
\author{Wolfgang A. Wall}
\ead{wall@lnm.mw.tum.de}
\author{Martin Kronbichler\corref{correspondingauthor1}}
\cortext[correspondingauthor1]{Corresponding author at: Institute for Computational Mechanics, Technical University of Munich, Boltzmannstr. 15, 85748 Garching, Germany. Tel.: +49 89 28915300; fax: +49 89 28915301}
\ead{kronbichler@lnm.mw.tum.de}
\address{Institute for Computational Mechanics, Technical University of Munich,\\ Boltzmannstr. 15, 85748 Garching, Germany}

\begin{abstract}
We present a robust and accurate discretization approach for incompressible turbulent flows based on high-order discontinuous Galerkin methods. The DG discretization of the incompressible Navier--Stokes equations uses the local Lax--Friedrichs flux for the convective term, the symmetric interior penalty method for the viscous term, and central fluxes for the velocity--pressure coupling terms. Stability of the discretization approach for under-resolved, turbulent flow problems is realized by a purely numerical stabilization approach. Consistent penalty terms that enforce the incompressibility constraint as well as inter-element continuity of the velocity field in a weak sense render the numerical method a robust discretization scheme in the under-resolved regime. The penalty parameters are derived by means of dimensional analysis using penalty factors of order~$1$. Applying these penalty terms in a postprocessing step leads to an efficient solution algorithm for turbulent flows. The proposed approach is applicable independently of the solution strategy used to solve the incompressible Navier--Stokes equations, i.e., it can be used for both projection-type solution methods as well as monolithic solution approaches. Since our approach is based on consistent penalty terms, it is by definition generic and provides optimal rates of convergence when applied to laminar flow problems. Robustness and accuracy are verified for the Orr--Sommerfeld stability problem, the Taylor--Green vortex problem, and turbulent channel flow. Moreover, the accuracy of high-order discretizations as compared to low-order discretizations is investigated for these flow problems. A comparison to state-of-the-art computational approaches for large-eddy simulation indicates that the proposed methods are highly attractive components for turbulent flow solvers.
\end{abstract}


\end{frontmatter}

\section{Introduction}\label{Intro}
Turbulent flows involve length and time scales ranging over several orders of magnitude rendering such problems a challenging discipline in computational fluid dynamics. Computational methods for turbulent flows can be grouped into three main categories: Reynolds-averaged Navier--Stokes (RANS) simulation, large-eddy simulation (LES), and direct numerical simulation (DNS). The accuracy and predictive capabilities of the numerical approach increase towards the latter category at the cost of increasing demand for computational resources. Large-eddy simulation has developed as a separate discipline since direct numerical simulation of complex flow problems is computationally infeasible in the near future. Increasing computational power and progress in numerical discretization schemes as well as iterative solvers for sparse linear systems of equations allow to apply large-eddy simulation to increasingly complex problems. The methods discussed in this work can be applied to LES and DNS but the focus is on under-resolved turbulent flows in the present paper. In large-eddy turbulence modeling, the numerical discretization scheme is applied in the under-resolved regime and must account for the dissipation of the unresolved scales, see for example~\cite{Sagaut2006} for an overview of different LES models. For the LES approach used here, the dissipation is realized entirely by the numerical discretization scheme so that our approach might be denoted as an implicit LES approach. We mention, however, that the approach is not specifically designed to mimic the properties of explicit subgrid-scale models. The spatial discretization is based on the discontinuous Galerkin (DG) method which is often considered an advantageous combination of finite volume and finite element discretization techniques~\cite{Hesthaven07} in the sense of geometric flexibility, stability in convection-dominated flows, and high-order accuracy.

\subsection{High-order DG methods for under-resolved turbulent flows: state-of-the-art}
Since the 1990s, discontinuous Galerkin methods have been proposed for the compressible Navier--Stokes equations, e.g, in~\cite{Bassi1997,Lomtev1999,Baumann1999,Bassi2000,Hartmann2005}. DG discretizations for the compressible Navier--Stokes equations have been applied to under-resolved turbulent flows and have been validated for LES and DNS computations of canonical turbulent flows such as turbulent channel flow in~\cite{Collis2002a,Ramakrishnan2004,Wei2011,Chapelier2014,Wiart15}, the Taylor--Green vortex problem in~\cite{Chapelier2014,Gassner2013,Wiart14}, and for geometrically more complex transitional and turbulent flow problems in~\cite{Uranga2011,Beck2014b}. A rationale for the suitability of high-order DG discretizations for under-resolved turbulent flows is provided by linear dispersion--diffusion analysis, see for example~\cite{Gassner2011,Moura2015}, and is used as a motivation for no-model large-eddy simulation approaches in~\cite{Wiart15,Gassner2013,Beck2014b} for the compressible Navier--Stokes equations and in~\cite{Moura2017} for the compressible Euler equations. LES subgrid-scale models such as the standard Smagorinsky and the dynamic Smagorinsky model have been used in combination with high-order DG discretizations in several publications~\cite{Ramakrishnan2004,Zhang2007,Sengupta2007,Abba2015,Chapelier2016,Plata2017}. A more sophisticated approach proposed very recently in~\cite{Flad2017} is also based on this strategy where it is argued that a dissipation free discretization scheme in combination with an explicit subgrid-scale model might be beneficial. However, it has not been demonstrated so far that physically motivated LES subgrid-scale models allow to reliably improve the accuracy of the results as compared to a no-model LES strategy over a wide range of the relevant parameters such as the spatial resolution, the Reynolds number, and the considered flow problem. In fact, the investigations in~\cite{VanDerBos2010} reveal that improving the accuracy by using classical LES models is a complicated issue and that optimal values of model constants depend on several parameters. Various numerical flux functions for the inviscid and viscous fluxes have been proposed, and only recently, a discussion has started regarding the properties of the numerical fluxes for high-order DG discretizations of the compressible Navier--Stokes equations applied to under-resolved turbulence~\cite{Moura2017,Flad2017,Beck2014a,Moura2017setting}.

For the incompressible Navier--Stokes equations, numerous discontinuous Galerkin methods have been proposed over the last two decades. A detailed summary of these methods would be exhaustive and we refer to~\cite{Krank2017,Fehn17} for a recent overview. The DG discretization used in the present paper follows~\cite{Fehn17,Shahbazi07}. We use the local Lax--Friedrichs flux for the convective term, the symmetric interior penalty Galerkin (SIPG) method for the viscous term, and central fluxes for the velocity--pressure coupling terms. Quadrilateral/hexahedral elements are considered as well as mixed-order polynomials for velocity and pressure for reasons of inf--sup stability. An important observation as compared to the compressible Navier--Stokes equations is that application of this basic DG discretization approach to under-resolved, turbulent flows is not straightforward and requires additional stabilization techniques. Only recently, DG discretizations of the incompressible Navier--Stokes equations have been proposed addressing the numerical solution of turbulent flow problems. The approach of~\cite{Marek15} is based on physical subgrid-scale modeling and uses the subgrid-scale model of Vreman~\cite{Vreman2004}. The method proposed in~\cite{Ferrer17} uses a DG discretization in two dimensions and a purely spectral Fourier approach in the third dimension. Stability for turbulent flow problems is realized by scaling the penalty parameter of the SIPG method of the viscous term and a spectral vanishing viscosity method in the Fourier direction. A no-model approach with an artificial compressibility numerical flux is used in~\cite{Bassi16} but is only applied to DNS of flow past a sphere. An efficient numerical approach for turbulent flows based on consistent penalty terms added to the weak formulation such as a divergence penalty term and a continuity penalty term is proposed in~\cite{Krank2017} in the context of the dual splitting projection scheme~\cite{Karniadakis1991}. These penalty terms provide additional control over the incompressibility constraint as well as the continuity of the velocity field between elements in a weak sense. This approach is based on ideas that have first been proposed in~\cite{Steinmoeller13,Joshi16} in the context of projection methods where these additional terms are interpreted as a means to stabilize the discrete pressure projection operator, i.e., the pressure solution is obtained by solving a Poisson equation and a divergence-free velocity field is obtained by projecting the intermediate velocity onto the space of solenoidal vectors. 

Note that the instabilities for DG discretizations of the dual splitting scheme reported in~\cite{Krank2017,Ferrer11,Ferrer14} that occur for small time step sizes and are related to the discontinuous Galerkin discretization of the velocity-pressure coupling terms have been solved recently in~\cite{Fehn17} by performing integration by parts of the velocity divergence term and pressure gradient term and defining suitable numerical fluxes as well as consistent boundary conditions for the intermediate velocity field. Since this new formulation has not been available in~\cite{Krank2017,Steinmoeller13,Joshi16}, the origin of the instabilities discussed in~\cite{Krank2017,Steinmoeller13,Joshi16} remains unclear. New insight regarding the stability of projection methods in combination with DG discretizations~\cite{Fehn17} motivates to reconsider the stabilization approach based on divergence and continuity penalty terms not as a means to stabilize spatially discretized projection operators~\cite{Krank2017,Steinmoeller13,Joshi16}, but as a general tool leading to a stable and robust discretization scheme for turbulent flows that can be applied independently of the applied solution strategy. Very recently, the stabilization approach using divergence and continuity penalty terms has also been justified theoretically in~\cite{Akbas2017} where this approach is considered as an analogue of grad--div stabilization for nonconforming discretizations. Moreover, the use of consistent divergence and continuity penalty terms in a discontinuous ($L_2$-conforming) setting can be interpreted as an enforcement of~$H(\mathrm{div})$-conformity and the velocity/pressure spaces of Raviart--Thomas elements in a weak sense. This situation defines the starting point of the present work which aims at gaining further insight into these stabilization terms.

\subsection{Desired properties for turbulent flow solvers}
In order to develop computational approaches for large-eddy simulations of incompressible flows we first summarize desired properties that we aim to fulfill when designing computational methods for the numerical solution of turbulent flow problems:

\begin{enumerate}
\item A key aspect is to obtain robustness and stability of the discretization method for coarse spatial resolutions and convection-dominated, under-resolved simulations. Moreover, the stability properties should be independent of the temporal discretization approach or the solution strategy used to solve the incompressible Navier--Stokes equations.\label{RobustnessAndStability}

\item The method should be accurate and the accuracy should only weakly depend on the model parameters. This property also means that the model should be able to accurately predict the solution without the need to fit model parameters to a specific flow configuration.\label{AccuracyAndSensitivity}

\item When applied to laminar flow problems, the turbulent flow solver should reproduce the exact solution with optimal rates of convergence in space.\label{LaminarFlowProblems}

\item The method should be generic so that it can be applied to arbitrary geometries, i.e., the method should not require quantities such as the wall distance and should not neccesitate spatial averaging in homogeneous directions or other techniques limiting its applicability to complex flow configurations.\label{GeneralityOfTurbulenceModel}

\item The turbulent flow solver should be efficient and should not introduce large computational overhead as compared to laminar flow solvers. In particular, this means that efficient preconditioners developed for laminar flow solvers can be adapted without the need to develop new preconditioners.\label{Efficiency}
\end{enumerate}

We note that the above requirements are the guiding principles of the present work and that alternative requirements with an emphasis on different properties may be defined as well. We found that the approach of~\cite{Marek15} using the model of Vreman and the approach outlined in~\cite{Ferrer12} based on the Smagorinsky model do not lead to a robust discretization scheme in combination with the DG discretization approach used here and that such an approach requires further numerical stabilization techniques that are discussed in this work.
%The need to adjust model parameters is a further disadvantage of that approach.
The approach of~\cite{Ferrer17} is not applicable to arbitrary 3D geometries and the excessive scaling of penalty factors appears to be rather inefficient when using state-of-the-art iterative solution techniques and preconditioners. Among the DG discretization methods for turbulent flows mentioned above, the approach proposed by the authors in~\cite{Krank2017} appears to be the most promising one due to its generality, robustness, accuracy, and computational efficiency. As explained above, this approach requires further investigation as well as rigorous numerical analysis as noted recently in~\cite{Akbas2017}. While an analysis of physical subgrid-scale models for high-order DG methods would clearly be interesting, the purpose of the present work is to analyze the basic numerical scheme in the context of turbulence.

\subsection{Objectives and novel contributions of the present work}
By focusing on a coupled solution approach for the incompressible Navier--Stokes equations that does not contain a discrete projection step onto the space of divergence-free vectors, we show numerically that the divergence and continuity penalty terms are essential components to obtain a stable discretization method in the convection-dominated, under-resolved regime. These penalty terms are motivated by considering the spatially discretized continuity equation and the rate of change of the kinetic energy for the spatially discretized system of equations. This approach might also be interpreted as a numerical subgrid-scale model for large-eddy simulation. Since this approach is based on consistent penalty terms added to the weak formulation, it constitutes a purely numerical approach for large-eddy simulation. As a consequence, this LES approach inherently fulfills properties~\ref{LaminarFlowProblems} and~\ref{GeneralityOfTurbulenceModel} defined above. While these penalty terms might be added to the momentum equation of the incompressible Navier--Stokes equations in case of a coupled solution approach, we propose to apply them in a postprocessing step. As a consequence, efficient preconditioners available for the iterative solution of incompressible, laminar flows can be directly applied to turbulent flow problems leading to an efficient solution algorithm while the complexity associated to the postprocessing of the velocity field is treated separately, i.e., this procedure aims at fulfilling property~\ref{Efficiency}. 

The main objective of the present work is to analyze the proposed approach with respect to the requirements~\ref{RobustnessAndStability} and~\ref{AccuracyAndSensitivity}. The stability of the discretization scheme is investigated for different solution strategies, i.e., we consider a monolithic solution approach as well as projection methods such as velocity-correction and pressure-correction schemes. The results presented in this work demonstrate that the purely divergence penalty based approach preferred in~\cite{Krank2017} is not sufficiently stable in general, but a robust discretization scheme is obtained by including the continuity penalty term enforcing mass conservation over interior element faces, which is in agreement with the recent theoretical considerations in~\cite{Akbas2017}. The accuracy of the proposed schemes is verified by comparing numerical results to accurate reference solutions or available DNS data. While high-order methods are way more accurate than low-order methods for the same number of unknowns in case of well-resolved, laminar flow problems with smooth solutions, the picture is less clear for highly under-resolved, turbulent flow problems. Throughout our numerical investigations we put an emphasis on the accuracy of high-order methods and detailed information is provided regarding the efficiency of high-order discontinuous Galerkin methods in the context of under-resolved turbulent flow problems. Furthermore, we demonstrate that the proposed methods are competitive to state-of-the-art LES approaches or allow to obtain more accurate results for the same number of unknowns. Although a detailed analysis of property~\ref{Efficiency} is beyond the scope of the present paper, we mention that the presented methods indeed lead to efficient solution algorithms. We present results for the 3D Taylor--Green vortex problem at~$\mathrm{Re}=1600$ computed with a high-order DG scheme on a mesh with~$1024^3$ nodes and~$3.7\cdot 10^9$ degrees of freedom demonstrating the high-performance capability of our approach. To the best of the authors' knowledge, the methodology presented in this paper is the first approach in the context of high-order, fully-discontinuous ($L_2$-conforming) Galerkin methods for the incompressible Navier--Stokes equations that provides a robust discretization scheme for under-resolved turbulent flows independently of the applied solution technique for the incompressible Navier--Stokes equations and that is generic in the sense that it can be applied to arbitrary three-dimensional geometries.

\subsection{Outline}

The outline of this paper is as follows. In Section~\ref{MathematicalModel}, the incompressible Navier--Stokes equations are described. The temporal discretization of the incompressible Navier--Stokes equations is briefly summarized in Section~\ref{TemporalDiscretization}. Aspects related to the spatial discontinuous Galerkin discretization as well as the stabilization approach for under-resolved flows are devoted to Section~\ref{SpatialDiscretizationAndTurbulenceModels}. In Section~\ref{NumericalResults}, numerical results are presented where we focus on benchmark problems such as the Orr--Sommerfeld stability problem, the 3D Taylor--Green vortex problem, and turbulent channel flow. A conclusion as well as an outlook is given in Section~\ref{Conclusion}.

\section{Mathematical model}\label{MathematicalModel}
We consider the numerical solution of the incompressible Navier--Stokes equations consisting of the momentum equation and the continuity equation in a domain~$\Omega \subset \mathbb{R}^d$
\begin{align}
\frac{\partial \bm{u}}{\partial t} + \nabla \cdot \bm{F}_{\mathrm{c}}(\bm{u}) - \nabla \cdot \bm{F}_{\mathrm{v}} (\bm{u}) + \nabla p &= \bm{f} \;\; \text{in}\; \Omega \times [0, T] \; ,\label{MomentumEquation}\\
\nabla \cdot \bm{u} &= 0  \;\; \text{in}\; \Omega \times [0, T] \; ,\label{ContinuityEquation}
\end{align}
where~$\bm{f}$ denotes the body force vector,~$\bm{u}$ the velocity vector, and~$p$ the kinematic pressure. The convective flux and the viscous flux are~$\bm{F}_{\mathrm{c}}(\bm{u}) = \bm{u}\otimes \bm{u}$ and~$\bm{F}_{\mathrm{v}} (\bm{u}) = \nu \Grad{\bm{u}}$, respectively, where~$\nu$ denotes the constant kinematic viscosity.
%and~$\bm{F}_{\mathrm{v}} (\bm{u}) = \nu \left(\Grad{\bm{u}} + (\Grad{\bm{u}})^T\right)$, respectively. For a constant kinematic viscosity,~$\nu=\mathrm{const}$, the divergence formulation is equivalent to the Laplace formulation,~$-\Div{\left(\nu \left(\Grad{\bm{u}} + (\Grad{\bm{u}})^T\right)\right)}= - \nu \nabla^2 \bm{u}$, which is derived by using the incompressibility constraint~\eqref{ContinuityEquation}. Introducing a parameter~$\gamma$, the viscous flux can be written as
%\begin{align}
%\bm{F}_{\mathrm{v}} (\bm{u}) = \nu \left(\Grad{\bm{u}} + \gamma \left(\Grad{\bm{u}}\right)^T\right)\; ,\label{FormulationsViscousTerm}
%\end{align}
%where~$\gamma=0$ corresponds to the Laplace formulation and~$\gamma=1$ to the divergence formulation of the viscous term. 
The domain boundary~$\Gamma = \partial \Omega = \boundary{D} \cup \boundary{N}$ consists of Dirichlet boundaries~$\boundary{D}$ and Neumann boundaries~$\boundary{N}$, where the Dirichlet and Neumann boundary conditions read
\begin{align}
\bm{u} &= \bm{g}\;\; \text{on} \; \boundary{D} \times [0, T]\; ,\label{DirichletBC}\\
\left(\bm{F}_{\mathrm{v}} (\bm{u})  - p \bm{I} \right) \cdot \bm{n} &= \bm{h}\;\; \text{on} \; \boundary{N} \times [0, T]\; .\label{NeumannBC_Coupled}
\end{align}
Note that the Neumann boundary condition~$\bm{h}$ is split in a viscous part~$\bm{h}_u$ and a pressure part~$g_p$ when using projection methods to solve the incompressible Navier--Stokes equations as explained in~\cite{Fehn17}. In this paper, we also analyze flow problems involving periodic boundaries, which are formally treated as interior boundaries. At time~$t=0$, an initial condition is prescribed for the velocity field,~$\bm{u}(\bm{x},t=0) = \bm{u}_0(\bm{x})$ in~$\Omega$, where~$\bm{u}_0(\bm{x})$ fulfills the divergence-free constraint~\eqref{ContinuityEquation} as well as the Dirichlet boundary condition~\eqref{DirichletBC}.

\section{Temporal discretization}\label{TemporalDiscretization}
In the present work, we focus on a monolithic solution approach where a coupled system of equations is solved for velocity and pressure. We mention that the aim is to develop numerical methods that are robust independently of the applied solution strategy, see property~\ref{RobustnessAndStability}. Hence, the stability properties of the proposed approach have been analyzed for different widely-used solution strategies for the incompressible Navier--Stokes equations including a monolithic solution approach on the one hand and projection-type solution methods such as the high-order dual splitting scheme~\cite{Karniadakis1991} and pressure-correction schemes~\cite{Guermond2004} on the other hand.
A detailed description of these solution strategies along with the respective boundary conditions and aspects related to the implementation of these methods in the context of high-order discontinuous Galerkin discretizations is given in~\cite{Fehn17}. While a summary of the temporal discretization scheme for the projection-type solution strategies is given in~\ref{TimeIntegrationProjectionMethods}, we describe the time integration scheme for the coupled solution approach in the following. In the present paper, the time integration scheme is based on BDF time integration and a semi-implicit formulation is used where the convective term is treated explicitly.

\subsection{Time integration scheme for coupled solution approach}\label{TemporalDiscretization_CoupledSolution}
In case of the coupled solution approach, the following coupled system of equations is solved in each time step to obtain the velocity solution~$\bm{u}^{n+1}$ and the pressure solution~$p^{n+1}$ at time~$t_{n+1}$
\begin{align}
\frac{\gamma_0 \bm{u}^{n+1}-\sum_{i=0}^{J-1}\left(\alpha_i\bm{u}^{n-i}\right)}{\Delta t}
- \Div{\bm{F}_{\mathrm{v}} (\bm{u}^{n+1})} + \Grad{p^{n+1}} &=  
- \sum_{i=0}^{J-1}\left(\beta_i \Div{\bm{F}_{\mathrm{c}}\left(\bm{u}^{n-i}\right)}\right) 
+ \bm{f}\left(t_{n+1}\right)\; ,\label{TemporalDiscretization_Coupled_Momentum}\\
\Div{\bm{u}^{n+1}} &= 0 \; ,\label{TemporalDiscretization_Coupled_Continuity}
\end{align}
where~$\gamma_0$ and~$\alpha_i$,~$i=0,...,J-1$ are coefficients of the BDF time integration scheme of order~$J$. An extrapolation scheme of order~$J$ with coefficients~$\beta_i$,~$i=0,...,J-1$ is used for the convective term. In the present work, the second order accurate BDF scheme ($J=2$) with~$\gamma_0=3/2$,~$\alpha_0 = 2$,~$\alpha_1=-1/2$ and extrapolation scheme with~$\beta_0=2,\beta_1=-1$ is considered. 

\section{Spatial discretization}\label{SpatialDiscretizationAndTurbulenceModels}
The spatial discretization approach is based on high-order discontinuous Galerkin methods. For a detailed description of the DG discretization used in this work we refer to~\cite{Fehn17} and references mentioned therein. In the following, we briefly summarize the main aspects of the spatial discretization approach while technicalities, e.g., regarding the imposition of boundary conditions, are avoided for reasons of brevity.
\subsection{Notation}
The computational domain~$\Omega_h = \bigcup_{e=1}^{N_{\text{el}}} \Omega_{e}$ with boundary~$\Gamma_h = \partial \Omega_h= \hboundary{D} \cup \hboundary{N}$ consists of~$N_{\text{el}}$ non-overlapping quadrilateral/hexahedral elements. The spaces of test and trial functions used to represent the discrete velocity solution~${\bm{u}_h(\bm{x},t)\in\mathcal{V}^{u}_h}$ and pressure solution~$p_h(\bm{x},t)\in \mathcal{V}^{p}_h$ are defined as
\begin{align}
\mathcal{V}^{u}_{h} &= \left\lbrace\bm{u}_h\in \left[L_2(\Omega_h)\right]^d\; : \; \bm{u}_h\left(\bm{x}(\boldsymbol{\xi})\right)\vert_{\Omega_{e}}= \tilde{\bm{u}}_h^e(\boldsymbol{\xi})\vert_{\tilde{\Omega}_{e}}\in \mathcal{V}^{u}_{h,e}=[\mathcal{P}_{k_u}(\tilde{\Omega}_{e})]^d\; ,\;\; \forall e=1,\ldots,N_{\text{el}} \right\rbrace\;\; ,\\
\mathcal{V}^{p}_{h} &= \left\lbrace p_h\in L_2(\Omega_h)\; : \; p_h\left(\bm{x}(\boldsymbol{\xi})\right)\vert_{\Omega_{e}} = \tilde{p}_h^e(\boldsymbol{\xi})\vert_{\tilde{\Omega}_{e}}\in \mathcal{V}^{p}_{h,e}=\mathcal{P}_{k_p}(\tilde{\Omega}_{e})\; ,\;\; \forall e=1,\ldots,N_{\text{el}} \right\rbrace\; ,
\end{align}
respectively, where~$\tilde{\Omega}_e=[0,1]^d$ denotes the reference element in reference coordinates~$\boldsymbol{\xi}$,~$\bm{x}(\boldsymbol{\xi}) : \tilde{\Omega}_e \rightarrow \Omega_e$  the mapping (of polynomial degree~$k_u$) from reference space to physical space, and~$\mathcal{P}_{k}(\tilde{\Omega}_{e})$  the space of polynomials of tensor degree~$\leq k$. A nodal approach is applied where the multidimensional shape functions are given as the tensor product of one-dimensional shape functions which are Lagrange polynomials using the Legendre--Gauss--Lobatto nodes as support points. In this work, mixed-order polynomials of degree~$(k_u,k_p) = (k,k-1)$ for velocity and pressure are used, see also the discussion in~\cite{Fehn17}. 

Volume and face integrals occurring in the weak formulation are written as~$\intele{v}{u} = \int_{\Omega_e} v \odot u \; \mathrm{d}\Omega$ and~$\inteleface{v}{u} = \int_{\partial \Omega_e} v \odot u \; \mathrm{d} \Gamma$, where the operator~$\odot$ symbolizes inner products, i.e.,~$v u$ for rank-0 tensors,~$\bm{v}\cdot\bm{u} = \sum_i v_i u_i$ for rank-1 tensors, and~$\bm{v} : \bm{u} = \sum_{i,j} v_{ij} u_{ij}$ for rank-2 tensors. The average operator~$\avg{\cdot}$ and the jump operators~$\jump{\cdot}$ and~$\jumporiented{\cdot}$ needed to define numerical fluxes on element faces are given as~$\avg{u} = (u^- + u^+)/2$,~$ \jump{u} = u^- \otimes \bm{n}^- + u^+ \otimes \bm{n}^+$, and~$\jumporiented{u}=u^- - u^+$, where~$(\cdot)^-$ denotes interior information,~$(\cdot)^+$ exterior information from the neighboring element, and~$\bm{n}$ the outward pointing unit normal vector.

Gaussian quadrature is used to numerically calculate volume and surface integrals occurring in the weak formulations summarized below. The number of one-dimensional quadrature points is chosen such that all integrals are calculated exactly in case of affine element geometries with constant Jacobian.

\subsection{Weak discontinuous Galerkin formulation}\label{WeakDGFormulation}
The weak discontinuous Galerkin formulation for the time-continuous system of equations~\eqref{MomentumEquation} and~\eqref{ContinuityEquation} is given as follows: Find~$\bm{u}_h\in\mathcal{V}^u_h$,~$p_h\in \mathcal{V}^{p}_h$ such that for all~$(\bm{v}_h, q_h) \in \mathcal{V}^{u}_{h,e} \times \mathcal{V}^{p}_{h,e}$ and for all elements~$e=1,...,N_{\text{el}}$
\begin{align}
\begin{split}
m^e_{h,u}\left(\bm{v}_h,\frac{\partial \bm{u}_h}{\partial t} \right)
+ v^e_h\left(\bm{v}_h,\bm{u}_h\right)
+ g^e_h\left(\bm{v}_h,p_h\right) +  c^e_h\left(\bm{v}_h,\bm{u}_h\right)
&= \intele{\bm{v}_h}{\bm{f}(t)} \; ,
\end{split} \label{WeakForm_CoupledSolution_Momentum_TimeContinuous}\\
-d^e_h(q_h,\bm{u}_h)&=
 0  \; .\label{WeakForm_CoupledSolution_Continuity_TimeContinuous}
\end{align}
In the above equations,~$m^e_{h,u}\left(\bm{v}_h,\bm{u}_h\right) = \intele{\bm{v}_h}{\bm{u}_h}$ is the (velocity) mass matrix term,~$c^e_h\left(\bm{v}_h,\bm{u}_h\right)$ denotes the convective term,~$v^e_h\left(\bm{v}_h,\bm{u}_h\right)$ the viscous term,~$g^e_h\left(\bm{v}_h,p_h\right)$ the pressure gradient term, and~$d^e_h\left(q_h,\bm{u}_h\right)$ the velocity divergence term. The weak formulation of the above operators is derived by performing integration by parts and defining suitable numerical flux functions, see~\cite{Fehn17} for a more detailed description. In this respect, the weak form of the convective term is
\begin{align}
c^e_h\left(\bm{v}_h,\bm{u}_h\right) = -\intele{\Grad{\bm{v}_h}}{\bm{F}_{\mathrm{c}}(\bm{u}_h)} + 
\inteleface{\bm{v}_h}{\bm{F}_{\mathrm{c}}^*(\bm{u}_h)   \cdot \bm{n}}\label{WeakForm_ConvectiveTerm}
\end{align}
where the local Lax--Friedrichs flux is used as numerical flux function
\begin{align}
\bm{F}_{\mathrm{c}}^*(\bm{u}_h)=\avg{\bm{F}_{\mathrm{c}}(\bm{u}_h)} + \frac{\Lambda}{2}\jump{\bm{u}_h} \;\; \text{with} \;\; \Lambda = \max \left( 2 \vert \bm{u}_h^{-} \cdot \bm{n}\vert , 2 \vert \bm{u}_h^{+} \cdot \bm{n}\vert\right) \; .\label{LaxFriedrichsFlux}
\end{align}
 For the pressure gradient term and the velocity divergence term central fluxes are used resulting in the following weak forms
\begin{align}
g^e_h\left(\bm{v}_h,p_h\right) &= -\intele{\Div{\bm{v}_h}}{p_h}+\inteleface{\bm{v}_h}{ \avg{p_h}\bm{n}}\; ,\\
d^e_h\left(q_h,\bm{u}_h\right) &= -\intele{\Grad{q_h}}{\bm{u}_h}+\inteleface{q_h}{\avg{\bm{u}_h}\cdot\bm{n}}\; .
\end{align}
The symmetric interior penalty Galerkin method is used to discretize the viscous term
\begin{align}
\begin{split}
v_{h}^{e}(\bm{v}_h,\bm{u}_h) = 
 \intele{\Grad{\bm{v}_h}}{\nu \Grad{\bm{u}_h}}
  - \inteleface{\Grad{\bm{v}_h}}{\frac{\nu}{2} \jump{\bm{u}_h}}
  - \inteleface{\bm{v}_h}{\nu \avg{\Grad{\bm{u}_h}}\cdot\bm{n}}
  + \inteleface{\bm{v}_h}{\nu\tau \jump{\bm{u}_h}\cdot\bm{n}}\; ,
\end{split}
\end{align}
where the definition of the interior penalty parameter~$\tau$ used in~\cite{Fehn17} is applied in this work.

\subsection{Stabilization approach for under-resolved flows}\label{NumericalLES}
Stability of the DG discretization scheme for under-resolved, turbulent flows is obtained by adding consistent penalty terms to the weak formulation. These penalty terms are a divergence penalty term~$a^e_{\mathrm{D}}(\bm{v}_h,\bm{u}_h)$ weakly enforcing the incompressibility constraint and a continuity penalty term $a^e_{\mathrm{C}}(\bm{v}_h,\bm{u}_h)$ weakly enforcing inter-element continuity of the velocity field, which are defined in detail in Section~\ref{DivAndContiPenaltyTerms}. Consequently, the methods discussed here constitute a purely numerical approach for LES of turbulent flows. These terms have been proposed in~\cite{Krank2017} in the context of the dual splitting scheme and are based on ideas first discussed in~\cite{Steinmoeller13,Joshi16} in the context of projection methods where similar terms have been used as a means to stabilize the spatially discretized pressure projection operator. Recalling the motivation used by the authors in~\cite{Krank2017}, the divergence and continuity penalty terms can be seen as a measure to improve mass conservation by considering the discretized continuity equation in the so-called strong formulation: Find~$\bm{u}_h\in\mathcal{V}^u_h$ such that
\begin{align}
-\intele{q_h}{\Div{\bm{u}_h}}+\inteleface{q_h}{\frac{1}{2}\jumporiented{\bm{u}_h}\cdot\bm{n}}  = 0\; ,\label{WeakFormContinuityEquationStrongFormulation}
\end{align}
for all~$q_h \in \mathcal{V}^{p}_{h,e}$ and for all elements~$e=1,...,N_{\text{el}}$. In the present work, we demonstrate that these penalty terms can be seen as a very general approach stabilizing the DG scheme in the under-resolved regime independently of the applied solution strategy. As argued recently in~\cite{Akbas2017}, these penalty terms can be interpreted as an analogue of grad--div stabilization applied to nonconforming discretization. In addition to the argument of improved mass conservation used above, we motivate these penalty terms by using an energy argument as detailed in the following. For this analysis, we assume vanishing body forces,~$\bm{f}=\bm{0}$, as well as periodic boundary conditions. The rate of change of the kinetic energy~$e_{\mathrm{k}}$ can be expressed in terms of the velocity mass matrix operator
\begin{align}
e_{\mathrm{k}} = \int_{\Omega_h} \frac{1}{2} \bm{u}_h\cdot \bm{u}_h\; \mathrm{d}\Omega
\;\;\leadsto \;\;
\frac{\mathrm{d}e_{\mathrm{k}}}{\mathrm{d} t} = m_{h,u}\left(\bm{u}_h,\frac{\partial\bm{u}_h}{\partial t}\right) = \sum_{e=1}^{N_{\mathrm{el}}} m^e_{h,u}\left(\bm{u}_h,\frac{\partial\bm{u}_h}{\partial t}\right)\; ,\label{KineticEnergyProductionRate1}
\end{align}
assuming that~$\Omega_h$ is time-invariant. Inserting the discretized momentum equation~\eqref{WeakForm_CoupledSolution_Momentum_TimeContinuous} into equation~\eqref{KineticEnergyProductionRate1} along with the above assumptions results in
\begin{align}
\frac{\mathrm{d}e_{\mathrm{k}}}{\mathrm{d} t} 
= - \sum_{e=1}^{N_{\mathrm{el}}} \left( c^e_h\left(\bm{u}_h,\bm{u}_h\right) + v^e_h\left(\bm{u}_h,\bm{u}_h\right)
+ g^e_h\left(\bm{u}_h,p_h\right)\right) = - \left(c_h\left(\bm{u}_h,\bm{u}_h\right) + v_h\left(\bm{u}_h,\bm{u}_h\right)
+ g_h\left(\bm{u}_h,p_h\right)\right) \; .
\end{align}
The discontinuous Galerkin formulation introduced above is symmetric with respect to the pressure gradient term and velocity divergence term,~$g_h\left(\bm{u}_h,p_h\right)=-d_h\left(p_h,\bm{u}_h\right)$. Moreover, we know that~$-d_h\left(p_h,\bm{u}_h\right)=0$ due to the discretized continuity equation~\eqref{WeakForm_CoupledSolution_Continuity_TimeContinuous} so that the pressure gradient term does not contribute to the rate of change of the kinetic energy. The SIPG discretization of the viscous term is positive definite and thus~$v_h\left(\bm{u}_h,\bm{u}_h\right)\geq 0$. Hence, the estimate translates to
\begin{align}
\frac{\mathrm{d}e_{\mathrm{k}}}{\mathrm{d} t} \leq - \sum_{e=1}^{N_{\mathrm{el}}} c^e_h\left(\bm{u}_h,\bm{u}_h\right) \; .
\end{align}
We first reformulate the convective term by integrating the first term on the right-hand side of equation~\eqref{WeakForm_ConvectiveTerm} by parts once again
\begin{align}
\begin{split}
c^e_h\left(\bm{u}_h,\bm{u}_h\right) &= \intele{\bm{u}_h}{\Div{\bm{F}_{\mathrm{c}}(\bm{u}_h)}}+ 
\inteleface{\bm{u}_h}{\left(\bm{F}_{\mathrm{c}}^*(\bm{u}_h)-\bm{F}_{\mathrm{c}}(\bm{u}_h)\right) \cdot \bm{n}} \\
&= \frac{1}{2}\intele{\Div{\bm{u}_h}}{\bm{u}_h\cdot \bm{u}_h}+\inteleface{\bm{u}_h}{\left(\bm{F}_{\mathrm{c}}^*(\bm{u}_h)-\frac{1}{2}\bm{F}_{\mathrm{c}}(\bm{u}_h)\right)   \cdot \bm{n}} \; ,
\end{split}
\end{align}
where the relation~$\intele{\bm{u}_h}{\Div{\bm{F}_{\mathrm{c}}(\bm{u}_h)}} = \frac{1}{2}\intele{\Div{\bm{u}_h}}{\bm{u}_h\cdot \bm{u}_h} + \frac{1}{2}\inteleface{\bm{u}_h}{\bm{F}_{\mathrm{c}}(\bm{u}_h)\cdot \bm{n}}$ has been used in the second step, which can also be derived by performing integration by parts. Inserting the Lax--Friedrichs flux, equation~\eqref{LaxFriedrichsFlux}, and summing over all elements yields
\begin{align}
\begin{split}
\frac{\mathrm{d}e_{\mathrm{k}}}{\mathrm{d} t} \leq &- \sum_{e=1}^{N_{\mathrm{el}}} \left(
\frac{1}{2}\intele{\Div{\bm{u}_h}}{\bm{u}_h\cdot \bm{u}_h} 
+ \inteleface{\bm{u}_h}{\frac{1}{2}\bm{F}_{\mathrm{c}}(\bm{u}_h^+)\cdot \bm{n}}
+\inteleface{\bm{u}_h\otimes \bm{n}}{\frac{\Lambda}{2}\jump{\bm{u}_h}}\right) \\
= &-\frac{1}{2}\intdomain{\Div{\bm{u}_h}}{\bm{u}_h\cdot \bm{u}_h} \\
&- \intinteriorfaces{\bm{u}^-_h}{\frac{1}{2}\bm{F}_{\mathrm{c}}(\bm{u}_h^+)\cdot \bm{n}^-}
- \intinteriorfaces{\bm{u}^+_h}{\frac{1}{2}\bm{F}_{\mathrm{c}}(\bm{u}_h^-)\cdot \bm{n}^+}
- \intinteriorfaces{\jump{\bm{u}_h}}{\frac{\Lambda}{2}\jump{\bm{u}_h}}  \; ,
 \end{split}
\end{align}
where~$\Gamma_{h}^{\mathrm{int}}$ denotes the set of all interior faces (note that~$\Gamma_h =\partial \Omega_h= \emptyset$ since we consider periodic boundary conditions). The second and the third term on the right-hand side of the above inequality can be further simplified by using the oriented jump operator~$ \jumporiented{\bm{u}_h}= \bm{u}_h^- - \bm{u}_h^+$ to arrive at the result
\begin{align}
\begin{split}
\frac{\mathrm{d}e_{\mathrm{k}}}{\mathrm{d} t} \leq & -\frac{1}{2}\intdomain{\Div{\bm{u}_h}}{\bm{u}_h\cdot \bm{u}_h} 
+\frac{1}{2} \intinteriorfaces{\jumporiented{\bm{u}_h}\cdot \bm{n}}{\bm{u}_h^-\cdot \bm{u}_h^+}
- \intinteriorfaces{\jump{\bm{u}_h}}{\frac{\Lambda}{2}\jump{\bm{u}_h}} \\
\leq 
& +\frac{1}{2}\intdomain{\vert \Div{\bm{u}_h}\vert}{\bm{u}_h\cdot \bm{u}_h} 
+\frac{1}{2} \intinteriorfaces{\vert\jumporiented{\bm{u}_h}\cdot \bm{n}\vert}{\vert\bm{u}_h^-\cdot \bm{u}_h^+\vert}
- \intinteriorfaces{\jump{\bm{u}_h}}{\frac{\Lambda}{2}\jump{\bm{u}_h}}\; .\label{EstimateDissipation}
 \end{split}
\end{align}
While the jump term corresponding to the Lax--Friedrichs flux shows a dissipative behavior, two potential sources of instabilities can be identified in the above equation: A violation of the divergence-free constraint as well as jumps of the normal component of the velocity across element faces might produce energy. To balance these terms we add consistent penalty terms to the weak formulation enforcing the divergence-free constraint as well as inter-element continuity of the velocity in the direction normal to the face. We note that these terms would vanish in case of~$H(\mathrm{div})$ conforming velocity elements (for which the normal component of the velocity is continuous across element faces) along with polynomial spaces providing exactly divergence-free velocity fields in the discrete case such as Raviart--Thomas elements combined with a discontinuous pressure one order lower. Such approaches are, however, restrictive with respect to the element geometry and can, e.g., only be applied in case of affine element geometries unless other measures such as a Piola transformation are applied. The jump penalty term weakly enforces~$H(\mathrm{div})$-conformity so that this stabilization can also be denoted as~$H(\mathrm{div})$ stabilization~\cite{Akbas2017}. Similarly, the divergence penalty term might be interpreted as a weak enforcement of exactly divergence-free velocity spaces and might be seen as a weak realization of Raviart--Thomas elements for the velocity. However, it is unclear whether the proposed penalty terms have, in addition to~$H(\mathrm{div})$-conforming elements  and Raviart--Thomas elements, a positive impact on the dissipation properties of the numerical discretization scheme regarding large-eddy simulation. Enforcing the divergence-free constraint and inter-element continuity of the velocity in a weak sense does not guarantee that the discretization scheme does not produce kinetic energy, but we give numerical evidence in this work that the penalty based approach is very effective. Moreover, the present approach is a general approach that can be applied to arbitrary element geometries. Adding these terms to the momentum equation of the coupled system of equations results in the weak form
\begin{align}
\begin{split}
m^e_{h,u}\left(\bm{v}_h,\frac{\partial \bm{u}_h}{\partial t} \right)
+ a^e_{\mathrm{D}}(\bm{v}_h,\bm{u}_h)
+ a^e_{\mathrm{C}}(\bm{v}_h,\bm{u}_h) 
+ v^e_h\left(\bm{v}_h,\bm{u}_h\right)
+ g^e_h\left(\bm{v}_h,p_h\right) +  c^e_h\left(\bm{v}_h,\bm{u}_h\right)
&= \intele{\bm{v}_h}{\bm{f}(t)} \; ,
\end{split}\label{WeakForm_Momentum_DivConti_TimeContinuous}\\
-d^e_h(q_h,\bm{u}_h)&=
 0  \; .
\end{align}
where~$a^e_{\mathrm{D}}(\bm{v}_h,\bm{u}_h)$ is the divergence penalty term and~$a^e_{\mathrm{C}}(\bm{v}_h,\bm{u}_h)$ the continuity penalty term. Taking into account these penalty terms results in the modified energy estimate
\begin{align}
\begin{split}
\frac{\mathrm{d}e_{\mathrm{k}}}{\mathrm{d} t} \leq 
& -a^e_{\mathrm{D}}(\bm{u}_h,\bm{u}_h) +\frac{1}{2}\intdomain{\vert \Div{\bm{u}_h}\vert}{\bm{u}_h\cdot \bm{u}_h} \\
&-a^e_{\mathrm{C}}(\bm{u}_h,\bm{u}_h)+\frac{1}{2} \intinteriorfaces{\vert\jumporiented{\bm{u}_h}\cdot \bm{n}\vert}{\vert\bm{u}_h^-\cdot \bm{u}_h^+\vert}
- \intinteriorfaces{\jump{\bm{u}_h}}{\frac{\Lambda}{2}\jump{\bm{u}_h}}\; ,\label{EstimateDissipationWithPenaltyTerms}
 \end{split}
\end{align}
where the divergence and continuity penalty terms are positive semi-definite (see Section~\ref{DivAndContiPenaltyTerms}) and yield a non-positive contribution to the kinetic energy evolution.
\subsubsection{Weak DG formulation for coupled solution approach}\label{WeakFormCoupled}
For reasons of computational efficiency, however, we suggest to apply the divergence and continuity penalty terms separately in a postprocessing step instead of adding these terms to the momentum equation. In a first step, an intermediate velocity~$\hat{\bm{u}}_h$ is calculated by solving the coupled system of equations for velocity and pressure unknowns
\begin{align}
\begin{split}
m^e_{h,u}\left(\bm{v}_h,\frac{\gamma_0 \hat{\bm{u}}_h-\sum_{i=0}^{J-1}\left(\alpha_i\bm{u}^{n-i}_h\right)}{\Delta t} \right)
+ v^e_h\left(\bm{v}_h,\hat{\bm{u}}_h\right)
+ g^e_h\left(\bm{v}_h,p^{n+1}_h\right) = 
&- \sum_{i=0}^{J-1}\left(\beta_i c^e_h\left(\bm{v}_h,\bm{u}^{n-i}_h\right)\right)\\
&+ b^e_h\left(\bm{v}_h,\bm{f}(t_{n+1})\right) \; ,
\end{split} \\
-d^e_h(q_h,\hat{\bm{u}}_h)=& \; 0  \; .
\end{align}
Subsequently, the divergence and continuity penalty terms are applied in a postprocessing step to obtain the final velocity~$\bm{u}^{n+1}_h$
\begin{align}
m_{h,u}^{e}(\bm{v}_h,\bm{u}^{n+1}_h) 
+ a^e_{\mathrm{D}}(\bm{v}_h,\bm{u}^{n+1}_h)
+ a^e_{\mathrm{C}}(\bm{v}_h,\bm{u}^{n+1}_h) 
= m_{h,u}^{e}\left(\bm{v}_h,\hat{\bm{u}}_h\right)\; .\label{WeakForm_CoupledSolver_Divergence_Continuity_Penalty}
\end{align}
While we did not observe noticeable differences between both approaches in terms of accuracy, the second approach using a postprocessing of the velocity field is advantageous in terms of computational costs, i.e., preconditioning strategies available for laminar incompressible flows can be directly applied to turbulent flows while the complexity associated to the postprocessing of the velocity field is treated separately~\footnote{A development of block-preconditioners for the coupled solution approach taking into account additional stabilization terms has been considered in~\cite{Heister2013} in the context of conforming finite element discretizations with grad--div stabilization.}. This procedure allows to obtain cost-effective solution algorithms for turbulent flows where the computational costs increase only moderately as compared to the respective laminar flow solvers.

The divergence penalty and the continuity penalty terms as well as solution and preconditioning strategies used to solve this system of equations are discussed in more detail in Section~\ref{DivAndContiPenaltyTerms}. For projection-type solution methods, the divergence and continuity penalty terms are applied in the projection step in which a divergence-free velocity field is calculated. A summary of the weak discontinuous Galerkin formulation for the projection-type solution strategies is provided in~\ref{WeakDGFormulationProjectionMethods}.

\subsubsection{Divergence and continuity penalty terms}\label{DivAndContiPenaltyTerms}
The divergence penalty term~$a^e_{\mathrm{D}}(\bm{v}_h,\bm{u}_h)$ has similarities with the grad--div stabilization term often used in continuous finite element methods, see for example~\cite{Olshanskii09}, and is defined as
\begin{align}
a^e_{\mathrm{D}}(\bm{v}_h,\bm{u}_h) = \intele{\Div{\bm{v}_h}}{\tau_{\mathrm{D}}\Div{\bm{u}_h}} \; .
\end{align}
The penalty parameter~$\tau_{\mathrm{D}}$ can be derived by means of dimensional analysis and is expressed in terms of a characteristic velocity, an effective element length and the time step size to obtain
% TODO: old definition of penalty parameter using || U_mean || -> problems with symmetric solutions, e.g., Taylor--Green vortex problem
%\textcolor{red}{
%\begin{align}
%\tau_{\mathrm{D},e}=\zeta_{\mathrm{D}} \; \Vert \overline{\bm{u}^{n+1,\mathrm{ex}}_h} \Vert \; \frac{h}{k_u + 1} \; \Delta t \; ,\label{DivergencePenaltyFactorOld}
%\end{align}}
%or (new improved formulation)
\begin{align}
\tau_{\mathrm{D},e}=\zeta_{\mathrm{D}} \; \overline{\Vert\bm{u}^{n+1,\mathrm{ex}}_h \Vert} \; \frac{h}{k_u + 1} \; \Delta t \; ,\label{DivergencePenaltyFactor}
\end{align} 
where~$\bm{u}^{n+1,\mathrm{ex}}_h=\sum_{i=0}^{J-1} \left(\beta_i \bm{u}^{n-i}_h\right)$ is an extrapolation of the velocity field of order~$J$,~$\overline{(\cdot)}$ an elementwise volume-averaged quantity, and~$h=V_e^{1/3}$ a characteristic element length where~$V_e$ is the element volume. A similar expression is obtained in~\cite{Olshanskii09} in the convection-dominated regime, where the grad--div stabilization term is motivated from the point of view of variational multiscale methods and residual-based subgrid modeling. The factor~$h/(k_u+1)$ in equation~\eqref{DivergencePenaltyFactor} is an effective element length scale taking into account shape functions of higher polynomial degree. The factor~$\Delta t$ has to be omitted in case the penalty term is added to the momentum equation~\eqref{WeakForm_CoupledSolution_Momentum_TimeContinuous} as shown in equation~\eqref{WeakForm_Momentum_DivConti_TimeContinuous}. In other words, the factor~$\Delta t$ appears in equation~\eqref{DivergencePenaltyFactor} since the momentum equation has been multiplied by the time step size. For all numerical computations presented in this work a penalty factor of~$\zeta_{\mathrm{D}}=1$ is used. The continuity penalty term~$a^e_{\mathrm{C}}(\bm{v}_h,\bm{u}_h)$ is defined as
\begin{align}
a^e_{\mathrm{C}}(\bm{v}_h,\bm{u}_h)=\intelefaceInterior{\bm{v}_h\cdot \bm{n}}{\tau_{\mathrm{C},f}\jumporiented{\bm{u}_h}\cdot \bm{n}} \; .
\end{align}
Note that we only penalize the normal component of the velocity in accordance with theoretical considerations in equations~\eqref{WeakFormContinuityEquationStrongFormulation} and~\eqref{EstimateDissipation}, see also~\cite{Akbas2017}. The penalty parameter~$\tau_{\mathrm{C},f}$ on an interior face~$f\subseteq \partial\Omega_{e}\setminus\Gamma_h$ is~$\tau_{\mathrm{C},f}= \avg{\tau_{\mathrm{C},e}}=\left(\tau_{\mathrm{C},e^-}+\tau_{\mathrm{C},e^+}\right)/2$. The elementwise continuity penalty factor is derived by means of dimensional analysis using the velocity~$\overline{\Vert \bm{u}^{n+1,\mathrm{ex}}_h \Vert}$ and the time step size~$\Delta t$
\begin{align}
\tau_{\mathrm{C},e}=\zeta_{\mathrm{C}}\;\overline{\Vert \bm{u}^{n+1,\mathrm{ex}}_h \Vert} \; \Delta t \; .
\end{align}
As for the divergence penalty term, the factor~$\Delta t$ in the above equation has to be omitted in case the continuity penalty term appears as an additional term in the momentum equation, see equation~\eqref{WeakForm_Momentum_DivConti_TimeContinuous}. A penalty factor of~$\zeta_{\mathrm{C}}=1$ is used for all numerical results shown in this work. We emphasize that the penalty terms are formulated in a way such that the physical units are consistent with the other terms of the Navier--Stokes equations.

In the following, we investigate the stability of a purely divergence penalty based postprocessing on the one hand and a postprocessing involving both divergence and continuity penalty terms on the other hand. While the focus is on the stability properties of both variants in the present paper, we mention that the two variants also differ in terms of computational costs. Without coupling to neighboring elements the divergence penalty term is a local (elementwise) operator. Accordingly, equation~\eqref{WeakForm_CoupledSolver_Divergence_Continuity_Penalty} and equivalently equations~\eqref{DualSplitting_Projection_WeakForm} and~\eqref{PressureCorrection_Projection_WeakForm} can be solved elementwise for~$\zeta_{\mathrm{C}}=0$ without the need to solve a global linear system of equations. The continuity penalty term, however, introduces a coupling of neighboring degrees of freedom and requires the solution of a global linear system of equations. For both variants, the inverse mass matrix represents an effective preconditioner due to the relatively small time step sizes that arise from the CFL condition for the turbulent flow problems considered here. The inverse mass matrix operator can be implemented in a matrix-free way~\cite{Kronbichler2016} with costs comparable to the forward application of the mass matrix or a diagonal mass matrix (the mass matrix operation is a memory-bound operation) so that this preconditioner is also an efficient preconditioner especially for high polynomial degrees. The purely divergence penalty based variant has been preferred in~\cite{Krank2017} for reasons of computational efficiency. The numerical results shown in Section~\ref{NumericalResults}, however, demonstrate that this variant is less robust and that both divergence and continuity penalty terms should be used to obtain a robust and accurate discretization scheme.

\subsection{CFL condition}
Due to the explicit treatment of the convective term, the time step size~$\Delta t$ is restricted according to the CFL (Courant--Friedrichs--Lewy) condition
\begin{align}
\Delta t = \frac{\mathrm{Cr}}{k_u^{1.5}}\frac{h_{\mathrm{min}}}{\Vert \bm {u} \Vert_{\mathrm{max}}} \; ,\label{CFL_Condition}
\end{align}
where~$\mathrm{Cr}$ denotes the Courant number,~$h_{\mathrm{min}}$ a characteristic length scale of the mesh calculated as the minimum vertex distance, and~$\Vert \bm {u} \Vert_{\mathrm{max}}$ an estimation of the maximum velocity. The factor~$1/k_u^{1.5}$ highlights that the time step size has to be reduced for increasing polynomial degrees, where we found experimentally that an exponent of~$1.5$ for the polynomial degree~$k_u$ allows to use a constant Courant number over a wide range of polynomial degrees for the flow problems considered in this work, see also the discussion in~\cite{Hesthaven07} regarding the exponent of the polynomial degree.

\section{Numerical results}\label{NumericalResults}
\subsection{Objectives}
In this section we present numerical results for several benchmark problems for incompressible turbulent flows. The aim of this paper is to investigate the proposed stabilized approach as compared to the standard DG discretization with respect to properties~\ref{RobustnessAndStability} and~\ref{AccuracyAndSensitivity}. Accordingly, the objectives are twofold:

\begin{itemize}
\item We rigorously analyze the robustness and stability of our approach as compared to the basic DG discretization without additional stabilization terms for several test cases such as the Orr-Sommerfeld stability problem, the Taylor--Green vortex problem, and turbulent channel flow where we investigate a wide range of spatial resolutions (refinement level~$l$ and polynomial degree~$k$) beginning with the coarsest possible spatial resolutions. This is a major difference and advancement to previous publications. A careful analysis of the proposed methods in the limit of coarse spatial resolutions is regarded as a necessity to evaluate and demonstrate the relevance of this turbulence approach. In our opinion, such an analysis is more illustrative than demonstrating practicability for a specific setup, e.g., a specific spatial resolution. 

\item While the improved efficiency of high-order methods as compared to low-order methods appears to be obvious for simple analytical test cases and well-resolved problems, the efficiency of high-order methods is still an open issue for more complex applications and under-resolved problems, see also the discussion in~\cite{Gassner2013,Wiart14}. For this reason, we analyze the accuracy of high-order discretizations as compared to low-order discretizations by comparing the accuracy of numerical results for the same number of unknowns and different polynomial degrees. A detailed efficiency analysis in terms of accuracy versus number of unknowns is performed for the flow problems investigated in the following. An investigation of the overall efficiency in terms of accuracy versus computational costs is beyond the scope of the present paper and the reader is referred to~\cite{Fehn18b} where these aspects are discussed in detail for the proposed discretization approach and under-resolved turbulent flows.
\end{itemize}

\subsection{Implementation}
The code is implemented in \texttt{C++} and makes use of the object-oriented finite element library~\texttt{deal.II}~\cite{dealII}. The incompressible Navier--Stokes solvers described above are implemented using a high-performance framework for generic finite element operator application developed in~\cite{Kronbichler12,Kronbichler2017,Kronbichler2017fast} that is based on a matrix-free implementation. The matrix-free implementation exploits sum-factorization on quadrilateral/hexahedral elements and uses vectorization over several elements. A key feature of this implementation is the fact that the computational costs for evaluating discretized finite element operators per number of unknowns is almost independent of the polynomial degree for a wide range of polynomial degrees,~$1\leq k \leq 10$, which is a basic requirement to obtain efficient high-order discretization schemes. The solution of linear systems of equations is based on state-of-the-art iterative solution techniques (Krylov methods) using efficient preconditioners such as geometric multigrid methods with polynomial, matrix-free smoothing ensuring mesh independent convergence and allowing to scale the alogrithm up to~$\mathcal{O}(10^5)$ processors as shown in~\cite{Krank2017}.

\subsection{Vortex problem: Assessment of consistency and optimal convergence}\label{VortexProblem}
As a prerequisite for the turbulent flow simulations considered below, we demonstrate numerically that optimal rates of convergence are obtained with respect to the temporal discretization and the spatial discretization for the standard formulation without penalty terms and for the stabilized formulation with divergence and continuity penalty terms applied in a postprocessing step, see property~\ref{LaminarFlowProblems}.

For this purpose, we perform convergence tests for the vortex problem analyzed in~\cite{Hesthaven07}. The analytical solution of the two-dimensional, unsteady incompressible Navier--Stokes equations with body force vector~$\bm{f}=\bm{0}$ is given as
\begin{align}
\begin{split}
\bm{u}(\bm{x},t) &=  \begin{pmatrix}
-\sin(2\pi x_2)\\
+\sin(2\pi x_1)
\end{pmatrix}
\exp\left(-4\nu\pi^2 t\right)\; ,\\
p(\bm{x},t) &=  -\cos(2\pi x_1)\cos(2\pi x_2)\exp\left(-8\nu \pi^2 t\right)\; .
\end{split}\label{AnalyticalSolutionVortex}
\end{align}
The domain~$\Omega=[-L/2,L/2]^2$ is a square of length~$L=1$, the time interval is~$0\leq t\leq T=1$, and the viscosity is~$\nu=0.025$. On the domain boundary, Dirichlet and Neumann boundary conditions are prescribed at the inflow part and the outflow part of the boundary, respectively, see also~\cite{Hesthaven07}. Initial conditions are obtained by interpolating the analytical solution and Dirichlet and Neumann boundary conditions are derived from the analytical solution. The computational domain is discretized using a uniform Cartesian grid with element length~$h=L/2^l$, where~$l$ denotes the refinement level. For this test case we use absolute solver tolerances of~$10^{-12}$ and relative solver tolerances of~$10^{-6}$.

% LAPLACE FORMULATION of viscous term: temporal convergence tests

%The Laplace formulation of the viscous term ($\gamma = 0$) is used for this test case.

\begin{figure}[!ht]
 \centering 
\includegraphics[width=0.75\textwidth]{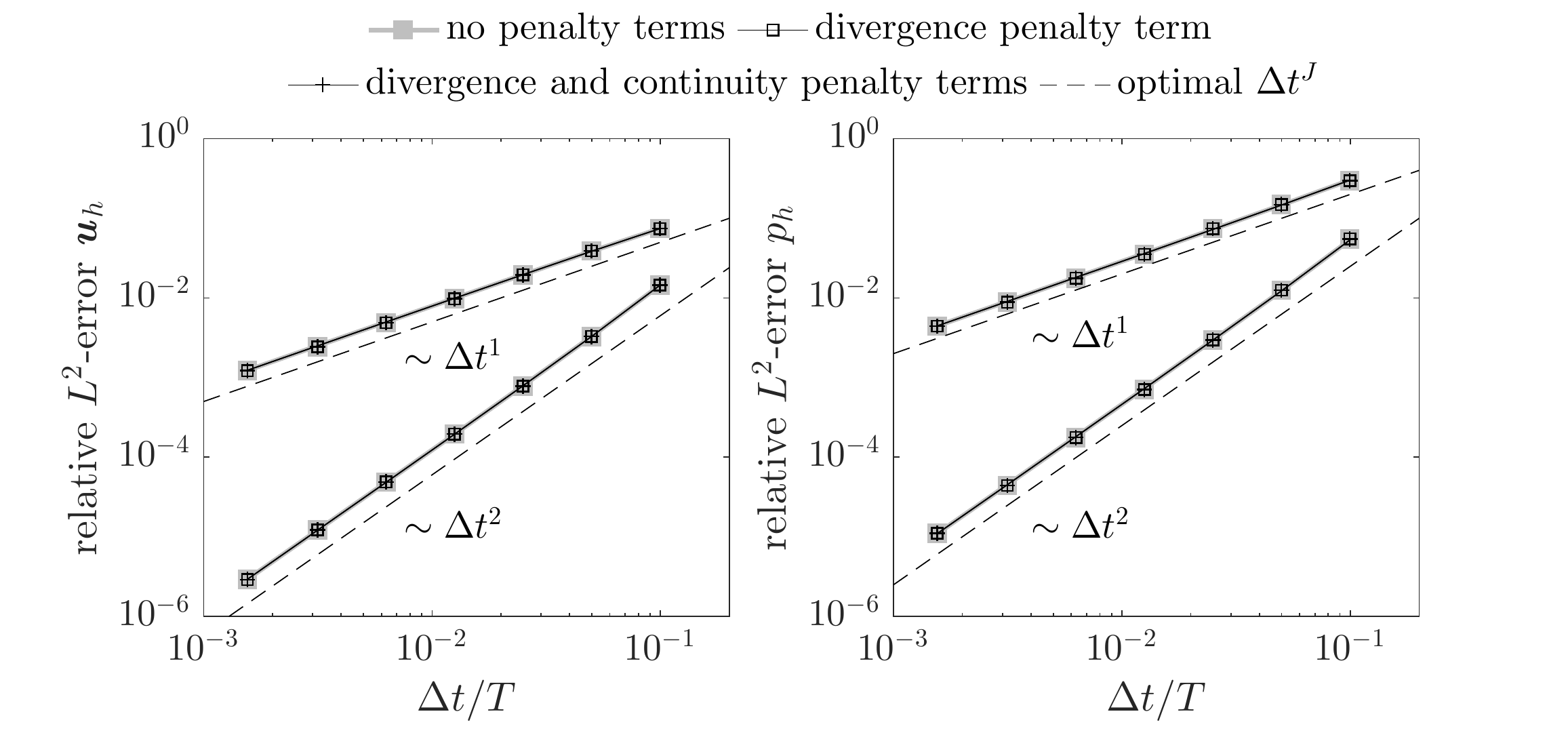}
\caption{Temporal convergence tests for vortex flow problem: assessment of optimal rates of convergence for BDF time integration schemes of order~$J=1,2$ and time step sizes~$\Delta t /T = 0.1/2^m$, where~$m=0,1,...,6$. The spatial resolution is~$l=3$ and~$(k_u,k_p)=(8,7)$.}
\label{fig:temporal_convergence_vortex_problem}
\end{figure}

% LAPLACE FORMULATION of viscous term: spatial convergence tests
\begin{figure}[!ht]
 \centering 
 \subfigure[Spatial convergence test: relative~$L^2$-errors of velocity and pressure as a function of~$h$.]{
	\includegraphics[width=0.75\textwidth]{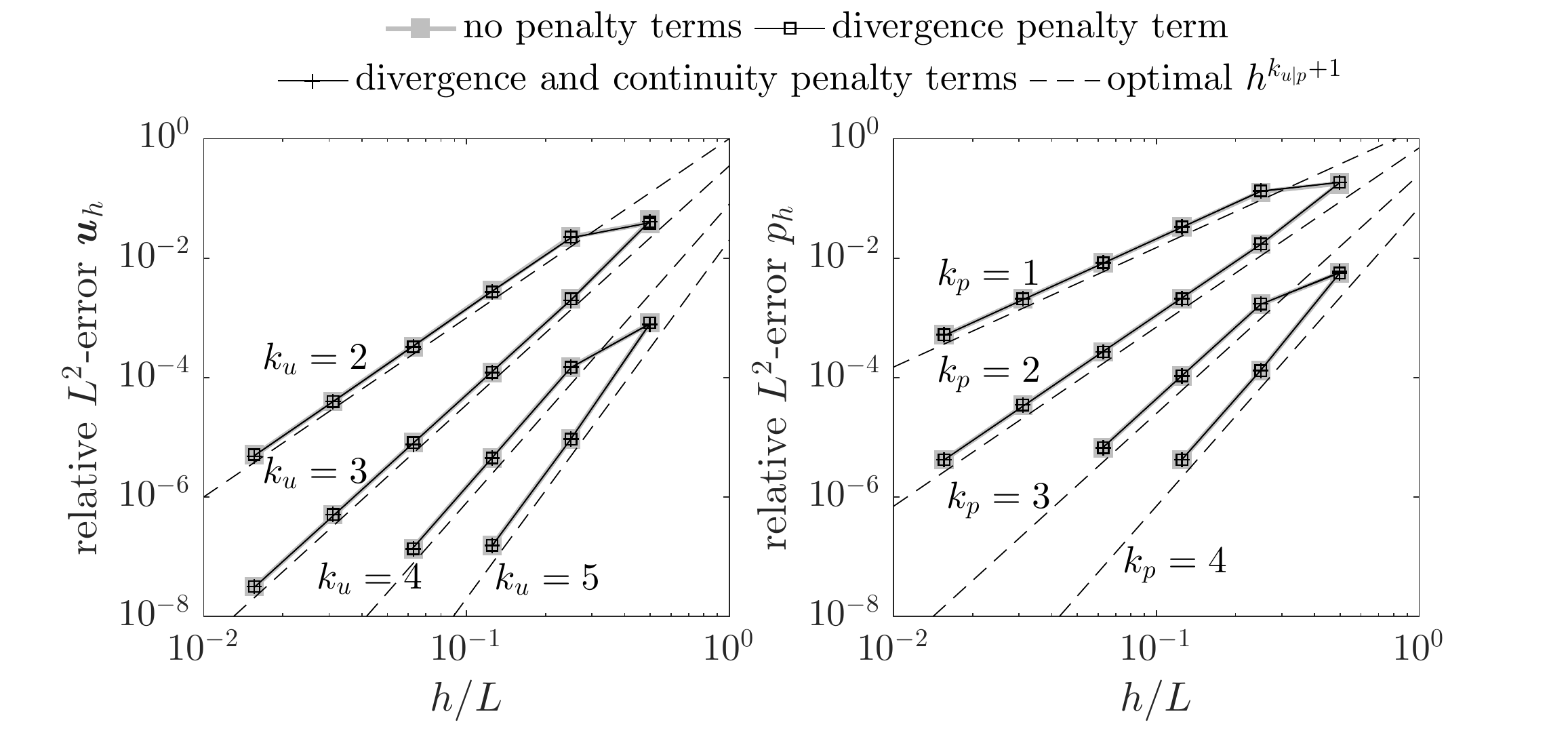}}
 \subfigure[Spatial convergence test: relative~$L^2$-errors of velocity and pressure as a function of~$N_{\mathrm{dofs}}$.]{
	\includegraphics[width=0.75\textwidth]{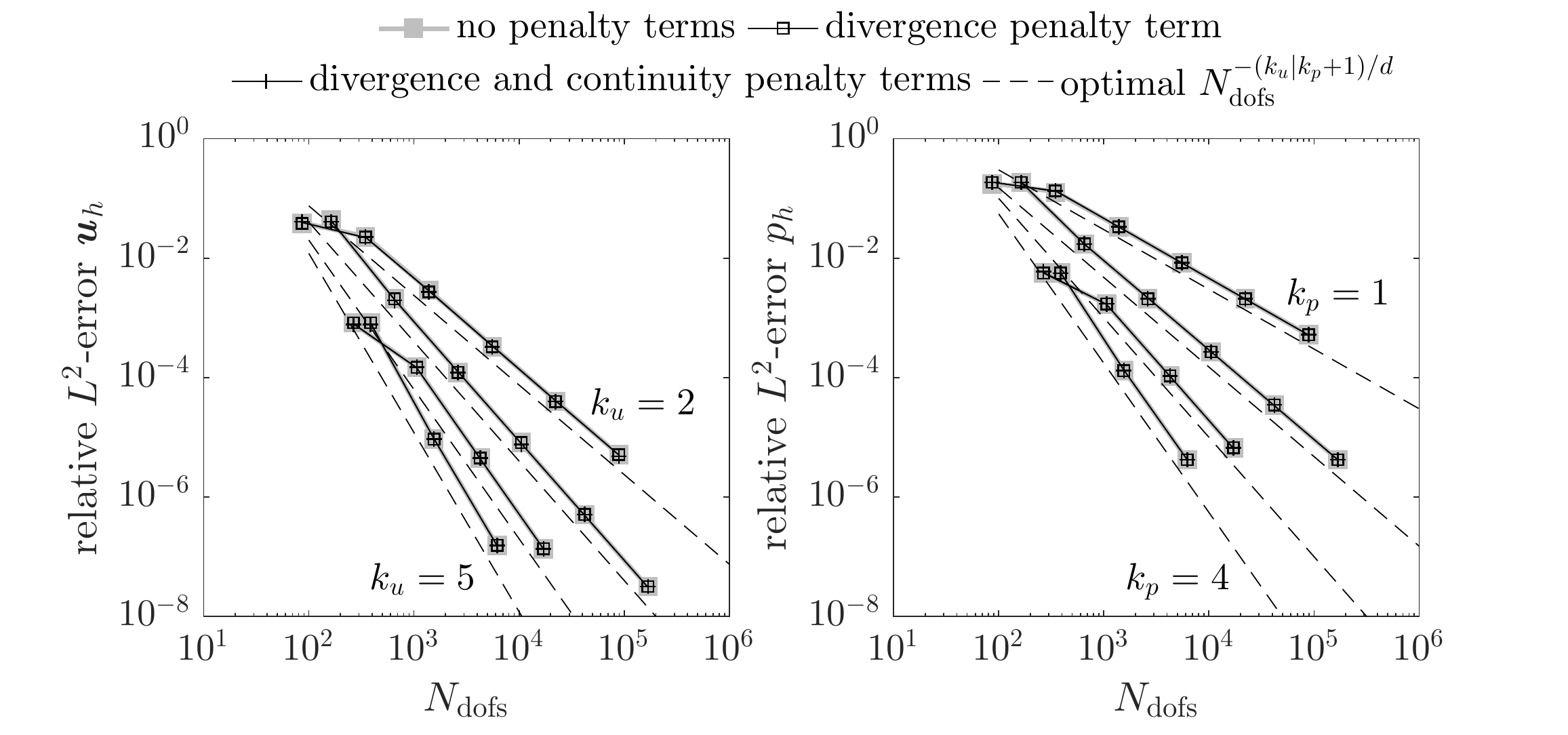}}
\caption{Spatial convergence tests for vortex flow problem: assessment of optimal rates of convergence for various refine levels~$l$ and polynomial degrees~$(k_u,k_p)=(k,k-1)$ with~$k=2,3,4,5$. The BDF2 scheme is used with a fix time step size of~$\Delta t /T = 5\cdot 10^{-5}$.}
\label{fig:spatial_convergence_vortex_problem}
\end{figure}

Figure~\ref{fig:temporal_convergence_vortex_problem} shows results of temporal convergence tests for the standard formulation without penalty terms, the formulation with divergence penalty term, and the formulation with divergence and continuity penalty terms. BDF schemes of order~$J=1,2$ are investigated and a high spatial resolution, refine level~$l=3$ and polynomial degrees~$(k_u,k_p)=(8,7)$, is used in order to make sure that the spatial discretization error is negligible. We mention that instabilities related to the CFL condition do not show up for this particular flow problem which might be due to the fact that the viscosity is comparably large and the fact that the vortex does not move but simply decays over time. The results are indistinguishable for the different formulations and experimental rates of convergence agree with the optimal rates of convergence of order~$\Delta t^J$ for all formulations.

Results of spatial convergence tests are shown in Figure~\ref{fig:spatial_convergence_vortex_problem} for polynomial degrees~$(k_u,k_p)=(k,k-1)$ with~$k=2,3,4,5$. The BDF2 scheme and a fix time step size of~$\Delta t /T = 5\cdot 10^{-5}$ is used so that the overall error is dominated by the spatial discretization error. Again, experimental rates of convergence agree with the optimal rates of convergence of order~$h^{k_{u|p}+1}$ for all formulations. Moreover, the error is shown as a function of the number of unknowns~$N_{\mathrm{dofs}}$ in order to evaluate the efficiency of high polynomial degrees. Here, efficiency is defined as the ratio of accuracy (inverse of error) and the number of unknowns. For this analytical test case, high-order methods are significantly more accurate than low order methods for the same number of unknowns. In the following, similar experiments are performed for more challenging turbulent flow problems.

The results of these convergence tests highlight that applying the divergence and continuity penalty terms in a postprocessing step does not lead to a deterioration of optimal rates of convergence. Hence, this approach is well suited to obtain an accurate DG solver that targets turbulent flows on the one hand but also reproduces the exact solution when applied to laminar flow problems.

\subsection{Orr--Sommerfeld stability problem: Investigation of stability}\label{OrrSommerfeld}
We analyze the stability of the proposed incompressible Navier--Stokes solvers for the Orr--Sommerfeld stability problem applied to the two-dimensional Poiseuille flow problem which has been analyzed, e.g., in~\cite{Shahbazi07,Fischer1997}. The computational domain is a rectangular channel with dimensions~$\left[0,L\right]\times\left[-H,H\right]$. No-slip boundary conditions are prescribed at~$x_2=\pm H$ and periodic boundary conditions in streamwise direction. Due to the periodic boundary conditions a constant body force,~$f_1= 2 \nu U_{\text{max}}/H^2$, has to be prescribed to sustain the mean flow which is given as~$U_1(x_2)/U_{\text{max}} = 1-\left(x_2/H\right)^2$. To obtain the initial solution for this problem one has to solve the Orr--Sommerfeld equation. The Orr--Sommerfeld equation is derived by superimposing a small disturbance upon the mean flow~$\bm{U}(\bm{x}) = (U_1(x_2),0)^T$ that fulfills the incompressible Navier--Stokes equations. The Navier--Stokes equations are subsequently linearized by assuming that the perturbation is small. To obtain the Orr--Sommerfeld equation, the following ansatz is used
\begin{align}
u_1(\bm{x},t) &= U_1(x_2) + \varepsilon\ \mathrm{Re}\left\lbrace \frac{\mathrm{d}\psi(x_2)}{\mathrm{d}x_2}\exp\left(i\left(\alpha x_1 - \omega t\right)\right) \right\rbrace\ ,\label{OrrSommerfeldVelocity1}\\
u_2(\bm{x},t) &= - \varepsilon\ \mathrm{Re}\left\lbrace i\alpha \psi(x_2)\exp\left(i\left(\alpha x_1 - \omega t\right)\right) \right\rbrace\ .\label{OrrSommerfeldVelocity2}
\end{align}
The perturbation velocity is based on the streamfunction~$\Psi(\bm{x},t) = \varepsilon\ \psi(x_2)\exp\left(i\left(\alpha x_1 - \omega t\right)\right)$ with wavenumber~$\alpha$, complex frequency~$\omega$, and perturbation amplitude~$\varepsilon\ll 1$. 
The Orr--Sommerfeld equation then reads
\begin{align*}
i\alpha \left[\left(U_1-\frac{\omega}{\alpha}\right)\left(\psi''-\alpha^2 \psi\right) - U^{\prime \prime}_1\psi \right] = \nu \left(\psi'''' - 2 \alpha^2 \psi'' + \alpha^4 \psi \right) \ ,
\end{align*}
where~$\left(\cdot\right)'=\frac{\mathrm{d \left(\cdot\right)}}{\mathrm{d}x_2}$. The Orr--Sommerfeld equation is a fourth-order homogeneous ordinary differential equation with variable coefficients. The boundary conditions are~$\psi(-H) = \psi(H) = 0$ and~$\psi'(-H) = \psi'(H) = 0$. This equation is typically solved for the complex frequency~$\omega=\omega_{\text{r}}+i\omega_{\text{i}}$ and~$\psi(x_2)$ by prescribing a wavenumber~$\alpha$, a Reynolds number or viscosity~$\nu$, and the mean flow~$U_1(x_2)$. Following~\cite{Fischer1997}, we use the parameters~$H=1$,~$U_{\text{max}}=1$,~$\mathrm{Re}=U_{\text{max}}H/\nu =7500$,~$\alpha=1$, and~$\varepsilon=10^{-5}$. The Orr--Sommerfeld equation is discretized using a spectral Galerkin ansatz with one finite element of high polynomial degree~$k$ (e.g.,~$k=200$) resulting in a generalized eigenvalue problem~$\bm{A} \bm{\Psi}=\lambda \bm{B}\bm{\Psi}$ for the eigenvalue~$\lambda=-i\omega$ and the eigenvector~$\bm{\Psi}=(\Psi_1,...,\Psi_{k+1})$. The initial solution prescribed for the Orr--Sommerfeld stability analysis performed below is given by equations~\eqref{OrrSommerfeldVelocity1} and~\eqref{OrrSommerfeldVelocity2} where~$\omega$ is the complex frequency corresponding to the only unstable eigensolution of the Orr--Sommerfeld equation,~$\omega_{\text{i}} > 0$, and~$\psi(x_2)=\sum_{i=1}^{k+1}N_i^k(x_2)\Psi_i$ is the interpolation of the corresponding eigenvector~$\bm{\Psi}$ which is normalized to a maximum value of 1,~$\max_i \vert\Psi_i\vert = 1$.
The length~$L$ of the computational domain equals the wavelength of the Tollmien--Schlichting (TS) waves,~$L=2\pi/\alpha$. The simulations are performed for the time interval~$0\leq t\leq T = 2 T_0$, where~$T_0=\alpha L / \omega_{\text{r}}$ is the time the TS waves need to travel through the computational domain.

The perturbation energy~$E=\int_{\Omega} \Vert \bm{u} - \bm{U} \Vert^2 \mathrm{d} \bm{x}$ grows exponentially in time according to linear stability theory,~$E(t)/E(t=0) = \exp(2\omega_{\text{i}} t)$. Similar to~\cite{Fischer1997} we use the quantity~$e(t)= \vert \exp(2\omega_{\text{i}} t) - E_h(t)/E_h(t=0)\vert$ as an error measure, where~$E_h$ is the perturbation energy calculated using the numerical solution~$\bm{u}_h$.

The computational domain is discretized using a uniform Cartesian grid where~$l$ denotes the number of refinements with~$N_{\text{el}}(l=0)=1$. The time step size is calculated according to the CFL condition~\eqref{CFL_Condition} with~$\mathrm{Cr} = 0.2$ and~$\Vert \bm {u} \Vert_{\mathrm{max}}=U_{\text{max}}$. The resulting time step sizes are small enough so that the error is dominated by the spatial discretization error. The coupled solution approach is used (BDF2 scheme) using small absolute and relative solver tolerances of~$10^{-14}$. 

%The Laplace formulation of the viscous term is used.

\begin{figure}[!ht]
 \centering 
 \includegraphics[width=0.9\textwidth]{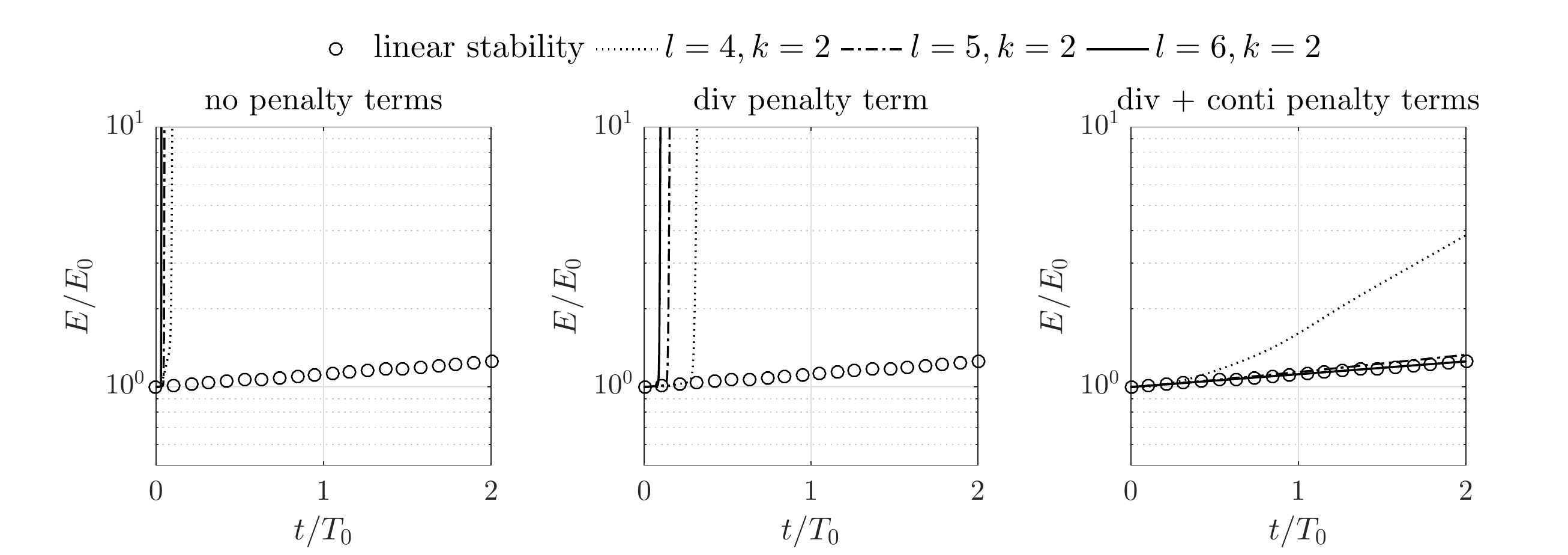}
\caption{Orr--Sommerfeld stability problem: Stability for polynomial degree~$k=2$ and refine levels~$l=4,5,6$ using the coupled solution approach.}
\label{fig:stability_orr_sommerfeld_problem}
\end{figure}

To carefully analyze the robustness and stability properties of our approach we perform simulations for various polynomial degrees~$k=2,4,6,8,10$ and refinement levels~$l=0,1,2,...$ beginning with very coarse spatial resolutions consisting of only one element. For~$k=2$, refinement levels of~$l=1,2,...$ are investigated since the perturbation energy would be zero for refinement level~$l=0$ at initial time~$t=0$. A comparison of the stability of the standard DG discretization without penalty terms and the stabilized approach including the divergence penalty term only as well as both the divergence and continuity penalty terms is shown in Figure~\ref{fig:stability_orr_sommerfeld_problem} for polynomial degree~$k=2$ and refinement levels~$l=4,5,6$. For the standard formulation without penalty terms, unphysical growth of the perturbation energy occurs as observed in~\cite{Shahbazi07}. Using the divergence penalty term we observe similar instabilities and the stability is only marginally improved as compared to the basic DG discretization. While such instabilities are not observed for the other polynomial degrees, these results highlight that the divergence penalty based postprocessing does not lead to a robust discretization scheme in general. Using both divergence and continuity penalty terms, the method is stable for all refinement levels and all polynomial degrees. Moreover, the results shown in Figure~\ref{fig:stability_orr_sommerfeld_problem} demonstrate that the solution tends towards the theoretical value for increasing spatial resolution. We also analyzed the stability for the dual splitting scheme and pressure-correction scheme and the results are in qualitative agreement with those shown here for the coupled solution approach.

\begin{figure}[!ht]
 \centering 
 \includegraphics[width=0.5\textwidth]{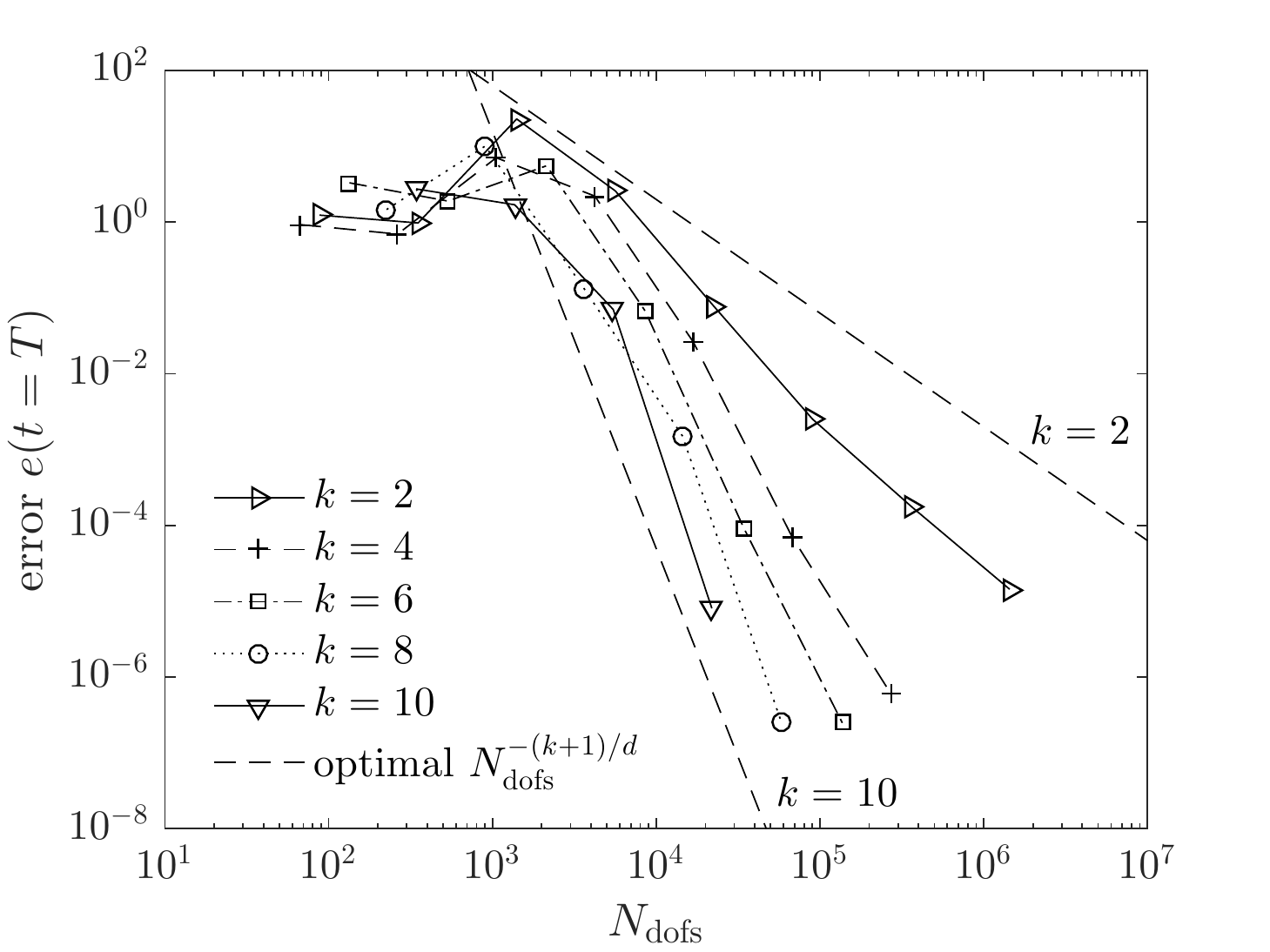}
\caption{Orr--Sommerfeld stability problem: Convergence tests for polynomial degrees~$k=2,4,6,8,10$ using the coupled solution approach.}
\label{fig:convergence_orr_sommerfeld_problem}
\end{figure}

To analyze the convergence properties for the Orr--Sommerfeld stability problem in more detail, Figure~\ref{fig:convergence_orr_sommerfeld_problem} shows the error as a function of the number of unknowns for various polynomial degrees and refinement levels where we use the formulation with both divergence and continuity penalty terms. Optimal rates of convergence are obtained for higher refinement levels for all polynomial degrees which is to be expected since the solution is relatively smooth for the Orr--Sommerfeld problem. For~$N_{\mathrm{dofs}}>10^3$, high-order methods are systematically more accurate than low-order methods for the same number of unknowns and the number of unknowns required to reach a certain level of accuracy can be reduced by a factor of~$10$ to~$100$ for high polynomial degrees as compared to the low-order method with~$k=2$.

\begin{remark}
Although we consider mixed-order polynomials in the present paper, we mention that we also observe an unphysical growth of the perturbation energy for equal-order polynomials on coarse meshes when using the basic DG discretization. This is in contrast to the results in~\cite{Shahbazi07}, where stability has been obtained for equal-order polynomials. This might be explained by the fact that the spatial resolutions considered here are significantly coarser than the ones analyzed in~\cite{Shahbazi07}, where a mesh consisting of~$128$ triangles is considered for polynomial degrees~$k=6,7,8$.
\end{remark}

\subsection{3D Taylor--Green vortex problem}\label{TaylorGreenVortexProblem}
The three-dimensional Taylor--Green vortex problem~\cite{Taylor1937} is a widely used benchmark problem for the numerical solution of turbulent flows. It is characterized by a simple, laminar initial field breaking down into complex turbulent flow structures. While this problem has already been studied in~\cite{Brachet1983,Brachet1991} using direct spectral numerical simulation, the Taylor--Green vortex problem still poses a challenging problem on modern computer hardware. In the context of high-order DG discretization methods, this test case has been analyzed in~\cite{Gassner2013,Wiart14} for compressible Navier--Stokes solvers and in~\cite{Piatkowski16} using an incompressible Navier--Stokes solver.

\subsubsection{Problem description}
\begin{figure}[!ht]
 \centering 
 \subfigure[$t=0$]{\includegraphics[width=0.32\textwidth]{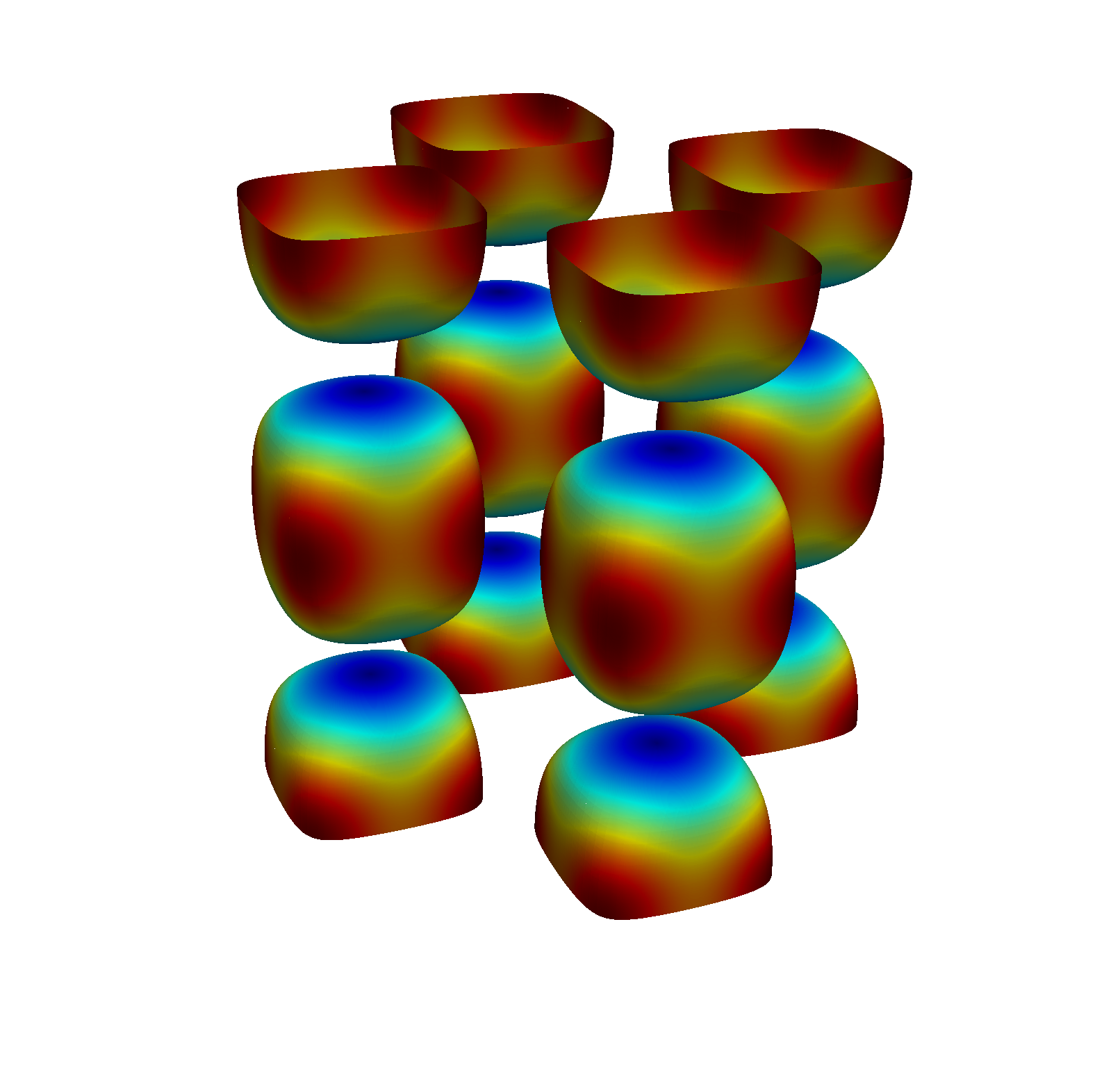}}
 \subfigure[$t=T/2$]{\includegraphics[width=0.32\textwidth]{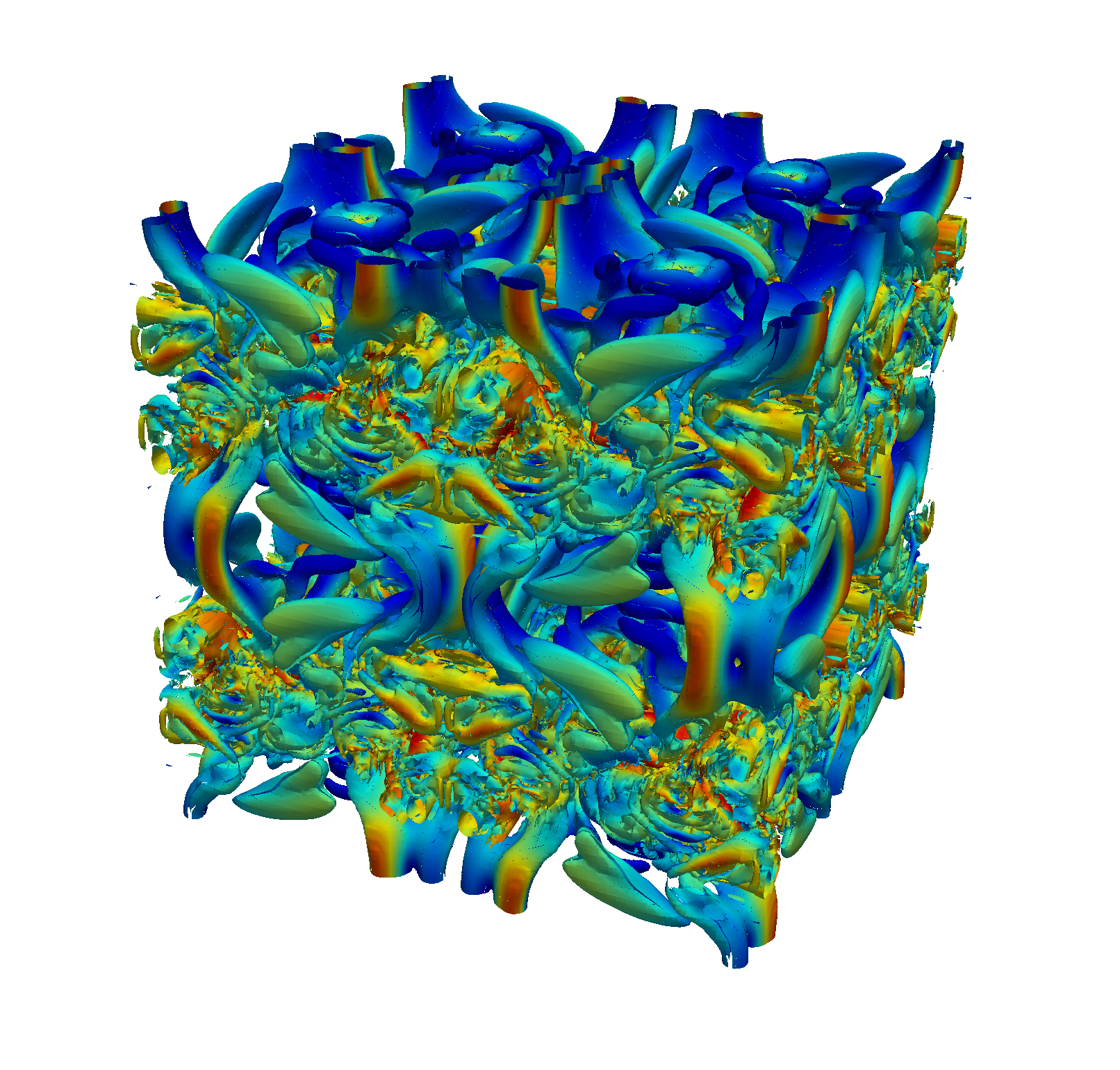}}
 \subfigure[$t=T$]{\includegraphics[width=0.32\textwidth]{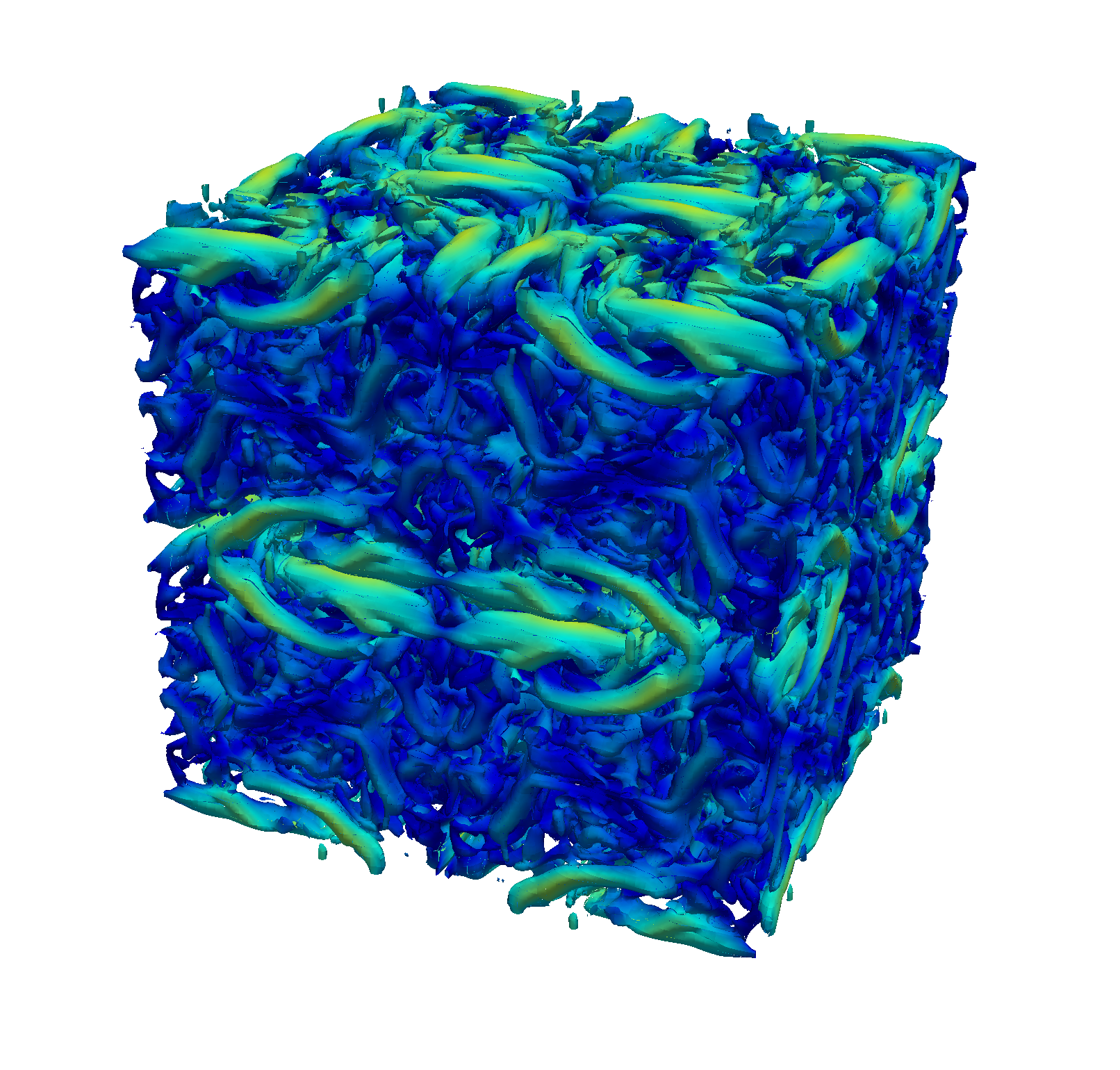}}
\caption{Visualization of Taylor--Green vortex problem for~$\mathrm{Re}=1600$: Iso-surfaces of Q-criterion (value of~$0.1 U_0/L$) colored by velocity magnitude (same color scale for all time instants, red indicates high velocity and blue low velocity). The numerical LES approach including divergence and continuity penalty terms is used and a spatial resolution of~$l=6$ and~$(k_u,k_p)=(3,2)$ corresponding to an effective resolution of~$256^3$ velocity degrees of freedom.}
\label{fig:TaylorGreen_Re1600_Visualization}
\end{figure}

The computational domain~$\Omega_h$ is a box~$\Omega_h = [-\pi L,\pi L]^3$ where~$L$ is a characteristic length scale. Periodic boundary conditions are prescribed in all coordinate directions and the body force vector is zero,~$\bm{f}=\bm{0}$. The initial solution for the three-dimensional Taylor--Green vortex problem is given as
\begin{align*}
u_1(\bm{x},t=0) &= U_0\sin\left(\frac{x_1}{L} \right)\cos\left(\frac{x_2}{L}\right)\cos\left(\frac{x_3}{L}\right)\; ,\\
u_2(\bm{x},t=0) &= -U_0\cos\left(\frac{x_1}{L} \right)\sin\left(\frac{x_2}{L}\right)\cos\left(\frac{x_3}{L}\right)\; ,\\
u_3(\bm{x},t=0) &= 0\; ,\\
p(\bm{x},t=0) &= p_0 + \frac{U_0^2}{16}\left(\cos\left(\frac{2x_1}{L}\right)+ \cos\left(\frac{2x_2}{L}\right)\right)\left(\cos\left(\frac{2x_3}{L}\right)+2 \right)\; .
\end{align*}
The Reynolds number is defined as~$\mathrm{Re}=\frac{U_0 L}{\nu}$. As in~\cite{Gassner2013,Wiart14,Piatkowski16}, we consider a Reynolds number of~$\mathrm{Re}=1600$. We use the parameters~$p_0=0$ and~$U_0=1$,~$L=1$ so that the viscosity is given as~$\nu=\frac{1}{\mathrm{Re}}$. The simulations are performed for the time interval~$0\leq t \leq T$ where we use an end time of~$T=20 T_0$ with~$T_0=\frac{L}{U_0}$. The mesh is discretized using a uniform Cartesian grid consisting of~$(2^l)^3$ elements where~$l$ denotes the level of refinement. For this test case, the coupled solution approach is used with absolute solver tolerances of~$10^{-12}$ and relative solver tolerances of~$10^{-6}$. A visualization of the solution at times~$t=0,T/2,T$ is shown in Figure~\ref{fig:TaylorGreen_Re1600_Visualization}.

\subsubsection{Investigation of stability}
% continuity penalty term: penalize normal component of velocity only

In a first step, we analyze the stability of the discretization approach by performing simulations over a wide range of refinement levels and polynomial degrees. We consider polynomial degrees of~$k=3,7,15$ since these values allow to construct spatial resolutions with the same number of velocity degrees of freedom when using a mesh with uniform refinements starting from one element. In addition, we consider polynomial degree~$k=2$ as a reference lower order method. The coarsest meshes considered in the following correspond to refine level~$l=0$, except for~$k=2$ where we start with~$l=1$ since the velocity field would be zero after interpolation of the initial condition for~$l=0$. The number of unknowns of the vector-valued velocity is~$ d (2^l(k+1))^d$. In the following, we use the quantity~$(2^l (k+1))^d$ as an effective mesh resolution and denote this quantity, for simplicity, as the number of velocity degrees of freedom. In terms of velocity degrees of freedom the coarsest meshes correspond to effective resolutions of~$6^3,4^3,8^3,16^3$ dofs for~$k=2,3,7,15$, respectively. The refinement level is limited by considering effective resolutions of at most~$256^3$ velocity degrees of freedom for~$k=3,7,15$ (refinement levels~$l=6,5,4$, respectively) and~$384^3$ velocity degrees of freedom for~$k=2$ (refinement level~$l=7$). To the best of our knowledge, such coarse spatial discretizations have not been analyzed before for the 3D Taylor--Green vortex problem in the context of high-order DG discretizations. For example, the spatial resolution are~$64^3$ velocity dofs in~\cite{Gassner2013}, at least~$192^3$ velocity dofs in~\cite{Chapelier2014,Wiart14}, and~$420^3$ velocity dofs in~\cite{Piatkowski16}. In contrast, we analyze the numerical robustness of the proposed methods not only for moderately under-resolved problems but in the limit of coarse spatial resolutions.

\begin{table}[!h]
\caption{3D Taylor--Green vortex problem for~$\mathrm{Re}=1600$: stability of the proposed discretization approach including divergence and continuity penalty terms as compared to standard DG discretization scheme without penalty terms. A successful completion of the simulation is indicated by \Checkmark, while instabilities leading to a crash of the simulation are denoted by \XSolidBrush. The sign '$-$' indicates that the specific spatial resolution is not considered for reasons explained in the text.}\label{Taylor_Green_Vortex_Stability}
\renewcommand{\arraystretch}{1.1}
\begin{center}
\begin{tabular}{ccccccccccc}
\hline
\multicolumn{5}{c}{No penalty terms} & &\multicolumn{5}{c}{Divergence and continuity penalty terms}\\
\cline{1-5} \cline{7-11}
$l$ & \multicolumn{4}{c}{Polynomial degree $(k_u,k_p)=(k,k-1)$} & & $l$ & \multicolumn{4}{c}{Polynomial degree $(k_u,k_p)=(k,k-1)$}\\
\cline{2-5} \cline{8-11}
& $k=2$ & $k=3$ & $k=7$ & $k=15$ & & & $k=2$ & $k=3$ & $k=7$ & $k=15$\\
\hline
0	 & $-$          & \Checkmark    & \XSolidBrush & \XSolidBrush &  & 0 & $-$        & \Checkmark & \Checkmark & \Checkmark \\
1	 & \Checkmark   & \Checkmark    & \XSolidBrush & \XSolidBrush &  & 1 & \Checkmark & \Checkmark & \Checkmark & \Checkmark \\
2	 & \Checkmark   & \Checkmark    & \XSolidBrush & \Checkmark   &  & 2 & \Checkmark & \Checkmark & \Checkmark & \Checkmark \\
3	 & \Checkmark   & \XSolidBrush  & \XSolidBrush & \Checkmark   &  & 3 & \Checkmark & \Checkmark & \Checkmark & \Checkmark \\
4	 & \Checkmark   & \XSolidBrush  & \Checkmark   & \Checkmark	  &  & 4 & \Checkmark & \Checkmark & \Checkmark & \Checkmark \\
5	 & \Checkmark   & \Checkmark    & \Checkmark   & $-$          &  & 5 & \Checkmark & \Checkmark & \Checkmark & $-$        \\
6	 & \Checkmark	& \Checkmark	& $-$      	   & $-$          &  & 6 & \Checkmark & \Checkmark & $-$        & $-$        \\
7	 & \Checkmark	& $-$       	& $-$          & $-$          &  & 7 & \Checkmark & $-$        & $-$        & $-$        \\
\hline
\end{tabular}
\end{center}
\renewcommand{\arraystretch}{1}
\end{table}

Results of this stability experiment are summarized in Table~\ref{Taylor_Green_Vortex_Stability} where we compare the stability of the basic DG discretization without penalty terms to the proposed formulation including both divergence and continuity penalty terms. For the stabilized approach, stability is obtained for all spatial resolutions indicated by \Checkmark, where the Courant numbers used for the simulations are in the range~$0.1 \leq \mathrm{Cr} \leq 0.4$. The basic DG discretization without penalty terms is unstable for several spatial resolutions which is indicated by \XSolidBrush. In general, smaller Courant numbers have to be used for the reference formulation as compared to the formulation with divergence and continuity penalty terms to ensure stability. Hence, to make sure that the observed instabilities are not related to a violation of the CFL condition, the unstable simulations have been repeated for several smaller time step sizes. Starting with the individual~$\mathrm{Cr}$ values used for the stabilized approach,~$\mathrm{Cr}=\mathrm{Cr}_{\mathrm{stabilized}}$, we reduce the time step size in factors of two until stability is obtained. A specific spatial resolution is only denoted as unstable if the simulation becomes unstable for all~$\mathrm{Cr}=\mathrm{Cr}_{\mathrm{stabilized}}/2^m$ with~$m=0,1,...,4$. To sum up, the results in Table~\ref{Taylor_Green_Vortex_Stability} demonstrate that the proposed stabilization approach performs significantly better in terms of robustness and stability. When using the divergence penalty term only, stability has been obtained for all spatial resolutions considered in Table~\ref{Taylor_Green_Vortex_Stability}, but, in general, both penalty terms are necessary in order to avoid a negative numerical dissipation in agreement with the estimates~\eqref{EstimateDissipation} and~\eqref{EstimateDissipationWithPenaltyTerms}, see also the discussion in the following section.

\subsubsection{Investigation of accuracy as well as efficiency of high-order methods}
The accuracy of the numerical results obtained for the Taylor--Green vortex problem is evaluated by calculating the kinetic energy as well as dissipation rates of the kinetic energy similar to the analysis performed in~\cite{Gassner2013,Wiart14} for high-order DG discretizations of the compressible Navier--Stokes equations. The total kinetic energy is defined as
\begin{align*}
E_{\mathrm{k}} = \frac{1}{V_{\Omega_h}}\int_{\Omega_h} \frac{1}{2} \bm{u}_h\cdot \bm{u}_h\; \mathrm{d}\Omega \; ,
\end{align*}
where the volume of the computational domain is~$V_{\Omega_h}=\int_{\Omega_h} 1\; \mathrm{d}\Omega$. These integrals are calculated numerically using Gaussian quadrature with~$k_u+1$ quadrature points. The kinetic energy dissipation rate~$-\frac{\mathrm{d}E_{\mathrm{k}}}{\mathrm{d}t}$ is calculated using a second order accurate central difference scheme~$-\frac{\mathrm{d}E_{\mathrm{k}}}{\mathrm{d}t}(t=t_i)= - \frac{E_{\mathrm{k}}^{i+1}-E_{\mathrm{k}}^{i-1}}{t_{i+1}-t_{i-1}}$ for interior time points~$i=1,...,N-1$ and a first order accurate, one-sided finite difference scheme for the end points~$i=0,N$, where~$N$ denotes the number of time steps.
The molecular energy dissipation rate~$\varepsilon$ (also denoted as the dissipation rate of the resolved scales) is calculated as
\begin{align*}
\varepsilon = \frac{\nu}{V_{\Omega_h}} \int_{\Omega_h} \Grad{\bm{u}_h} : \Grad{\bm{u}_h}\; \mathrm{d}\Omega \; ,
\end{align*}
again using Gaussian quadrature as described above to calculate the integrals numerically. Using the kinetic energy dissipation rate~$-\frac{\mathrm{d}E_{\mathrm{k}}}{\mathrm{d}t}$ and the molecular dissipation rate~$\varepsilon$, the numerical dissipation of the discretization scheme is~$-\frac{\mathrm{d}E_{\mathrm{k}}}{\mathrm{d}t}-\varepsilon$.

% (a) NO PENALTY, and (b) DIV + NORMAL CONTI PENALTY, effective resolution 64^3
\begin{figure}[!ht]
 \centering 
 \subfigure[Standard DG discretization without penalty terms.]{
	\includegraphics[width=1.0\textwidth]{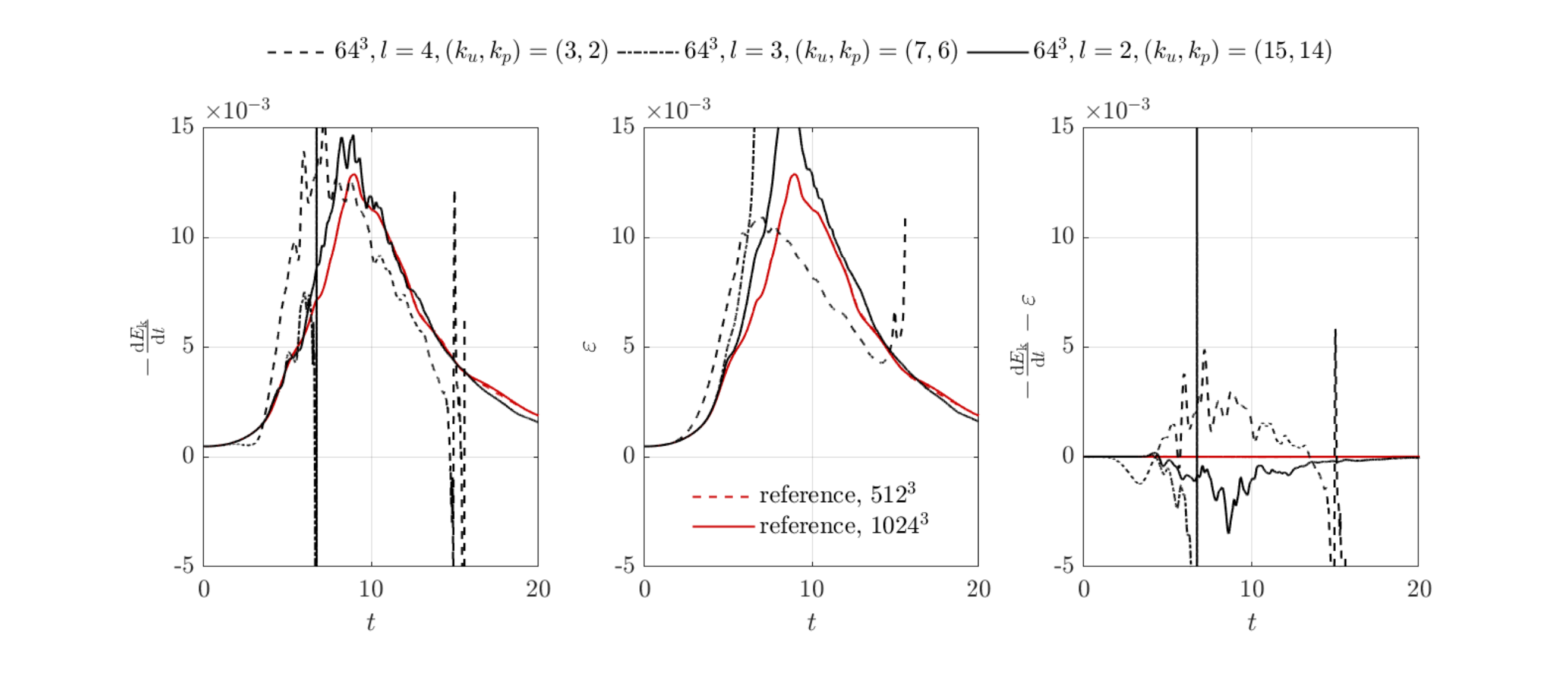}}
 \subfigure[Stabilized approach including both divergence and continuity penalty terms.]{
	\includegraphics[width=1.0\textwidth]{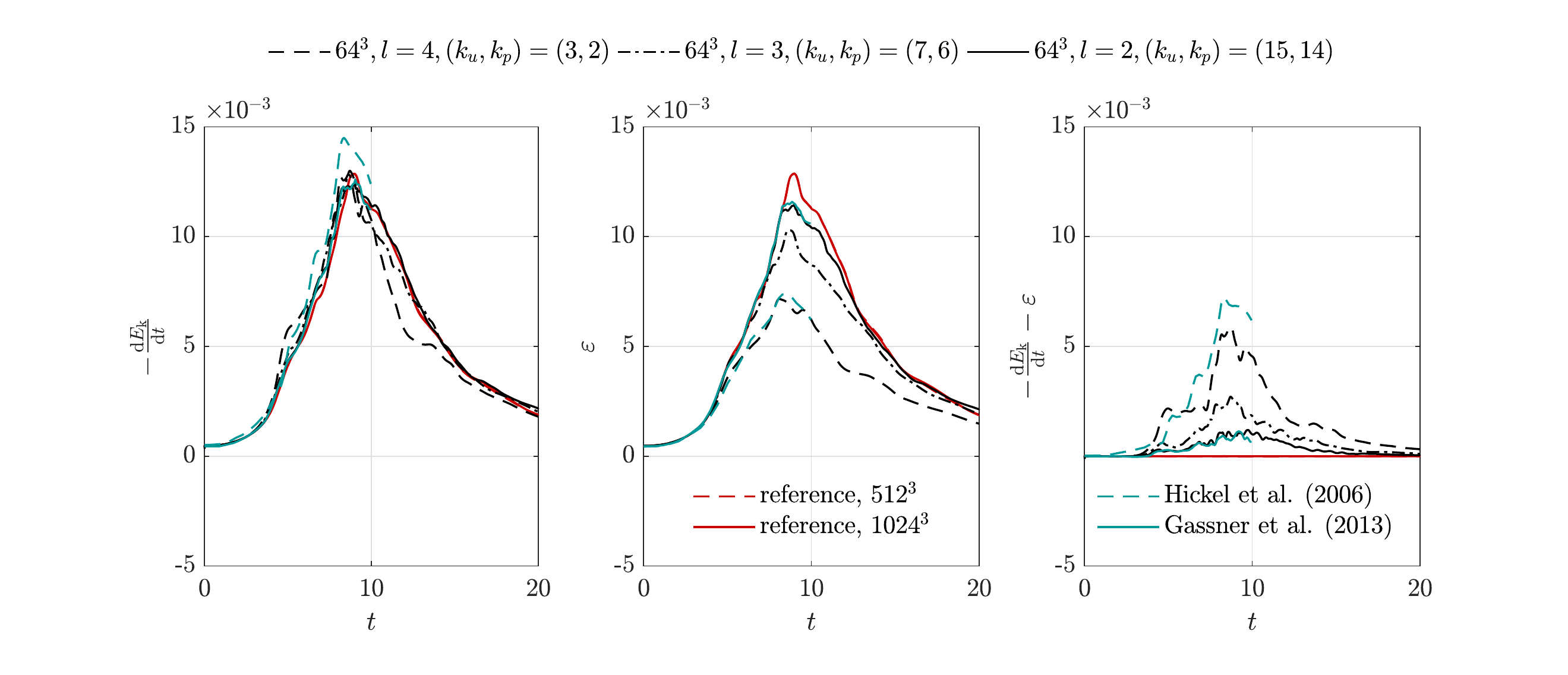}}
\caption{3D Taylor--Green vortex problem for~$\mathrm{Re}=1600$: Rate of change of kinetic energy, molecular dissipation, and numerical dissipation as a function of time using an effective resolution of~$64^3$ velocity degrees of freedom for polynomial degrees~$k=3,7,15$. The reference solution is computed on a mesh with~$l=8$ and~$k=3$ ($1024^3$ velocity dofs).}
\label{fig:3D_Taylor_Green_Kinetic_Energy_Dissipation}
\end{figure}

% DIV + NORMAL CONTI PENALTY
\begin{figure}[!ht]
 \centering 
 \includegraphics[width=0.9\textwidth]{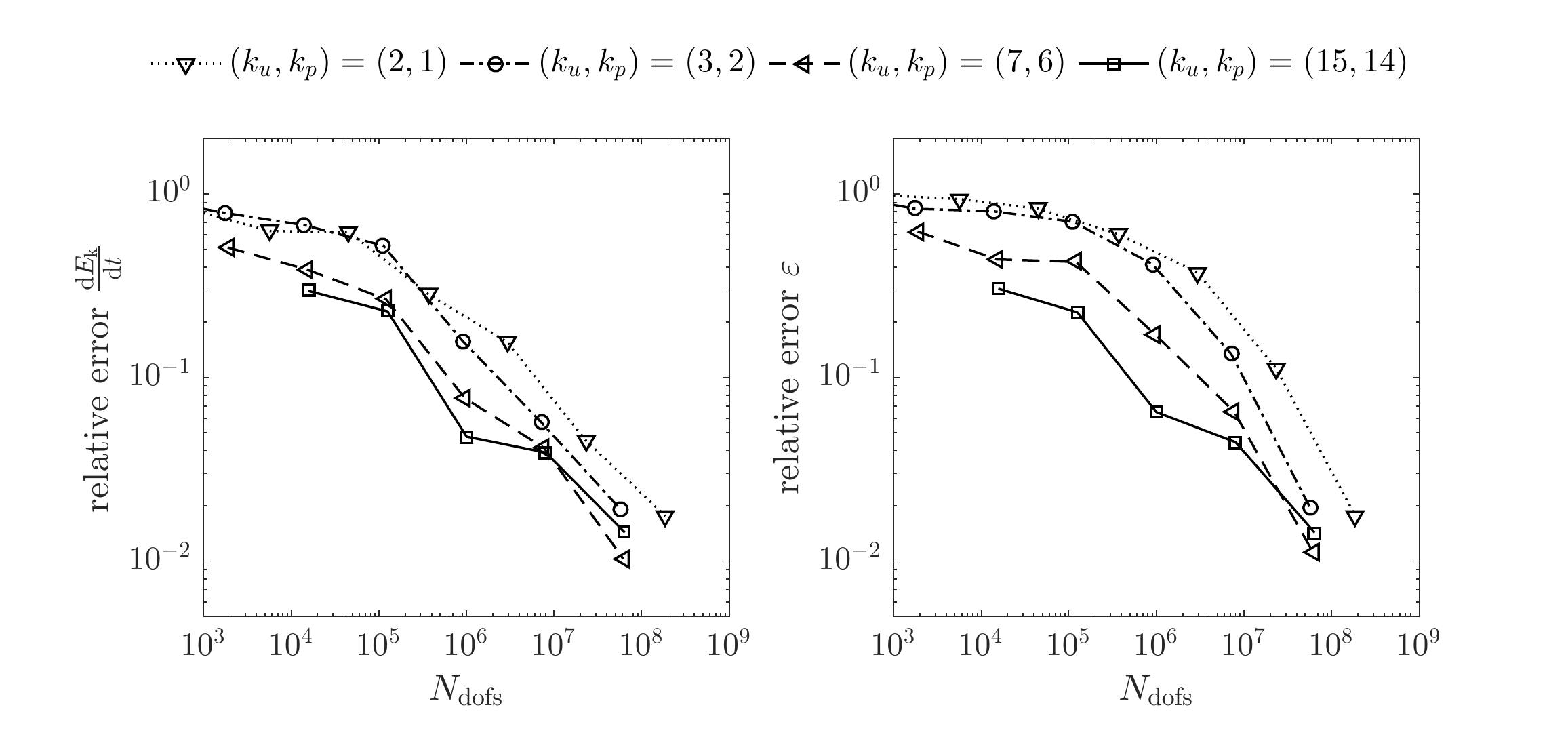}
\caption{3D Taylor--Green vortex problem for~$\mathrm{Re}=1600$: Efficiency of high-order discretizations in terms of accuracy versus number of unknowns for the proposed approach including divergence and continuity penalty terms. The reference solution used to calculate the errors is computed on a mesh with~$l=8$ and~$k=3$ ($1024^3$ velocity dofs).}
\label{fig:3D_Taylor_Green_Accuracy}
\end{figure}

Numerical results for the kinetic energy dissipation rate, the molecular dissipation, and the numerical dissipation are shown in Figure~\ref{fig:3D_Taylor_Green_Kinetic_Energy_Dissipation} where we compare the results obtained for the standard DG discretization and the stabilized approach including both divergence and continuity penalty terms. The spatial resolution is~$64^3$ velocity dofs for all polynomial degrees~$k=3,7,15$ and the results are compared to two reference solutions computed on meshes with refine levels~$l=7,8$ and polynomial degree~$k=3$ ($512^3,1024^3$ velocity degrees of freedom). To the best of our knowledge, these are the first results for such a fine resolution for the Taylor--Green vortex problem using a high-order DG code. The two reference solutions are almost indistinguishable so that these simulations can be considered as converged. For the reference solutions, the numerical dissipation is close to zero and the kinetic energy dissipation is completely realized by the molecular dissipation highlighting that the relevant flow structures are well-resolved for this spatial resolution. For the standard DG discretization without penalty terms, the numerical dissipation takes large negative values for all polynomial degrees. Instabilities occur for polynomial degrees~$k=3,7$, while stability is obtained for~$k=15$, see also the results in Table~\ref{Taylor_Green_Vortex_Stability}. These results highlight that the basic DG discretization scheme is not a robust and accurate method for under-resolved problems. Adding the consistent divergence and continuity penalty terms, stability is obtained for all polynomial degrees. Moreover, we observe that the numerical dissipation is positive~\footnote{Please note that positive is here used in an inexact way in the sense that the divergence and continuity penalty terms do not ensure that the numerical discretization scheme is strictly dissipative from a mathematical point of view. However, our numerical results show that small negative values of the numerical dissipation have a small impact as compared to the dissipation introduced by the stabilization terms.} for all polynomial degrees. An interesting fact is that high-order methods are systematically more accurate with respect to the molecular dissipation for the same number of unknowns. This leads to the conclusion that a larger part of the spectrum is resolved as compared to lower polynomial degrees. At the same time, the numerical dissipation decreases for increasing polynomial degree. Considering the kinetic energy dissipation rate, the improved resolution properties of high-order methods are less obvious. For example, the maximum dissipation rate is captured very well for all polynomial degrees. Thus, the missing molecular dissipation for lower polynomial degrees due to underresolution is, to a large extent, compensated by the increased numerical dissipation of lower polynomial degrees. This motivates to analyze the efficiency of high-order polynomial degrees in more detail in the following. Moreover, we compare our approach to the ALDM-LES approach of Hickel et al.~\cite{Hickel2006} for the same effective resolution of~$64^3$ and the compressible high-order DG, implicit LES computations of Gassner et al.~\cite{Gassner2013} for refine level~$l=2$ and polynomial degree~$k=15$, also resulting in an effective resolution of~$64^3$. For~$k=15$, our results agree very well with the compressible high-order DG results~\cite{Gassner2013}. The molecular dissipation of the ALDM model~\cite{Hickel2006} is similar to our~$k=3$ computations, while the maximum dissipation rate~$-\frac{\mathrm{d}E_{\mathrm{k}}}{\mathrm{d}t}$ is overpredicted by the ALDM model, which also shows the highest numerical dissipation. Hence, our results are clearly competitive to state-of-the-art LES methods in terms of accuracy.
Using the dual splitting scheme for efficient time integration and Courant numbers close to the critical Courant number, the computational costs (wall time~$t_{\mathrm{wall}}$ times number of cores~$N_{\mathrm{cores}}$) on an Intel Haswell system are~$0.50$,~$0.72$,~$2.8$ CPUh for polynomial degrees~$k=3,7,15$ and an effective resolution of~$64^3$. Compared to performance numbers published in~\cite{Gassner2013,Wiart14} for state-of-the-art high-order DG solvers for the compressible Navier--Stokes equations, these numbers suggest that our approach allows to reduce the computational costs by at least one order of magnitude for the same spatial resolution parameters. A detailed investigation of the computational efficiency of our approach is beyond the scope of this work and is detailed in a seperate publication~\cite{Fehn18b}.

In order to evaluate the efficiency of high polynomial degrees quantitatively, we define the relative~$L^2$-error norms for the kinetic energy dissipation rate~$\frac{\mathrm{d}E_{\mathrm{k}}}{\mathrm{d}t}$ and the molecular dissipation~$\varepsilon$
\begin{align*}
e_{E_{\mathrm{k}}}^2 = \frac{\int_{t=0}^{T} \left(  \frac{\mathrm{d}E_{\mathrm{k}}(t)}{\mathrm{d}t} - \frac{\mathrm{d}E_{\mathrm{k},\mathrm{ref}}(t)}{\mathrm{d}t} \right)^{2} \mathrm{d}t}{\int_{t=0}^{T} \left( \frac{\mathrm{d}E_{\mathrm{k},\mathrm{ref}}(t)}{\mathrm{d}t} \right)^{2} \mathrm{d}t} \;\; , \;\; 
e_{\varepsilon}^2 = \frac{\int_{t=0}^{T} \left(  \varepsilon(t) - \varepsilon_{\mathrm{ref}}(t) \right)^{2} \mathrm{d}t}{\int_{t=0}^{T} \left( \varepsilon_{\mathrm{ref}}(t) \right)^{2} \mathrm{d}t}   \;\; .
\end{align*}
The above integrals are calculated numerically using the trapezoidal rule where the data points (corresponding to the results written after each time step) for coarser spatial resolutions are first interpolated to the data points of the reference solution before evaluating the trapezoidal rule with the time step size of the reference simulation as step size. The reference solution with~$1024^3$ dofs is used to calculate the errors.

Numerical results of this analysis are shown in Figure~\ref{fig:3D_Taylor_Green_Accuracy} for the stabilized approach. To make sure that the error is not significantly influenced by an inaccurate reference solution, the errors have been analyzed for both reference solutions with~$512^3$ and~$1024^3$ dofs leading to very similar results. Moreover, it was found that time step sizes close to the CFL condition ensure that the temporal discretization error is small as compared to the overall error. In the context of high-order DG methods for the compressible Navier--Stokes equations, it was found in~\cite{Gassner2013} that high-order polynomials are superior by comparing results for~$k=15$ to a low order method with~$k=1$, while the authors of~\cite{Wiart14} conclude that the accuracy can not be improved significantly for polynomial degrees larger than~$k=3$ for coarse grids. Polynomials of degree~$k=3$ and~$k=5$ are analyzed in~\cite{Chapelier2014} where it is found that more accurate results can be obtained for the higher polynomial degree. The present results provide a more detailed view of this aspect in the context of DG discretizations of the incompressible Navier--Stokes equations considered here. The results are systematically more accurate for increasing polynomial degree. We observe that the gain in efficiency of high-order methods is reduced for the kinetic energy dissipation rate~$\frac{\mathrm{d}E_{\mathrm{k}}}{\mathrm{d}t}$ as compared to the molecular dissipation~$\varepsilon$, which is in agreement with the results shown in Figure~\ref{fig:3D_Taylor_Green_Kinetic_Energy_Dissipation}. For~$k=15$, the~$\frac{\mathrm{d}E_{\mathrm{k}}}{\mathrm{d}t}$ error can not be significantly improved as compared to~$k=7$.

\subsection{Turbulent channel flow}
Finally, we consider the turbulent channel flow problem which is a widely used benchmark problem to validate LES models for turbulent, wall-bounded flows. In the context of high-order discontinuous Galerkin discretizations, this test case has been analyzed in~\cite{Ramakrishnan2004,Chapelier2014,Wiart15,Wiart15b} for the compressible Navier--Stokes equations and in~\cite{Krank2017} for the incompressible Navier--Stokes equations.
\subsubsection{Problem description}
The computational domain~$\Omega_h$ is a rectangular box with physical dimensions~$(L_1,L_2,L_3)=(2\pi \delta,2\delta, \pi \delta)$ where~$\delta$   denotes the channel half-width. On the walls located at~$x_2=\pm \delta$, no-slip Dirichlet boundary conditions are prescribed,~$\bm{g}=\bm{0}$, while periodic boundaries are used in the streamwise ($x_1$) and spanwise ($x_3$) directions. For an improved resolution of large velocity gradients close to the no-slip boundaries a mesh stretching is applied in~$x_2$-direction, where we use the hyperbolic mesh stretching function~$f: [0,1]\rightarrow [-\delta,\delta]$ defined in~\cite{Krank2017}
\begin{align}
x_2 \mapsto f(x_2) = \delta \frac{\tanh(C(2x_2-1))}{\tanh(C)} \; .
\end{align}
The parameter~$C$ defines the mesh stretching and we use a value of~$C=1.8$ for all turbulent channel flow simulations in the following. Due to the homogeneity of the flow in streamwise and spanwise directions, elements are distributed equidistantly in~$x_1$- and~$x_3$-directions.

The friction Reynolds number is~$\mathrm{Re}_{\tau}=u_{\tau}\delta/\nu$, where~$u_{\tau}=\sqrt{\tau_{\mathrm{w}}/\rho}$ denotes the wall friction velocity defined as a function of the wall shear stress~$\tau_{\mathrm{w}}$ and the density~$\rho$. The body force vector~$\bm{f}$ driving the flow acts in~$x_1$-direction,~$\bm{f}=(f_1,0,0)$. Defining~$\rho=1$,~$\delta = 1$, and~$f_1=1$, a balance of forces in~$x_1$-direction implies~$\tau_{\mathrm{w}}=1$ and~$u_{\tau}=1$, so that the viscosity~$\nu$ is given as~$\nu=1/\mathrm{Re}_{\tau}$.

To evaluate the accuracy of the results, the profiles in wall normal direction of the mean velocity~$\left<u_1\right>$, the root--mean--square values~$\mathrm{rms}(u_i)=\left<{u_i^{\prime}}^2 \right>^{\frac{1}{2}} $,~$i=1,..,d$, and the Reynolds shear stress~$\left< u_1^{\prime} u_2^{\prime}\right>$ are considered, where statistical averages are denoted as~$\left< \cdot \right>$ and fluctuations as~$(\cdot)^{\prime}=(\cdot)-\left<\cdot\right>$. Velocities and components of the Reynolds stress tensor are normalized using the numerically calculated friction velocity~$u_{\tau}$ leading to~$u_1^+=\left<u_1\right>/u_{\tau}$,~$(u_i^{\prime})^+ = \mathrm{rms}(u_i)/u_{\tau}$, and~$(u_1^{\prime}u_2^{\prime})^+=\left< u_1^{\prime} u_2^{\prime}\right>/u_{\tau}^2$. The dimensionless wall normal coordinate is~$x_2^+ = (x_2+1)/l^+$ with~$l^+=\nu/u_{\tau}$. The simulations are performed for the time interval~$0\leq t \leq 50$. The sampling of statistical data is performed over a time period of~$\Delta t_{\mathrm{sampling}}=20$ ($t_{\mathrm{sampling,start}}=30$ and~$t_{\mathrm{sampling,end}}=50$) and results are sampled every~$10^{\mathrm{th}}$ time step. Note that the same numerical setup in terms of grid stretch factor and penalty factors of the divergence and continuity penalty terms are used for all mesh refinement levels and for all Reynolds numbers.

The time step size is calculated according to equation~\eqref{CFL_Condition} with~$\Vert \bm {u} \Vert_{\mathrm{max}}=22$. A Courant number of~$\mathrm{Cr}=1$ is used in general and smaller~$\mathrm{Cr}$ numbers if necessary to ensure stability. We found that the temporal discretization error is negligible for time step sizes smaller than the critical value according to the CFL condition. Different solution strategies for the incompressible Navier--Stokes equations are considered in this section and absolute solver tolerances of~$10^{-12}$ as well as relative solver tolerances of~$10^{-6}$ are used for all linear solvers.

The results are compared to accurate DNS reference data of~\cite{Moser99} for~$\mathrm{Re}_{\tau}=180$ and of~\cite{delAlamo2004,Hoyas2008} for~$\mathrm{Re}_{\tau}=950$ denoted as DNS MKM99 and DNS AJZM04, respectively, in the following.  Moreover, we compare our results for the turbulent channel flow problem to the~$\mathrm{AVM}^4$ turbulence model~\cite{Rasthofer2013} and the ALDM model~\cite{Hickel2007} in order to evaluate the accuracy of the presented methods. The~$\mathrm{AVM}^4$ model is a sophisticated turbulence method in the context of variational multiscale methods using a multifractal approach where the spatial discretization is based on continuous, stabilized, low-order finite element methods. The ALDM model is a state-of-the-art implicit LES approach in the context of finite volume schemes.

\subsubsection{Analysis of stability for~$\mathrm{Re}_{\tau}=180$}
In a first step, the stabilization terms discussed in Section~\ref{NumericalLES} are analyzed and terms needed to obtain a robust and accurate discretization scheme for the turbulent channel flow problem are identified. Since the relevant effects and instabilities already occur for small Reynolds numbers, we select a friction Reynolds number of~$\mathrm{Re}_{\tau}=180$ for this analysis. Moreover, since we expect that instabilities particularly arise in the under-resolved regime, we choose a very coarse spatial resolution. The mesh refinement level is~$l=2$ and a polynomial degree of~$k=3$ is used so that the effective mesh resolution in terms of velocity degrees of freedom per spatial dimension is~$(k_u+1)2^l=16$. A Courant number of~$\mathrm{Cr}=1$ is used in case that the discretization is stable but smaller Courant numbers have to be used in case of insufficient discretization schemes that lead to large velocity oscillations in order to avoid conflicts with the CFL condition. The divergence and continuity error measures defined in~\cite{Krank2017}
\begin{align}
\varepsilon_{\mathrm{D}}(\bm{u}_h) = \frac{L\int_{\Omega_h} \vert \Div{\bm{u}_h} \vert \; \mathrm{d}\Omega}{\int_{\Omega_h} \Vert \bm{u}_h \Vert \; \mathrm{d}\Omega} \;\; , \;\; 
\varepsilon_{\mathrm{C}}(\bm{u}_h) = \frac{\int_{\partial\Omega_{h}\setminus\Gamma_h } \vert \jumporiented{\bm{u}_h}\cdot \bm{n} \vert \; \mathrm{d}\Gamma}{ \int_{\partial\Omega_{h}\setminus\Gamma_h }  \vert \avg{\bm{u}_h}\cdot \bm{n} \vert \; \mathrm{d}\Gamma}
\end{align}
have proven effective in analyzing the stability properties for turbulent flows in the context of high-order discontinuous Galerkin methods. In the above equations,~$L$ is a characteristic length scale (given as~$L=\delta$ for the turbulent channel flow problem) and~$\partial\Omega_{h}\setminus\Gamma_h$ denotes all interior element faces.

% USE A TABLE WITH AVERAGED QUANTITIES INSTEAD OF figures SHOWING THE TEMPORAL EVOLUTION OF THE DIVERGENCE AND CONTINUITY ERRORS
\begin{table}[!h]
\caption{Turbulent channel flow for~$\mathrm{Re}_{\tau}=180$: Analysis of divergence error~$\varepsilon_{\mathrm{D}}$ and continuity error~$\varepsilon_{\mathrm{C}}$ (mean values) for different solution strategies. The spatial resolution is~$l=2$ and~$(k_u,k_p)=(3,2)$ corresponding to an effective resolution of~$16^3$ velocity degrees of freedom.}\label{Re180_stability_analysis_divergence_and_continuity_errors}
\renewcommand{\arraystretch}{1.1}
\begin{center}

% continuity penalty term: penalize normal component of velocity only
\begin{tabular}{lllllllll}
\hline
 & \multicolumn{2}{l}{No penalty terms}& &\multicolumn{2}{l}{Div penalty term}& &\multicolumn{2}{l}{Div + conti penalty terms}\\
\cline{2-3} \cline{5-6} \cline{8-9}
 & $\varepsilon_{\mathrm{D}}$ & $\varepsilon_{\mathrm{C}}$ & & $\varepsilon_{\mathrm{D}}$ & $\varepsilon_{\mathrm{C}}$ & & $\varepsilon_{\mathrm{D}}$ & $\varepsilon_{\mathrm{C}}$\\
\hline
Coupled solution approach  & 1.831 & 0.338 & & 0.056 & 0.329 & & 0.021 & 0.015\\
Dual splitting scheme      & 0.472 & 0.154 & & 0.029 & 0.065 & & 0.020 & 0.015\\
Pressure-correction scheme & 1.852 & 0.340 & & 0.056 & 0.332 & & 0.020 & 0.015\\
\hline
\end{tabular}
\end{center}
\renewcommand{\arraystretch}{1}
\end{table}

% continuity penalty term: penalize normal component of velocity only
\begin{figure}[!ht]
 \centering 
 \subfigure[Coupled solution approach]{\includegraphics[width=0.75\textwidth]{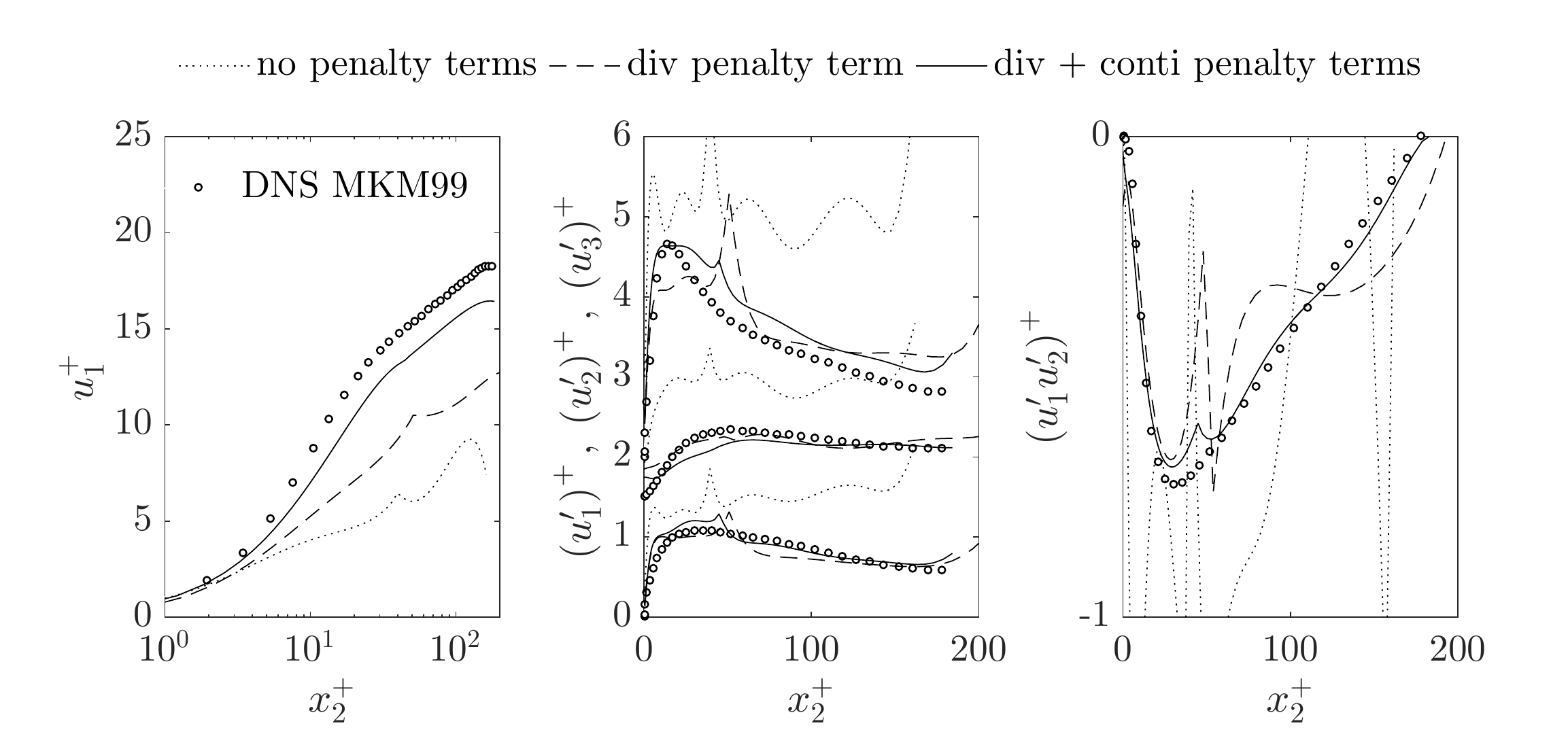}}
 \subfigure[Dual splitting scheme]{\includegraphics[width=0.75\textwidth]{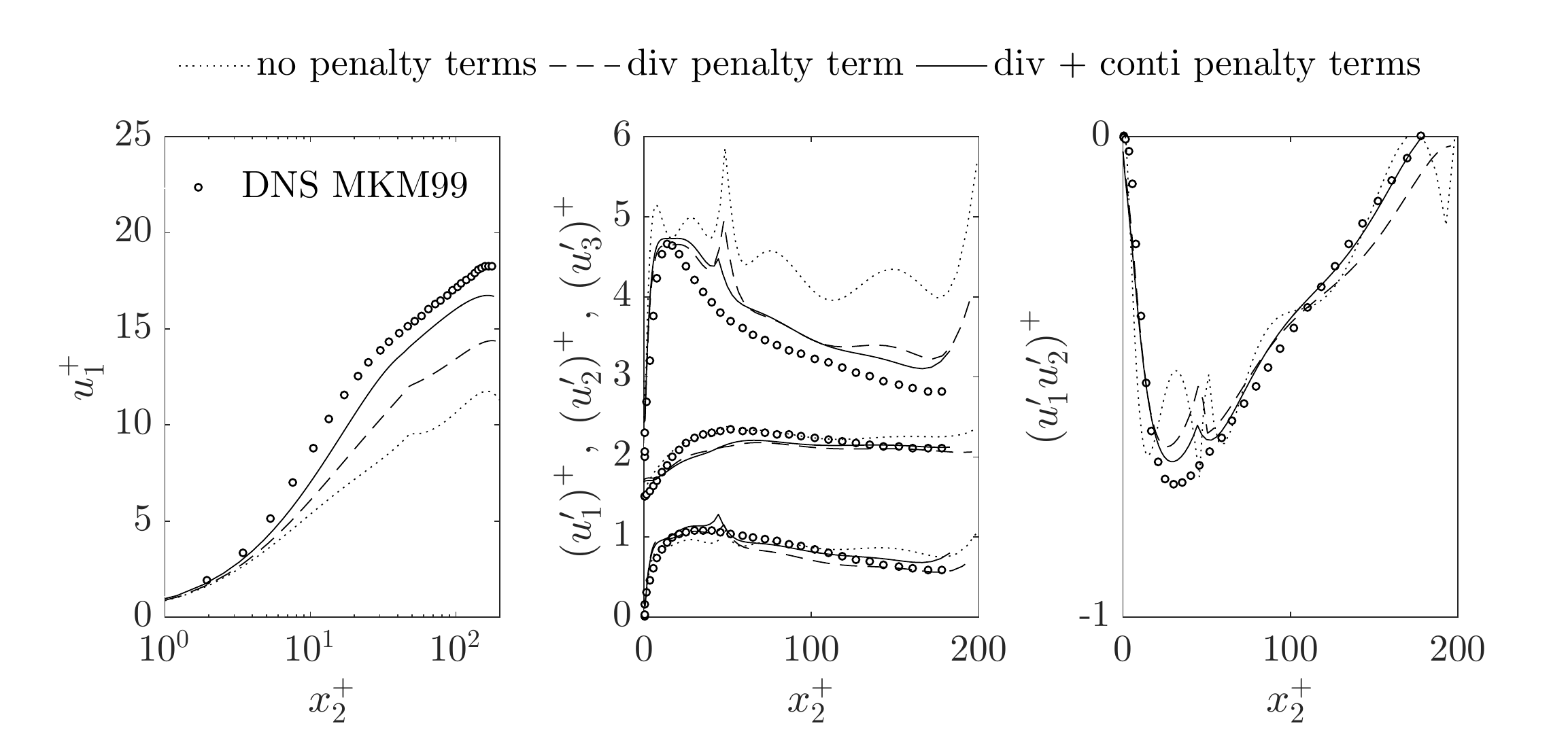}}
 \subfigure[Pressure-correction scheme]{\includegraphics[width=0.75\textwidth]{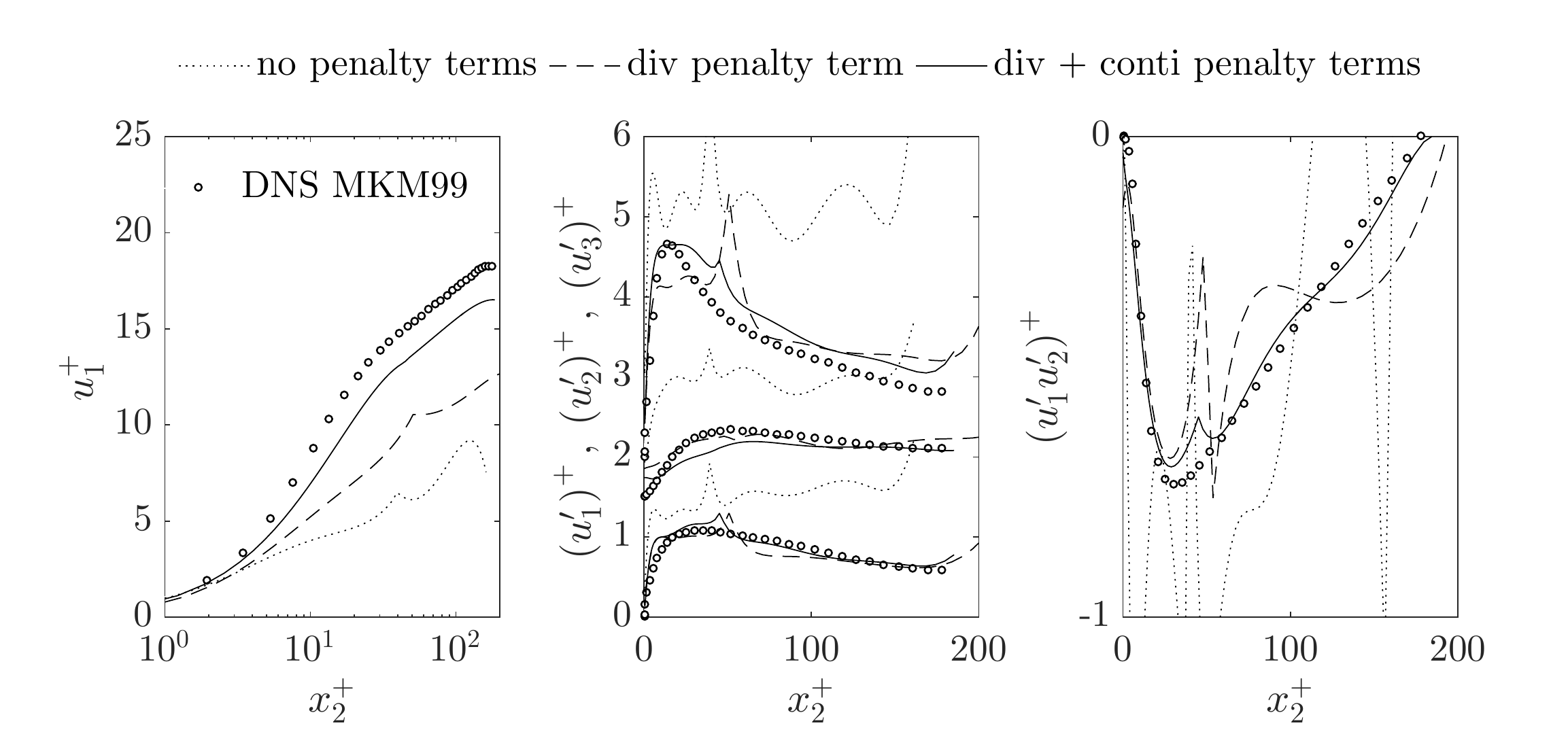}}
\caption{Turbulent channel flow for~$\mathrm{Re}_{\tau}=180$: Mean velocity profile and components of Reynolds stress tensor for different solution strategies. The spatial resolution is~$l=2$ and~$(k_u,k_p)=(3,2)$ corresponding to an effective resolution of~$16^3$ velocity degrees of freedom.}
\label{fig:Re180_stability_analysis_profiles_solution_strategies}
\end{figure}

Time averaged quantities of the divergence and continuity errors are listed in Table~\ref{Re180_stability_analysis_divergence_and_continuity_errors} for different formulations and for different solution strategies such as the coupled solution approach, the dual splitting scheme, and the pressure-correction scheme. The corresponding statistical results for this turbulent flow problem are presented in Figure~\ref{fig:Re180_stability_analysis_profiles_solution_strategies}. As a reference method we consider the standard DG formulation described in Section~\ref{WeakDGFormulation} without additional penalty terms. This formulation shows large divergence and continuity errors of order~$\mathcal{O}(1)$ indicating that this formulation does not lead to a stable and accurate discretization scheme for turbulent flow problems. In fact, the profiles shown in Figure~\ref{fig:Re180_stability_analysis_profiles_solution_strategies} exhibit large oscillations for all solution strategies so that this approach is insufficient for turbulent flow simulations. Including the divergence penalty term significantly reduces the divergence error but the continuity error remains high for the coupled solution approach and the pressure-correction scheme. Interestingly, including the divergence penalty term does not only reduce the divergence error but also the continuity error in case of the dual splitting scheme. However, to obtain small continuity errors for all solution strategies one also has to include the continuity penalty term. Using both penalty terms leads to a robust discretization scheme where the divergence and continuity errors are small for all solution techniques. The errors shown in Table~\ref{Re180_stability_analysis_divergence_and_continuity_errors} are in line with the statistical data presented in Figure~\ref{fig:Re180_stability_analysis_profiles_solution_strategies}. The results for the dual splitting scheme also explain why the purely divergence penalty based approach has been preferred over the postprocessing including both divergence and continuity penalty terms in~\cite{Krank2017} where the dual splitting scheme is used for discretization in time. Given the fact that the spatial resolution is very coarse, the profiles of the mean streamwise velocity, the rms-values, and the Reynolds shear stress are captured very well.

According to these results, we summarize that both the divergence penalty term and the continuity penalty term are identified as necessary ingredients to obtain a stable and accurate discontinuous Galerkin discretization for turbulent flow problems independently of the applied solution strategy. 
%While the continuity penalty term might not be necessary for increasingly fine spatial resolutions, it is an indispensable component in the limit of coarse spatial resolutions. 
For this reason, we only consider formulation~\eqref{WeakForm_CoupledSolver_Divergence_Continuity_Penalty} including both divergence and continuity penalty terms in the following. 

\subsubsection{Convergence test for~$\mathrm{Re}_{\tau}=180$}
\begin{figure}[!ht]
 \centering 
 \subfigure[Visualization of mesh for refinement levels~$l=2,3,4$ (from left to right) with~$4^3,8^3,16^3$ elements, respectively.]{
  \includegraphics[width=0.33\textwidth]{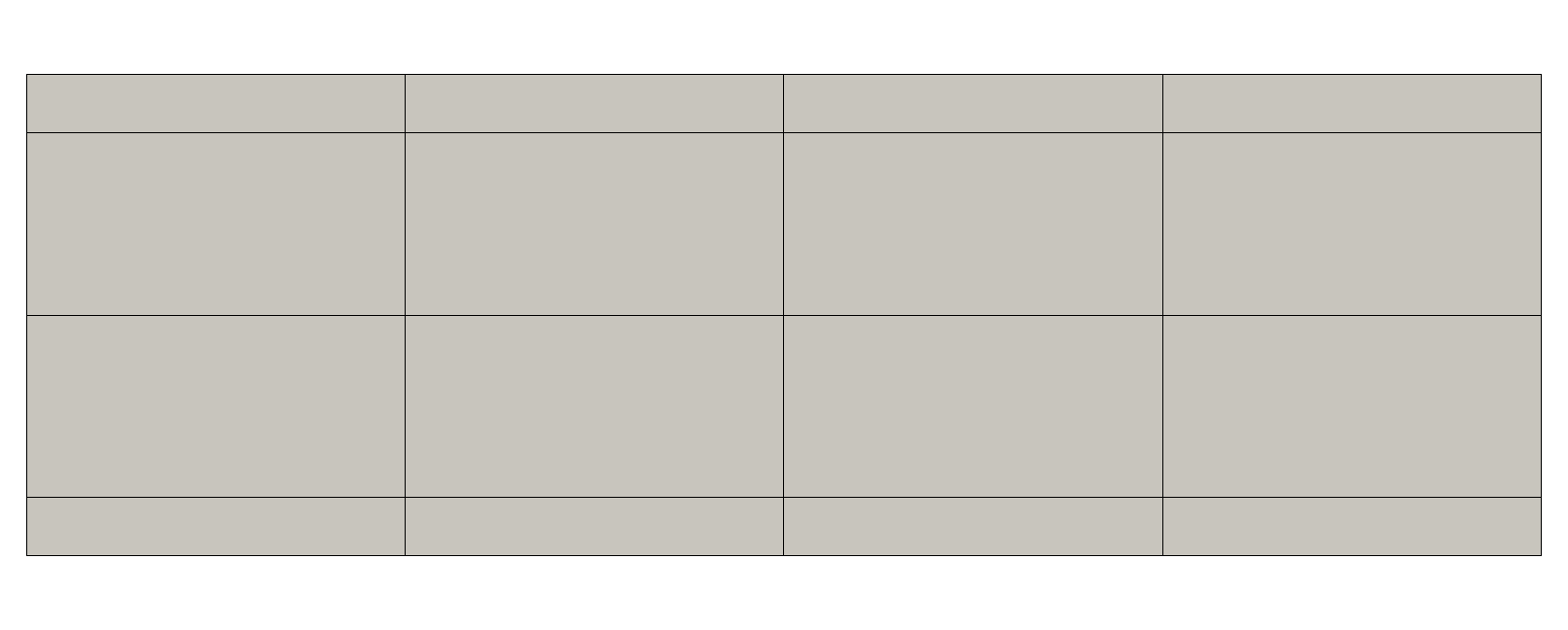}
  \includegraphics[width=0.33\textwidth]{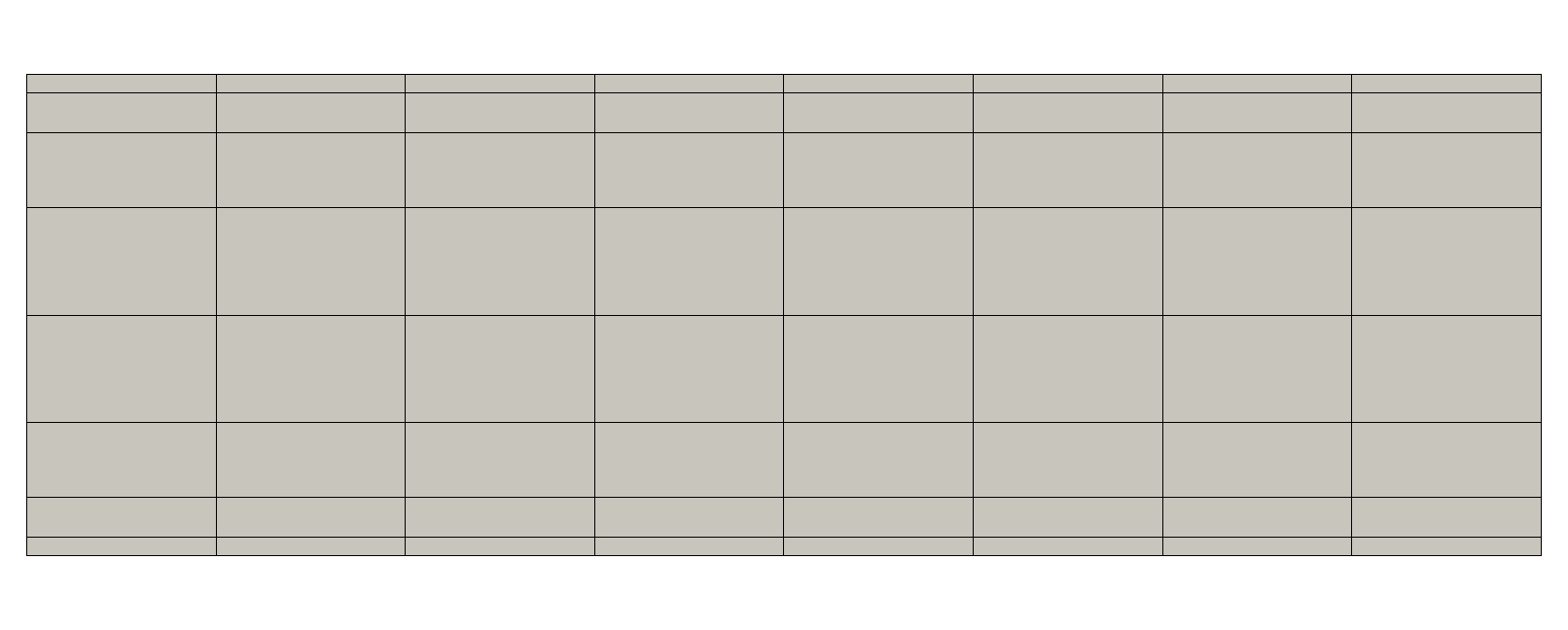}
   \includegraphics[width=0.33\textwidth]{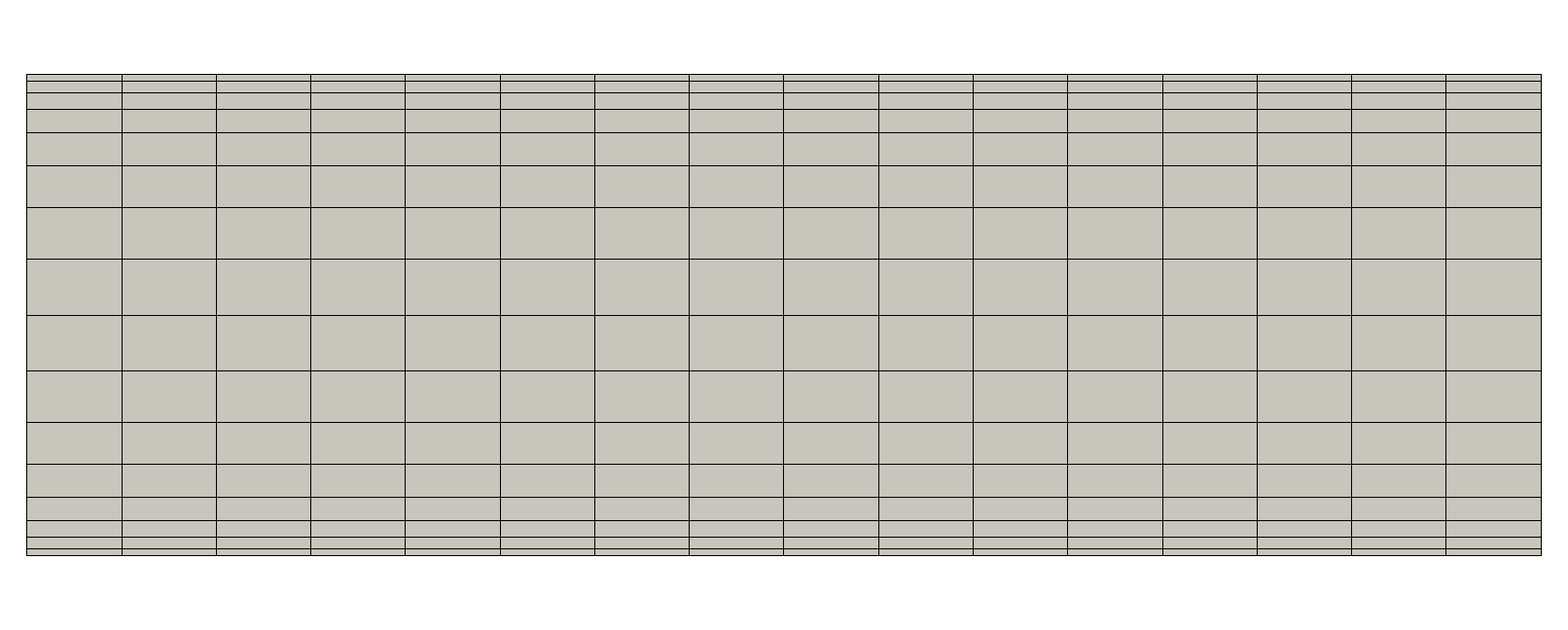}}
 \subfigure[Contour plots of velocity magnitude for refinement levels~$l=2,3,4$ (from left to right) and~$(k_u,k_p)=(3,2)$.]{
  \includegraphics[width=0.33\textwidth]{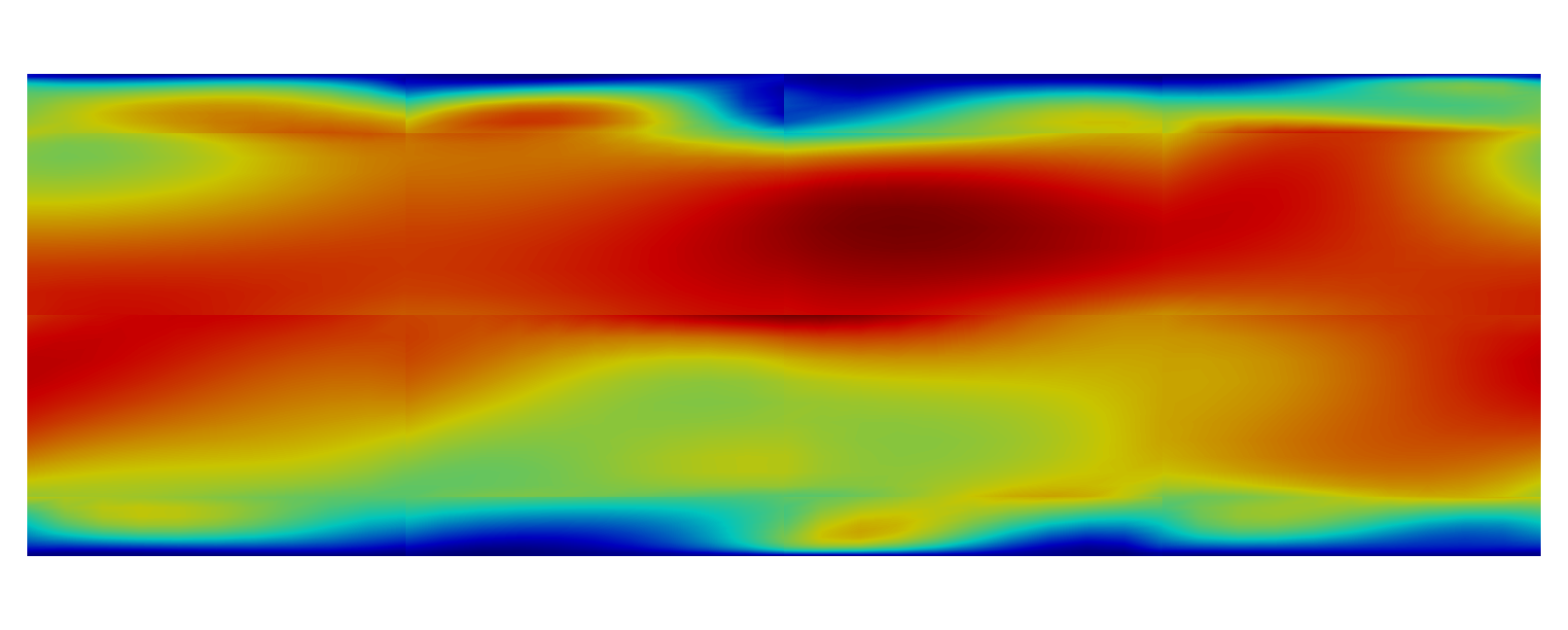}
 \includegraphics[width=0.33\textwidth]{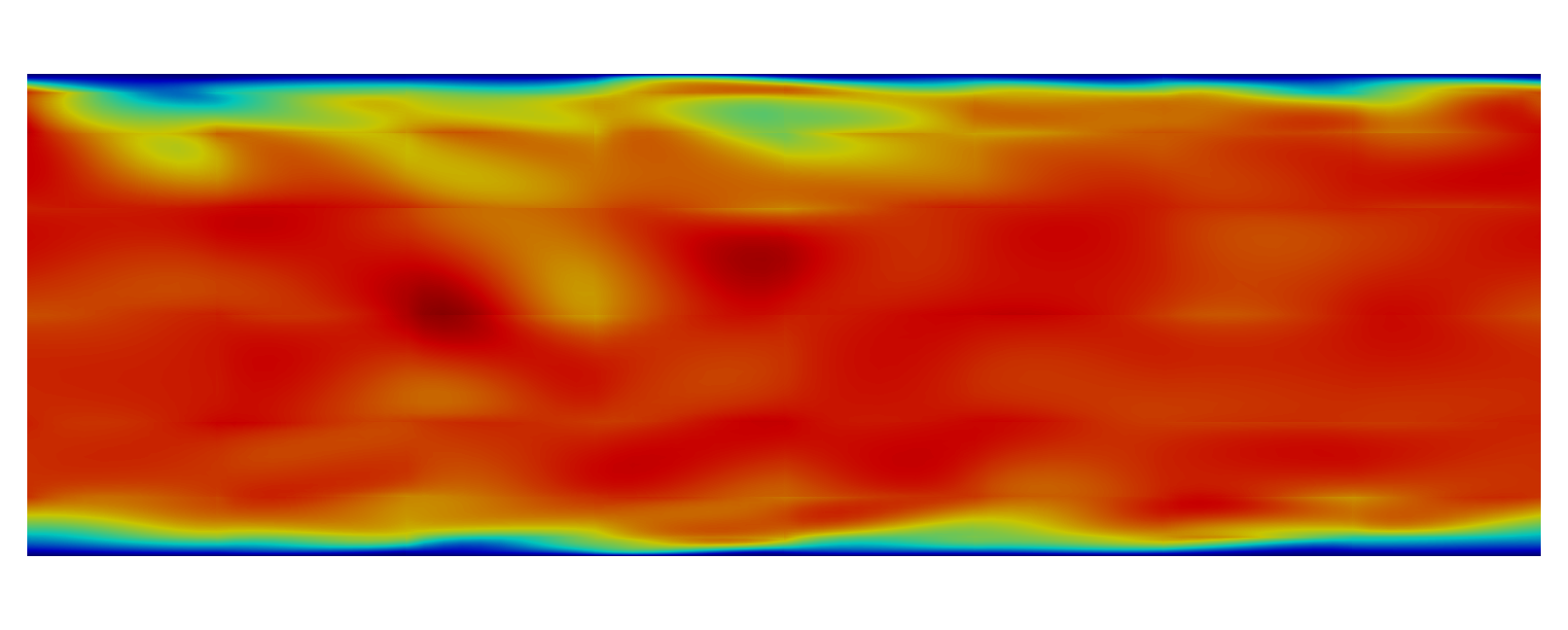}
  \includegraphics[width=0.33\textwidth]{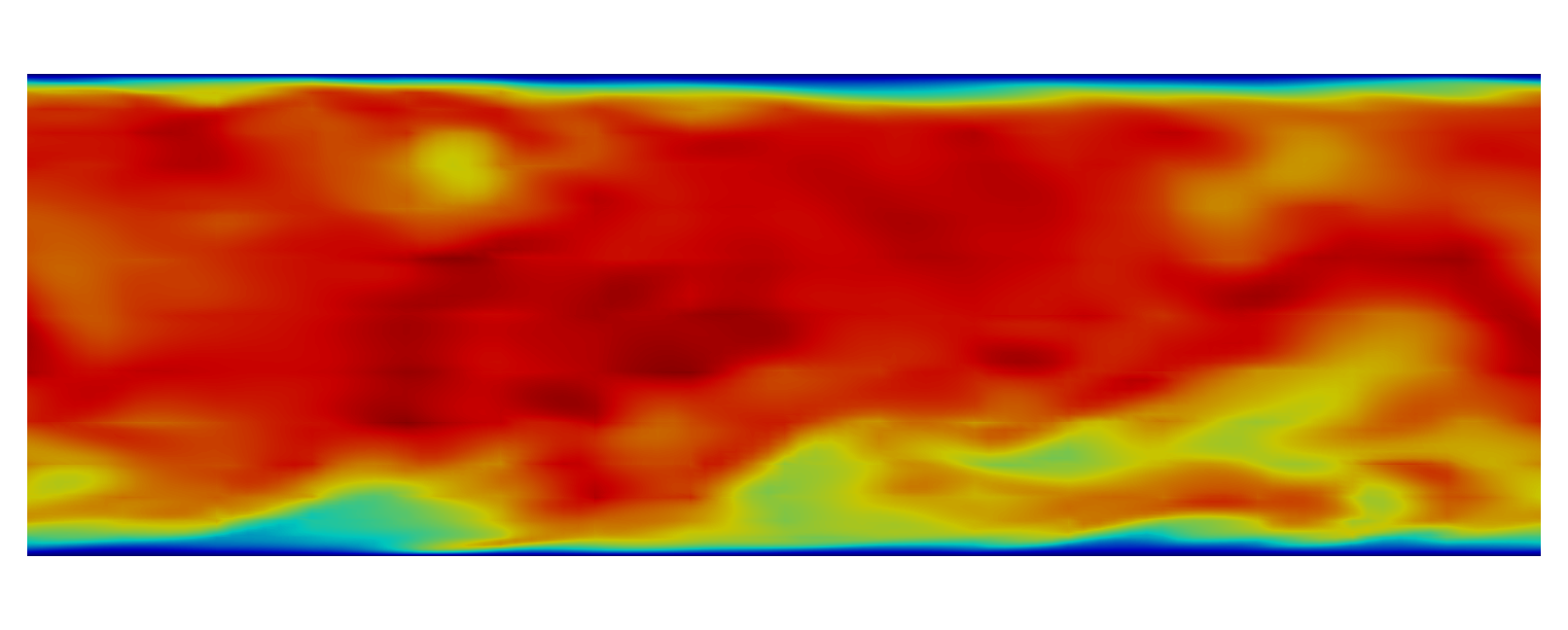}}
\caption{Turbulent channel flow for~$\mathrm{Re}_{\tau}=180$: Visualization of mesh stretching as well as contour plots of instantaneous velocity magnitude for different spatial resolutions. The coupled solution approach is used and the formulation with divergence and continuity penalty terms.}
\label{fig:Re180_mesh_contour_plots}
\end{figure}

We perform an~$h$-convergence test for a Reynolds number of~$\mathrm{Re}_{\tau}=180$ using the coupled solution approach along with the stabilized approach including both divergence and continuity penalty terms. The polynomial degree is~$k=3$ and the refine level is increased from~$l=2$ to~$l=4$ so that the effective resolution in terms of velocity degrees of freedom increases from~$16^3$ to~$64^3$. The mesh as well as contour plots of the velocity magnitude are displayed in Figure~\ref{fig:Re180_mesh_contour_plots} for the different spatial resolutions.

\begin{figure}[!ht]
 \centering 
	\includegraphics[width=1.0\textwidth]{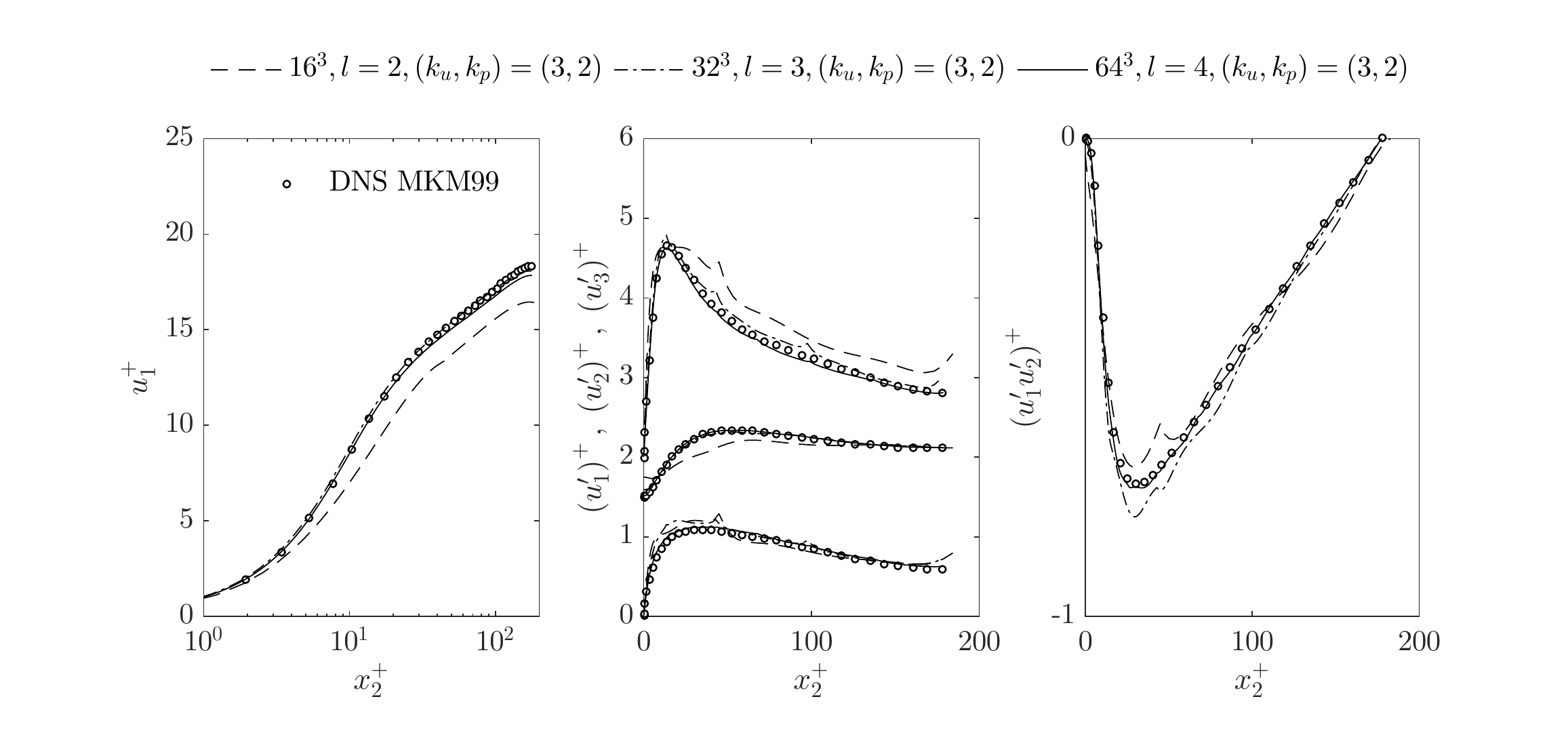}
\caption{Turbulent channel for~$\mathrm{Re}_{\tau}=180$: $h$-convergence test using the coupled solution approach and the formulation with both divergence and continuity penalty terms. The polynomial degree is~$k=3$ and the refinement level is increased from~$l=2$ to~$l=4$ corresponding to effective resolutions of~$16^3$ to~$64^3$ velocity dofs.}
\label{fig:turbulent_channel_Re180_convergence_coupled_solver_div_conti}
\end{figure}

We intentionally include a setup where the resolution of the numerical scheme is insufficient to obtain agreement with reference results in order to show when and how the results degenerate. Quantitative results of this convergence test are shown in Figure~\ref{fig:turbulent_channel_Re180_convergence_coupled_solver_div_conti}. These results demonstrate that the numerical solution tends to the DNS reference solution for increasing spatial resolution. For an effective resolution of~$64^3$ velocity degrees of freedom, the results agree very well with the DNS data and only the mean streamwise velocity is slightly underpredicted for this spatial resolution. The results for~$l=3,4$ appear to be comparable in terms of accuracy as those presented in~\cite{Ramakrishnan2004} where an implicit LES approach is used for DG discretizations of the compressible Navier--Stokes equations and where comparable spatial resolutions are considered for~$\mathrm{Re}_{\tau}=180$. A direct numerical simulation for~$\mathrm{Re}_{\tau}=180$ using a much finer mesh with an effective resolution of~$168^3$ has been performed in~\cite{Wiart15} and a simulation with an effective resolution of~$126\times 132\times 126$ in~\cite{Chapelier2014} for high-order DG discretizations of the compressible Navier--Stokes equations.

\subsubsection{Efficiency of high-order methods for~$\mathrm{Re}_{\tau}=180$}
In this section we investigate the question whether the accuracy of the results obtained for a given number of unknowns can be improved by increasing the polynomial degree of the shape functions. For this analysis, we consider an effective spatial resolution of~$32^3$ velocity degrees of freedom and compare the results obtained for polynomial degrees~$k=3,7,15$. Since the flow is clearly under-resolved for such coarse discretizations it is unclear whether the superior accuracy of high-order methods observed, e.g., for the two-dimensional vortex flow problem in Section~\ref{VortexProblem} and the Orr--Sommerfeld problem in Section~\ref{OrrSommerfeld} can also be observed for the turbulent channel flow problem.

% comparison to AVM^4 and ALDM (Hickel et al.)
\begin{figure}[!ht]
 \centering 
\includegraphics[width=1.0\textwidth]{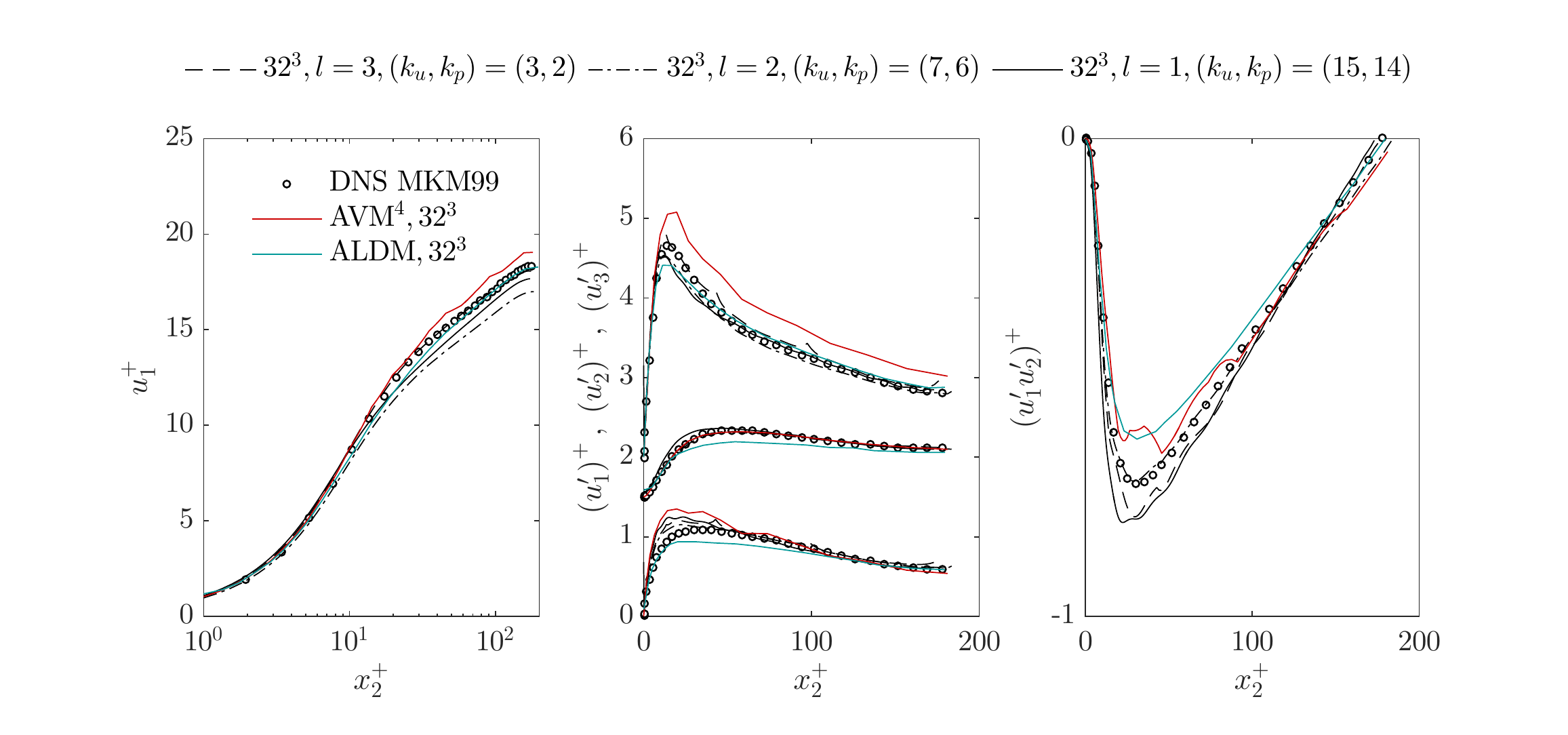}
\caption{Turbulent channel for~$\mathrm{Re}_{\tau}=180$: Investigation of efficiency of high-order methods using the coupled solution approach and the formulation with both divergence and continuity penalty terms. An effective resolution of~$32^3$ velocity degrees of freedom is considered which is realized by the combinations~$l=3$ and $(k_u,k_p)=(3,2)$,~$l=2$ and $(k_u,k_p)=(7,6)$, and~$l=1$ and $(k_u,k_p)=(15,14)$ of refine level~$l$ and polynomial degrees~$(k_u,k_p)$. The results are compared to the~$\mathrm{AVM}^4$ and ALDM turbulence models for the same effective resolution.}
\label{fig:turbulent_channel_Re180_eff_high_order_coupled_solver_div_conti}
\end{figure}

Numerical results for the turbulent channel flow problem at~$\mathrm{Re}_{\tau}=180$ are shown in Figure~\ref{fig:turbulent_channel_Re180_eff_high_order_coupled_solver_div_conti}. For all polynomial degrees, the statistical quantities are predicted correctly and the results achieve a similar level of accuracy as the results for the~$\mathrm{AVM}^4$ and the ALDM turbulence models. However, taking both mean velocity profiles and the fluctuations into account, no clear advantage in terms of accuracy of very high-order methods with polynomial degree~$k=7,15$ as compared to the moderate polynomial degree~$k=3$ can be observed. A possible explanation for this behavior could be that small flow structures are in fact better resolved for high polynomial degrees (as observed for the Taylor--Green vortex problem in Section~\ref{TaylorGreenVortexProblem}, see also the results in~\cite{Gassner2013}) but that this effect is counter-balanced by the increased numerical dissipation of lower order methods such that the macroscopic behavior in terms of statistical quantities of the flow is comparable for polynomial degrees ranging from~$k=3$ to~$k=15$. This aspect requires further investigation and could be subject of future considerations.

\subsubsection{Convergence test for~$\mathrm{Re}_{\tau}=950$}
\begin{figure}[!ht]
 \centering 
\subfigure[ $h$-convergence test for refinement levels~$l=3$ to~$l=5$ and polynomial degree~$k=3$ corresponding to effective resolutions of~$32^3$ to~$128^3$ velocity dofs.]{
	\includegraphics[width=1.0\textwidth]{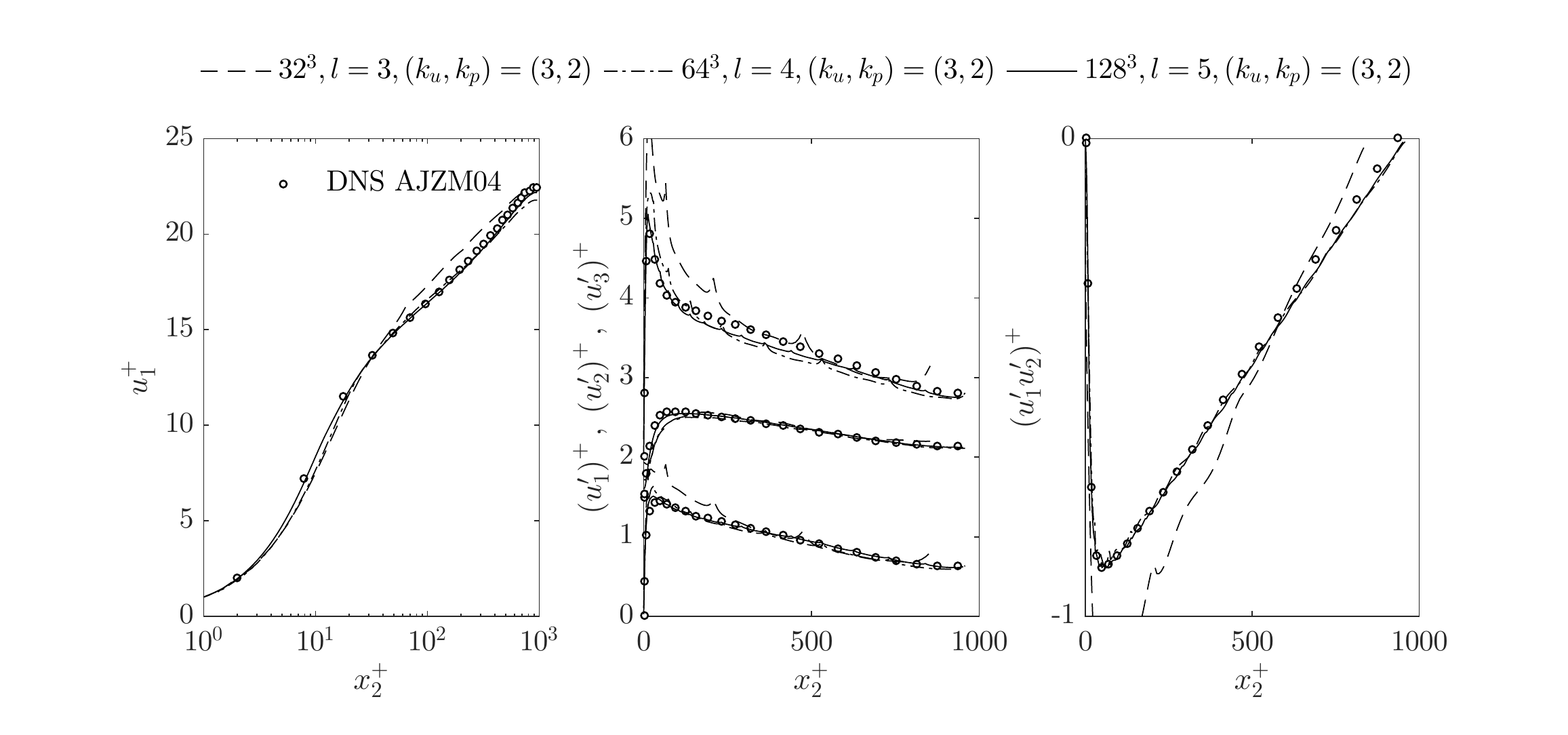}}
\subfigure[Comparison to the~$\mathrm{AVM}^4$ and ALDM turbulence models for the same effective resolution of~$128^3$ dofs.]{
	\includegraphics[width=1.0\textwidth]{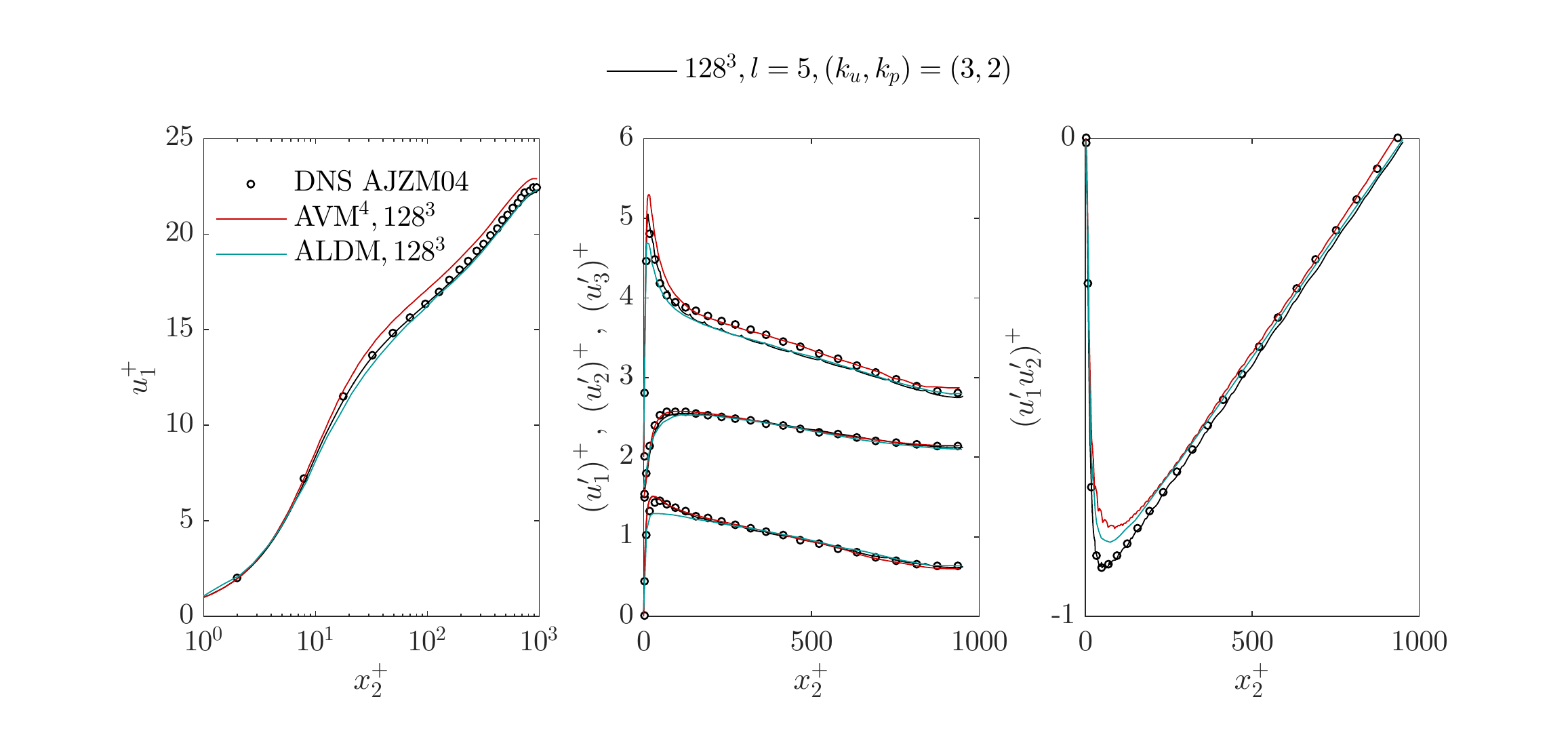}}
\caption{Turbulent channel for~$\mathrm{Re}_{\tau}=950$: Results for coupled solution approach and the formulation with both divergence and continuity penalty terms. An $h$-convergence test is performed and the results are compared to state-of-the-art LES models.}
\label{fig:turbulent_channel_Re950_convergence_coupled_solver_div_conti}
\end{figure}

Finally, we perform simulations of the turbulent channel flow problem for~$\mathrm{Re}_{\tau}=950$. Numerical results of an~$h$-convergence test using polynomial degree~$k=3$ and refinement levels~$l=3$ to~$l=5$ corresponding to effective mesh resolutions of~$32^3$ to~$128^3$ velocity degrees of freedom are presented in Figure~\ref{fig:turbulent_channel_Re950_convergence_coupled_solver_div_conti} where the results are compared to DNS reference data as well as to the~$\mathrm{AVM}^4$ and ALDM turbulence models. As for the~$\mathrm{Re}_{\tau}=180$ case, we include results for a setup where the resolution of the numerical scheme is insufficient in order to demonstrate the accuracy of the scheme in the highly under-resolved regime.

% comparison to AVM^4 and ALDM (Hickel et al.)

The mean streamwise velocity profile is predicted very accurately for all spatial resolutions. While the maxima of~$(u_1^{\prime})^+, (u_3^{\prime})^+$, and~$(u_1^{\prime}u_2^{\prime})^+$ are overpredicted for~$l=3$, the fluctuations are captured very accurately for the finer spatial resolutions~$l=4,5$. Compared to the~$\mathrm{AVM}^4$ and ALDM turbulence models with an effective resolution of~$128^3$ dofs, the present approach achieves a similar level of accuracy or is slightly more accurate than the reference results for~$\mathrm{AVM}^4$ and ALDM. For example, the maxima of the Reynolds stresses are underpredicted by the ALDM model, while the mean velocity is slightly too large for the~$\mathrm{AVM}^4$ model. Often, significantly finer spatial resolutions are required to obtain good agreement with the DNS reference data, see for example the approach that has been proposed recently in~\cite{Li2017} in the context of finite volume methods that automatically adjusts the dissipation of the Roe scheme and where the same Reynolds numbers of~$\mathrm{Re}_{\tau}=180$ and~$950$ are simulated. An effective resolution of~$256\times 128\times 256$ is used in~\cite{Wiart15} for DG discretizations of the compressible Navier--Stokes equations rendering a direct comparison of the results difficult. However, our results for refine level~$l=4$ (effective resolution of~$64^3$) are computed on a mesh that is even coarser than the wall-modelled ILES in~\cite{Wiart15b} where an effective resolution of~$84 \times 80 \times 64$ and polynomials of degree~$k=3$ are used. While the two simulations achieve a similar level of accuracy for the mean streamwise velocity profile, our results are significantly more accurate with respect to the Reynolds stresses.

Based on these result, we conclude that the high-order discontinuous Galerkin discretization along with divergence and continuity penalty terms is a very promising and efficient approach for large-eddy simulation of incompressible flows that is competitive to state-of-the-art LES approaches. The fact that the present approach is a purely numerical approach can be advantageous compared to physically motivated approaches for which model parameters have to be calibrated to a specific flow problem. Unlike the~$\mathrm{AVM}^4$ and ALDM models, our approach does not include a correction term for wall-bounded flows rendering the present methods a generic numerical approach for turbulent flow problems.

\section{Conclusion and outlook}\label{Conclusion}
We have discussed numerical methods stabilizing high-order DG discretizations based on the local Lax--Friedrichs flux for the convective term and the symmetric interior penalty method for the viscous term for under-resolved incompressible flow. The stabilization approach is based on consistent penalty terms added to the weak formulation, i.e., a divergence penalty term enforcing the incompressibility constraint as well as a continuity penalty term enforcing inter-element continuity of the velocity field. This stabilization approach can be interpreted as a purely numerical large-eddy turbulence model. The approach discussed in this work has very attractive properties that can be summarized as follows: Using both divergence and continuity penalty terms, we have shown by means of numerical investigation that a robust turbulent flow solver is obtained that is stable for coarse spatial resolutions. The numerical results and the comparison to state-of-the-art turbulence models has shown that the present approach is also very accurate. All results have been obtained using a fixed penalty factor of~$1$ for the divergence penalty term and the continuity penalty term. Although not explicitly shown here, the results are insensitive with respect to the penalty parameters which can be a major advantage of this approach as compared to physical LES models. Since the approach is based on consistent penalty terms, it is by definition generic, can be applied to arbitrary geometries, and reproduces the exact solution when applied to laminar flow problems. The results for the Taylor--Green vortex problem suggest that the proposed methods are well suited to accurately predict laminar--turbulent transition. Applying the divergence and continuity penalty terms in a post-processing step, the increase in computational costs as compared to a laminar flow solver is moderate, rendering our solver an efficient computational method for under-resolved turbulent flows. A detailed investigation of the computational efficiency of our high-performance matrix-free solver is discussed in a seperate publication~\cite{Fehn18b}.

In the present paper, the analysis is restricted to the local Lax--Friedrichs flux used to discretize the convective term written in divergence formulation. Although it can be expected that the discretization of the convective term is an essential component regarding the numerical dissipation behavior of the discretization scheme, the impact of the DG discretization of the convective term on the accuracy of the method is unclear and requires more detailed investigation. Hence, future work could focus on alternative formulations of the convective term such as the convective formulation or skew-symmetric formulations as well as alternative numerical fluxes for the convective term and their interaction with the stabilization approach.

\appendix
\section{Time integration scheme for projection-type solution strategies}\label{TimeIntegrationProjectionMethods}
In this section we summarize the temporal discretization with velocity-correction schemes (where we consider the high-order dual splitting scheme) and pressure-correction schemes. Both solution techniques are projection methods, i.e., a Poisson equation has to be solved for the pressure while a divergence-free velocity field is obtained by projecting the intermediate velocity field onto the space of divergence-free vectors. For the velocity, an unsteady (convection--)diffusion problem has to be solved. These approaches are attractive since the computational costs per time step might be significantly reduced as compared to a coupled solution approach. For a more detailed discussion of these methods in the context of high-order discontinuous Galerkin discretizations the reader is referred to~\cite{Fehn17}.
\subsection{High-order dual splitting scheme}
Using the high-order dual splitting scheme~\cite{Karniadakis1991}, the incompressible Navier--Stokes equations are split into four substeps, where the convective term, the pressure gradient term, and the viscous term are treated separately in different substeps of the projection scheme
\begin{align}
\frac{\gamma_0\hat{\bm{u}}-\sum_{i=0}^{J-1}\left(\alpha_i\bm{u}^{n-i}\right)}{\Delta t} &= 
- \sum_{i=0}^{J-1}\left(\beta_i \Div{\bm{F}_{\mathrm{c}}\left(\bm{u}^{n-i}\right)}\right)
+ \bm{f}\left(t_{n+1}\right)\; ,\label{DualSplitting_ConvectiveStep}\\
-\nabla^2 p^{n+1} &= -\frac{\gamma_0 }{\Delta t}\Div{\hat{\bm{u}}} \; ,\label{DualSplitting_PressureStep}\\
\hat{\hat{\bm{u}}} &= \hat{\bm{u}} - \frac{\Delta t}{\gamma_0} \Grad{p^{n+1}}\; ,\label{DualSplitting_ProjectionStep}\\
\frac{\gamma_0 }{\Delta t} \bm{u}^{n+1}  -  \Div{\bm{F}_{\mathrm{v}}\left(\bm{u}^{n+1}\right)} &=
\frac{\gamma_0 }{\Delta t}\hat{\hat{\bm{u}}} \; .\label{DualSplitting_ViscousStep}
\end{align}
An intermediate velocity field~$\hat{\bm{u}}$ is calculated in the first substep where the convective term and the body force term form the right-hand side of the equation. The pressure~$p^{n+1}$ is calculated in the second substep by solving a Poisson equation. A second intermediate velocity~$\hat{\hat{\bm{u}}}$ is obtained by projecting~$\hat{\bm{u}}$ onto the space of divergence-free vectors. Finally, the viscous term is considered and the final velocity~$\bm{u}^{n+1}$ is obtained by solving a Helmholtz-like equation. 

\subsection{Pressure-correction scheme}
For the pressure-correction scheme~\cite{Guermond2004}, the solution of each time step consists of the following four substeps, where the convective term and viscous term are treated in the same substep
\begin{align}
\frac{\gamma_0\hat{\bm{u}}-\sum_{i=0}^{J-1}\left(\alpha_i\bm{u}^{n-i}\right)}{\Delta t} 
-\Div{\bm{F}_{\mathrm{v}}\left(\hat{\bm{u}}\right)} &= 
- \sum_{i=0}^{J-1}\left(\beta_i \Div{\bm{F}_{\mathrm{c}}\left(\bm{u}^{n-i}\right)}\right)
- \sum_{i=0}^{J_p-1}\left(\beta_i \Grad{p^{n-i}}\right)
+ \bm{f}\left(t_{n+1}\right)\; ,\label{PressureCorrection_MomentumStep}\\
-\nabla^2 \phi^{n+1} &= -\frac{\gamma_0 }{\Delta t}\Div{\hat{\bm{u}}} \; ,\label{PressureCorrection_PressurePoissonEquation}\\
p^{n+1} &= \phi^{n+1} + \sum_{i=0}^{J_p-1}\left(\beta_i p^{n-i}\right) - \chi \nu \Div{\hat{\bm{u}}}\; ,\label{PressureCorrection_PressureUpdate}\\
\bm{u}^{n+1} &= \hat{\bm{u}} - \frac{\Delta t}{\gamma_0} \Grad{\phi^{n+1}}\; .\label{PressureCorrection_Projection}
\end{align}
In the first substep, equation~\eqref{PressureCorrection_MomentumStep}, an intermediate velocity field~$\hat{\bm{u}}$ is obtained by solving the momentum equation, where an extrapolation of the pressure gradient term is used on the right-hand side in case of the incremental pressure-correction scheme,~$J_p\geq 1$. Subsequently, a Poisson equation has to be solved for the pressure increment~$\phi^{n+1}$ in equation~\eqref{PressureCorrection_PressurePoissonEquation}. Equation~\eqref{PressureCorrection_PressureUpdate} shows that~$\phi^{n+1}$ represents the pressure increment with an additional divergence-correction term in case of the rotational formulation of the algorithm,~$\chi=1$, while this term is omitted for the standard formulation,~$\chi=0$. Finally, the velocity is projected onto the space of divergence-free vectors in equation~\eqref{PressureCorrection_Projection}. In the present work, the incremental~($J_p=1$) pressure-correction scheme in rotational form~($\chi=1$) is used.

\section{Weak discontinuous Galerkin formulation for projection-type solution strategies}\label{WeakDGFormulationProjectionMethods}
In this section, we present the weak discontinuous Galerkin discretization for the high-order dual splitting scheme and the pressure-correction scheme including the proposed divergence and continuity penalty terms. For a more detailed discussion, e.g., regarding the imposition of boundary conditions, the reader is referred to~\cite{Fehn17}.
\subsection{High-order dual splitting scheme}\label{WeakFormDualSplitting}
For the dual splitting scheme, the divergence and continuity penalty terms are applied in the projection step~\eqref{DualSplitting_ProjectionStep} since~$\hat{\hat{\bm{u}}}$ is the velocity that should be divergence-free, see also~\cite{Krank2017}. We arrive at the following weak DG formulation: Find~$\hat{\bm{u}}_h,\hat{\hat{\bm{u}}}_h ,\bm{u}_h^{n+1}\in\mathcal{V}^u_h$ and~$p_h^{n+1}\in\mathcal{V}^p_h$ such that for all~$\bm{v}_h \in \mathcal{V}^{u}_{h,e}$,~$q_h \in \mathcal{V}^{p}_{h,e}$ and for all elements~$e=1,...,N_{\text{el}}$
\begin{align}
m^{e}_{h,u}\left(\bm{v}_h,\frac{\gamma_0 \hat{\bm{u}}_h-\sum_{i=0}^{J-1}\left(\alpha_i\bm{u}^{n-i}_h\right)}{\Delta t} \right)
&= 
- \sum_{i=0}^{J-1} \left(\beta_i c^e_h\left(\bm{v}_h,\bm{u}^{n-i}_h\right)\right)
+ \intele{\bm{v}_h}{\bm{f}(t_{n+1})} \; ,
\label{DualSplitting_ConvectiveStep_WeakForm}\\
l_{h}^{e}\left(q_h,p_h^{n+1}\right) &= - \frac{\gamma_0}{\Delta t} d_{h}^{e}\left(q_h,\hat{\bm{u}}_h\right)
\; ,
\label{DualSplitting_Pressure_WeakForm}\\
m_{h,u}^{e}(\bm{v}_h,\hat{\hat{\bm{u}}}_h)+ a^e_{\mathrm{D}}(\bm{v}_h,\hat{\hat{\bm{u}}}_h) + a^e_{\mathrm{C}}(\bm{v}_h,\hat{\hat{\bm{u}}}_h)  &= m_{h,u}^{e}\left(\bm{v}_h,\hat{\bm{u}}_h\right)-\frac{\Delta t}{\gamma_0}g_h^{e}\left(\bm{v}_h,p_h^{n+1}\right)\; ,\label{DualSplitting_Projection_WeakForm}\\
m^{e}_{h,u}\left(\bm{v}_h,\frac{\gamma_0}{\Delta t} \bm{u}_h^{n+1} \right) 
+ v^{e}_{h}\left(\bm{v}_h,\bm{u}_h^{n+1}\right)
&= 
m^{e}_{h,u}\left(\bm{v}_h,\frac{\gamma_0}{\Delta t}\hat{\hat{\bm{u}}}_h \right)
\; .
\label{DualSplitting_ViscousStep_WeakForm}
\end{align}
In the above equations,~$l_{h}^{e}$ denotes the discontinuous Galerkin discretization of the negative Laplace operator which is given as
\begin{align}
l_h^e\left(q_h,p_h\right) = \intele{\Grad{q_h}}{\Grad{p_h}}
-\inteleface{\Grad{q_h}}{\frac{1}{2}\jump{p_h}}
- \inteleface{q_h}{\avg{\Grad{p_h}}\cdot\bm{n}}
+ \inteleface{q_h}{\tau\jump{p_h}\cdot\bm{n}}	\; .\label{WeakFormulationLaplace}
\end{align}
The weak formulation of the velocity mass matrix operator~$m^e_{h,u}\left(\bm{v}_h,\bm{u}_h\right)$, the convective term~$c^e_h\left(\bm{v}_h,\bm{u}_h\right)$, the viscous term~$v^e_h\left(\bm{v}_h,\bm{u}_h\right)$, the pressure gradient term~$g^e_h\left(\bm{v}_h,p_h\right)$, and the velocity divergence term~$d^e_h\left(q_h,\bm{u}_h\right)$ is specified in Section~\ref{WeakDGFormulation}. The divergence penalty term~$a^e_{\mathrm{D}}(\bm{v}_h,\bm{u}_h)$ and the continuity penalty term~$a^e_{\mathrm{C}}(\bm{v}_h,\bm{u}_h)$ are described in Section~\ref{DivAndContiPenaltyTerms}.

\subsection{Pressure-correction scheme}\label{PressureCorrectionWeakForm}
As for the dual splitting scheme, the divergence and continuity penalty terms are added to the projection step of the splitting scheme to obtain the weak formulation: Find~$\hat{\bm{u}}_h,\bm{u}_h^{n+1}\in\mathcal{V}^u_h$ and~$\phi_h^{n+1},p_h^{n+1}\in\mathcal{V}^p_h$  such that for all~$\bm{v}_h \in \mathcal{V}^{u}_{h,e}$,~$q_h \in \mathcal{V}^{p}_{h,e}$ and for all elements~$e=1,...,N_{\text{el}}$
\begin{align}
\begin{split}
m^{e}_{h,u}\left(\bm{v}_h,\frac{\gamma_0 \hat{\bm{u}}_h-\sum_{i=0}^{J-1}\left(\alpha_i\bm{u}^{n-i}_h\right)}{\Delta t} \right)
+ v^e_h\left(\bm{v}_h,\hat{\bm{u}}_h\right)\;\;\;\; &\\
+ \sum_{i=0}^{J_p-1} \left(\beta_i g^e_h\left(\bm{v}_h,p^{n-i}_h\right) \right)
 =& - \sum_{i=0}^{J-1} \left(\beta_i c^e_h\left(\bm{v}_h,\bm{u}^{n-i}_h\right)\right) + \intele{\bm{v}_h}{\bm{f}(t_{n+1})}  \; ,
\end{split} \label{PressureCorrection_MomentumImplicit_Nonlinear_WeakForm}\\
l_{h}^{e}\left(q_h,\phi_h^{n+1}\right) =&
- \frac{\gamma_0}{\Delta t} d_{h}^{e}\left(q_h,\hat{\bm{u}}_h\right)\; ,
\label{PressureCorrection_PressureStep_WeakForm}\\
\begin{split}
m_{h,p}^{e}\left(q_h,p_h^{n+1}\right) =& + m_{h,p}^{e}\left(q_h,\phi_h^{n+1} + \sum_{i=0}^{J_p-1}\left(\beta_i p_h^{n-i}\right)\right) \\
&- \chi \nu\; d_{h}^{e}\left(q_h,\hat{\bm{u}}_h\right) \; ,
\end{split}
\label{PressureCorrection_PressureUpdate_WeakForm}\\
m_{h,u}^{e}\left(\bm{v}_h,\bm{u}_h^{n+1}\right) + a^e_{\mathrm{D}}(\bm{v}_h,\bm{u}^{n+1}_h) + a^e_{\mathrm{C}}(\bm{v}_h,\bm{u}^{n+1}_h) 
=& + m_{h,u}^{e}\left(\bm{v}_h,\hat{\bm{u}}_h\right)-\frac{\Delta t}{\gamma_0}g_h^{e}\left(\bm{v}_h,\phi_h^{n+1}\right)\; .\label{PressureCorrection_Projection_WeakForm}
\end{align}
The weak discontinuous Galerkin formulation of the pressure mass matrix term is denoted as~$m_{h,p}^{e}$ and is given as
\begin{align}
m_{h,p}^{e}\left(q_h,p_h\right) = \intele{q_h}{p_h} \; .
\end{align}
The weak formulation of the negative Laplace operator is specified in equation~\eqref{WeakFormulationLaplace}. Moreover, the weak formulation of the velocity mass matrix operator~$m^e_{h,u}\left(\bm{v}_h,\bm{u}_h\right)$, the convective term~$c^e_h\left(\bm{v}_h,\bm{u}_h\right)$, the viscous term~$v^e_h\left(\bm{v}_h,\bm{u}_h\right)$, the pressure gradient term~$g^e_h\left(\bm{v}_h,p_h\right)$, and the velocity divergence term~$d^e_h\left(q_h,\bm{u}_h\right)$ is specified in Section~\ref{WeakDGFormulation}. The divergence penalty term~$a^e_{\mathrm{D}}(\bm{v}_h,\bm{u}_h)$ and the continuity penalty term~$a^e_{\mathrm{C}}(\bm{v}_h,\bm{u}_h)$ are described in Section~\ref{DivAndContiPenaltyTerms}.

\section*{Acknowledgments}
The research presented in this paper was partly funded by the German Research Foundation (DFG) under the project ``High-order discontinuous Galerkin for the EXA-scale'' (ExaDG) within the priority program ``Software for Exascale Computing'' (SPPEXA), grant agreement no. KR4661/2-1 and WA1521/18-1.

\bibliography{paper}

\end{document}